\documentclass[11pt]{article}

\usepackage{jcappub}
\usepackage[T1]{fontenc} 
\usepackage{amssymb}
\usepackage{graphicx}                            
\usepackage{amsmath} 
\usepackage{perpage} 
\usepackage{color}
\usepackage{cleveref}

\newcommand{\sech}{\mathop{\rm sech}\nolimits}

\newcommand{\ket}[1]{\left|#1\right\rangle}

\newcommand{\abs}[1]{\left| #1 \right|} 

\newcommand{\ev}{\epsilon_v}
\newcommand{\ed}{\epsilon_\delta}
\newcommand{\eD}{\epsilon_\Delta}
\newcommand{\ephi}{\epsilon_\phi}
\newcommand*\colvec[3][]{\begin{pmatrix}\ifx\relax#1\relax\else#1\\\fi#2\\#3\end{pmatrix}}

\newcommand{\beq}{\begin{equation}}
\newcommand{\beqn}{\begin{eqnarray}}
\newcommand{\eeq}{\end{equation}}
\newcommand{\eeqn}{\end{eqnarray}}
\subheader{MIT-CTP 4585}
\title{\boldmath Self-Scattering for Dark Matter with an Excited State}
\author[ a,\,b]{Katelin Schutz,}
\author[\,a]{Tracy R. Slatyer}
\affiliation[a]{Center for Theoretical Physics, Massachusetts Institute of Technology\\ Cambridge, MA 02139, USA }
\affiliation[b]{Department of Physics, UC Berkeley\\  Berkeley, CA 94720, USA}

\emailAdd{kschutz@berkeley.edu}
\emailAdd{tslatyer@mit.edu}
\abstract{Self-interacting dark matter scenarios have recently attracted much attention, as a possible means to alleviate the tension between N-body simulations and observations of the dark matter distribution on galactic and sub-galactic scales. The presence of internal structure for the dark matter --- for example, a nearly-degenerate state in the spectrum that could decay, or be collisionally excited or de-excited --- has also been proposed as a possible means to address these discrepancies. Such internal structure can be a source of interesting signatures in direct and indirect dark matter searches, for example providing a novel explanation for the 3.5 keV line recently observed in galaxies and galaxy clusters. We analyze a simple model of dark matter self-scattering including a nearly-degenerate excited state, and develop an accurate analytic approximation for the elastic and inelastic $s$-wave cross sections, which is valid outside the perturbative regime provided the particle velocity is sufficiently low (this condition is also required for the $s$-wave to dominate over higher partial waves). We anticipate our results will be useful in incorporating inelastic self-scattering into N-body simulations, in order to study the quantitative impact of nearly-degenerate states in the dark matter spectrum on galactic structure and dynamics, and in computing the indirect signatures of multi-state dark matter.}  
\begin{document}
\maketitle
\flushbottom
\section{Introduction}
The verification of the existence of dark matter (DM) on wildly disparate scales is one of the greatest triumphs of modern astrophysics. It also poses one of the greatest outstanding questions of modern particle physics, due to the lack of a viable DM candidate within the Standard Model (SM). As one of the most promising potential windows onto new fundamental physics, there are many efforts underway to probe the particle nature of DM and its interactions with the SM, but to date it has only been detected through its gravitational properties.

The formation and dynamics of DM structures can be highly sensitive to the microphysics of DM. There are stringent constraints on DM interactions with SM particles, but the interactions of DM within its own ``dark sector'' (at most weakly coupled to the SM) are far less constrained. Increasingly detailed probes of the distribution of dark matter in halos may allow us to explore and constrain such interactions, independent of the coupling of the dark sector to known particles. It is therefore important to understand the potential observational signatures of dark sector physics.

Already, several apparent discrepancies have been identified between observations and the predictions of collisionless cold DM (CCDM) simulations. These include the ``core-cusp'' problem \cite{2010AdAst2010E...5D,Oh:2010mc, Walker:2011zu}, the ``missing satellite'' problem \cite{1999ApJ...522...82K, 1999ApJ...524L..19M, 2007ApJ...657..262D, 2008MNRAS.391.1685S}, and the ``too big to fail'' problem \cite{BoylanKolchin:2011de, 2014arXiv1404.5313G}. In general, observed dwarf galaxies appear to be less abundant, less massive and less centrally concentrated than predicted from CCDM simulations. 

It is possible these discrepancies reflect the gravitational influence of baryonic matter, which is not included in CCDM-only simulations \cite{Governato:2012fa, 2014ApJ...786...87B, Zolotov:2012xd, Brooks:2012ah, Arraki21022014, DelPopolo:2014yta}. However, whether baryonic effects can explain all the observations remains an open question, and the discrepancies might also be clues to DM microphysics. DM self-interaction (via some novel dark sector physics) can alleviate the discrepancies by increasing the scattering of DM particles out of the dense central regions of halos \cite{2000PhRvL..84.3760S}; recent simulations have confirmed the efficacy of this process \cite{Vogelsberger:2012ku, Zavala:2012us, 2013MNRAS.430..105P, Rocha:2012jg}. Consistency with existing limits on dark matter self-interaction from large scale structures (which have much greater virial velocities than dwarf galaxies) is most easily achieved if the scattering cross-section has a velocity dependence (growing larger at low velocities); however, constant cross sections in the range $\sigma/m_\chi \sim 0.1-1$ cm$^2$/g remain viable \cite{Zavala:2012us}.

Nearly-degenerate excited states are another simple modification to DM physics with potentially striking observational signatures, especially in the presence of non-negligible self-interaction. Exothermic scatterings from an excited state would deliver velocity ``kicks'' to DM particles, which could be comparable to or larger than the escape velocity in bound DM structures, especially in the slow-moving environs of dwarf galaxies. In this way inelastic DM scattering can dilute dense cusps or dissipate dwarf halos entirely \cite{Loeb:2010gj}. (Velocity kicks from late-time \emph{decays} of a metastable excited state have been considered in e.g. \cite{SanchezSalcedo:2003pb, Abdelqader:2008wa, Peter:2010au, Peter:2010sz, Peter:2010jy, Bell:2010qt}.) Inelastic DM scattering from either DM or baryonic matter, exothermic or endothermic, can also have interesting signatures in direct and indirect searches for DM (e.g. \cite{Smith:2001hy, Graham:2010ca, McCullough:2013jma} for direct detection, \cite{Finkbeiner:2007kk,Cline:2010kv, Finkbeiner:2014sja} for indirect detection).

Such scenarios are easily constructed from a theoretical perspective. If DM is charged under some new dark gauge symmetry which is broken, then the states in the dark matter multiplet can naturally acquire a small mass splitting, regardless of whether the gauge group is Abelian or non-Abelian (e.g. \cite{ArkaniHamed:2008qn, Baumgart:2009tn, Cheung:2009qd, Chen:2009ab}). In any case where the mediator of the DM self-interaction is light, the scattering cross section is automatically enhanced at low velocities, ensuring a larger impact on dwarf galaxies relative to Milky-Way-size galaxies and clusters.

Previous studies of self-interacting DM (both analytic and simulation-based) have focused on the case where only degenerate dark matter states participate in the interaction (i.e. either DM is a single Majorana fermion, or a Dirac fermion with particle-particle and particle-antiparticle scatterings). Even in situations where the scattering is purely elastic (i.e. there is no transition to a state of different mass), as we will show in this article, the presence of a nearly-degenerate state in the spectrum can significantly modify the large resonances present in the low-velocity scattering rate. However, the addition of a second state adds at least one additional parameter to the problem (the mass splitting), making numerical analysis computationally expensive and, in parts of parameter space, unstable. 

The main goal of this article is to work out an analytic approximation for the DM-DM scattering cross section induced by an off-diagonal Yukawa interaction, in the presence of a single excited state. This corresponds to the case where the dark gauge group is $U(1)$ and provides a simple and illustrative toy model for inelastic dark matter self-scattering more generally. As we will show, our expressions give good agreement with numerically solving the appropriate multi-state Schr{\"o}dinger equation, but are very quick to compute and provide intuitive insight into the scattering behavior. Our results are applicable to the low-velocity regime where $s$-wave scattering dominates the total cross section; we defer study of the ``classical'' regime, where many partial waves contribute, to later work. We identify regions of parameter space with particular relevance to dwarf-galaxy-sized halos, and consequently to the discrepancies described above. We also determine regions of parameter space which could potentially provide a DM explanation for the observed 3.5 keV spectral line from clusters \cite{Bulbul:2014sua, Boyarsky:2014jta} via collisional excitation followed by decay, as in the ``XrayDM'' scenario \cite{Finkbeiner:2014sja}.

We begin, in Section~\ref{sec:model}, by describing our toy model and approach to computing the scattering cross sections (the method follows that of \cite{Slatyer:2009vg}). We discuss our analytic approximate scattering cross sections, and their behavior in several limiting regimes of interest, in Section~\ref{sec:scat}. We simultaneously verify the validity of our approximate solutions by direct comparison to the numerical results. In Section \ref{astropheno} we provide first simple estimates of the parameter space where such scatterings could affect the internal structure and dynamics of dwarf galaxies, or yield the appropriate upscattering cross section to account for the 3.5 keV line in the XrayDM scenario. Concluding remarks follow in Section \ref{conclusions}.
\section{Dark Matter with Inelastic Scattering}
\label{sec:model}
\subsection{A Simple Model}
\label{sec:modelintro}

We consider the case of a Yukawa-like interaction coupling two states with some small mass splitting, $\delta$. We take the dark matter to be a pseudo-Dirac fermion charged under a dark $U(1)$. At high energies, where the $U(1)$ symmetry is unbroken, the dark matter is a charged Dirac fermion; at low energies, a small Majorana mass splits the Dirac fermion into two nearly-degenerate Majorana states $\chi_1$ and $\chi_2$. Since these states cannot carry conserved charge, their couplings with the $U(1)$ vector boson $A_D$ are purely off-diagonal: that is, there is no vertex of the schematic form $\chi_1 \chi_1 A_D$ or $\chi_2 \chi_2 A_D$, only $\chi_1 \chi_2 A_D$.

We will not specify the high-energy physics that gives rise to the Majorana masses, since we restrict ourselves to the extremely non-relativistic environment relevant for present-day DM scattering. The phenomenology of such models in the context of dark matter annihilation and inelastic scattering has been discussed in e.g. \cite{Smith:2001hy, Finkbeiner:2007kk, ArkaniHamed:2008qn, Slatyer:2009vg, Finkbeiner:2010sm}. Inelastic scattering between states with small mass splittings can also naturally be generated in other contexts, for example composite dark matter \cite{Alves:2009nf} and sneutrino dark matter \cite{An:2011uq}.

Since we work always in the non-relativistic limit, the interaction between the DM-like states can be characterized entirely in terms of a matrix potential that couples two-body states. The scattering cross sections can then be calculated using the methods of quantum mechanics (we refer the reader to e.g. \cite{Cassel:2009wt} for a  field-theoretic discussion using the Bethe-Salpeter formalism, and \cite{Hisano:2004ds} for a derivation using effective field theory). The potential matrix coupling the $\ket{11}$ and $\ket{22}$ two-body states (corresponding to both particles being in the ground state or both particles being in the excited state) is \cite{Slatyer:2009vg}: 
\begin{equation} \mathcal{V}(r) = 
\left( \begin{matrix}
0 & -\hbar c \alpha \frac{e^{-m_\phi c r/\hbar}}{r} \\
-\hbar c \alpha \frac{e^{-m_\phi c r/\hbar}}{r} & 2\delta c^2 \end{matrix}\right)  
\end{equation}
where $\alpha$ is the coupling between the dark matter and the mediator, $m_\phi$ is the mass of the mediator, $\delta$ is the mass splitting between the ground and excited states, $r$ is the separation between the particles, and the first row corresponds to the ground state $\ket{1 1}$. The two-body Schr{\"o}dinger equation for the relative motion (factoring out the overall free motion of the system as usual) then takes the form:

\begin{equation} \hbar^2 \frac{\nabla^2 \Psi(\vec{r})}{m_\chi} = \left( \mathcal{V}(r) - m_\chi v^2 \right) \Psi(\vec{r}).\end{equation}
Here $m_\chi$ is the mass of the dark matter, $v$ is the individual velocity of either of the dark matter particles in the center-of-mass frame (half the relative velocity), and $\Psi(\vec{r})$ is the wavefunction.

As mentioned above, the interaction between the ground state, $\ket{1}$ and  the excited state, $\ket{2}$, is purely off-diagonal. This is an automatic consequence of taking the force carrier to be a vector, as the mass eigenstates are $45^\circ$ rotations of the high-energy gauge eigenstates, and do not carry a conserved charge. As a result, two particles initially in the same state (ground or excited) can only scatter into the two-body states where they are both in the ground state, or both in the excited state. Such scatterings are described by the off-diagonal terms in the $\mathcal{V}(r)$ matrix. If the initial state is $\ket{1 2}$ (i.e. one particle is in the ground state and the other in the excited state) then their scattering decouples from the other two-body states and is elastic, with the final state being $\ket{1 2}$ or $\ket{2 1}$. This case can be treated by the existing methods in the literature (e.g. \cite{Tulin:2013teo}).

In this article we will treat only $s$-wave scattering (the methods developed in \cite{Slatyer:2009vg} do not generalize to higher partial waves). Accordingly we will make the ansatz of spherical symmetry in solving the Schr{\"o}dinger equation (more formally, as discussed in \cite{Slatyer:2009vg}, we expand in partial waves and only keep the $\ell=0$ term), and write  $r \Psi(\vec{r}) = \psi(r)$. 

We define the dimensionless parameters:
\begin{equation}
\begin{aligned}
&\epsilon_v \equiv \frac{v}{c \alpha},
&\epsilon_\delta \equiv \sqrt{\frac{2 \delta}{m_\chi \alpha^2}},
&&\epsilon_\phi \equiv \frac{m_\phi}{m_\chi \alpha}
\end{aligned}
\end{equation}
so that rescaling $r$ by $\alpha m_\chi c/ \hbar$ gives the $s$-wave Schr{\"o}dinger equation:
\begin{equation}
\label{schrod}
\psi ''(r) = \left( \begin{matrix}
-\epsilon_v^2 & - \frac{e^{-\epsilon_\phi r}}{r} \\
- \frac{e^{-\epsilon_\phi r}}{r}& \epsilon_\delta^2 - \epsilon_v^2 \end{matrix}\right) \psi(r) \equiv \bar{V}(r) \psi(r) \equiv \left( \begin{matrix}
-\epsilon_v^2 & - V(r) \\
- V(r) & \epsilon_\delta^2 - \epsilon_v^2 \end{matrix}\right) \psi(r).
\end{equation}
In the last line we have defined the matrix potential $\bar{V}(r)$ and the scalar Yukawa potential $V(r) = e^{-\ephi r}/r$.

\subsection{Calculating Approximate Wavefunctions}
\label{sec:modelwavefn}

The eigenvalues $\lambda_\pm$ and eigenvectors $\psi_\pm$ of the matrix $\bar{V}(r)$ are \\
\begin{equation}
\begin{aligned}
& \lambda_\pm = - \epsilon_v^2 + \frac{\epsilon_\delta^2}{2} \pm \sqrt{\frac{\epsilon_\delta^2}{2} + \frac{e^{- 2 \epsilon_\phi r}}{r^2}}, & \psi_\pm =  \frac{1}{\sqrt{2}} \colvec{\mp \sqrt{1 \mp  \frac{1}{\sqrt{1 + (4 e^{-2 \epsilon_\phi r})/(r^2 \epsilon_\delta^4)}}}}{\sqrt{1 \pm \frac{1}{\sqrt{1 + (4 e^{-2 \epsilon_\phi r})/(r^2 \epsilon_\delta^4)}}}}.
\end{aligned}
\end{equation}
There is a transition in the behavior of the eigenvalues and eigenvectors when $\frac{e^{-\epsilon_\phi r}}{r} \sim \frac{\epsilon_\delta^2}{2}$. In the regime where $\frac{e^{-\epsilon_\phi r}}{r} \gg \frac{\epsilon_\delta^2}{2}$, the eigenvalues and eigenvectors can be approximated as
\begin{equation}\begin{aligned}
& \lambda_\pm \approx -\ev^2 + \frac{\ed^2 }{2} \pm \frac{e^{-\ephi r}}{r}, &&& \psi_\pm \approx \frac{1}{\sqrt{2}} \colvec{\,\mp1\,}{1}.
\end{aligned} \label{eq:approxev}\end{equation}
When  $\frac{e^{-\epsilon_\phi r}}{r} \ll \frac{\epsilon_\delta^2}{2}$, the eigenvalues and eigenvectors can be approximated as
\begin{equation}\begin{aligned}
& \lambda_+ \approx - \ev^2 + \ed^2 + \frac{e^{-2 \ephi r}}{r^2 \ed^2}, & \lambda_-\approx -\ev^2 -  \frac{e^{-2 \ephi r}}{r^2 \ed^2},
&& \psi_+\approx \colvec{\,0\,}{1}, && \psi_- \approx \colvec{\,1\,}{0}.
\end{aligned}  \end{equation}
In other words, at small radii the symmetry is restored and the energy eigenstates become the gauge eigenstates, as at high energies; at large radii the energy eigenstates are the mass eigenstates.
 
At small and large $r$, the diagonalization of the potential matrix is roughly independent of $r$. If $\phi_\pm''(r) = \lambda_\pm \, \phi_\pm(r)$, then $\phi_\pm(r) \vec{\psi}_\pm$ is an approximate solution to the matrix Schr{\"o}dinger equation in those regimes, and we can use standard techniques --- in particular, the WKB approximation --- to solve for the scalar wavefunctions $\phi_\pm(r)$. However, in the transition region where $\frac{e^{-\epsilon_\phi r}}{r} \sim \frac{\epsilon_\delta^2}{2}$, the eigenvectors will vary as a function of $r$.

Provided that $\ed^2/2 \lesssim \ephi$, this transition occurs at $r \gtrsim 1/\ephi$, where the exponential behavior of the potential dominates: in this case we can use an \emph{exact} solution for the two-state exponential potential (taken from \cite{PhysRevA.49.265}) to cross the transition region and match the WKB solution at small $r$ to the large-$r$ asymptotic solution. Where $\ed^2/2 > \ephi$, we can still use this approach, but the exponential potential is a poor approximation to the true $V(r)$ at the radius corresponding to the transition region, and so the results are less reliable.

\begin{table}
\center{\begin{tabular}{|c|c|}
\hline \multicolumn{2}{|c|}{}\\ \multicolumn{2}{|c|}{{\bf Regimes for the wavefunction}} \\ \multicolumn{2}{|c|}{}\\
\hline &\\
$V(r) \approx 1/r$ for small $r$ (exact solution) & $~V(r) \gg \epsilon_\delta^2/2$, $V(r) \gtrsim \ev^2$, $~\ell = 0~$\\&\\  \hline &\\
WKB approximation for intermediate $r$&  $V(r) \gtrsim \epsilon_\phi^2, \ed^2/2 $ \\ &\\\hline &\\
$V(r) \approx V_0 e^{- \mu r}$ for large $r$ (exact solution) & $r \gtrsim 1/\ephi$ \\&\\ \hline &\\
Match onto large-$r$ asymptotic eigenstates & $~V(r) \ll \epsilon_\delta^2/2$ \\&\\ \hline  \multicolumn{2}{|c|}{}\\ \multicolumn{2}{|c|}{{\bf Conditions on \boldmath$\ev,~ \ephi,~\text{and}~\ed$}} \\ \multicolumn{2}{|c|}{}\\
\hline & \\
$~~$ 2nd and 3rd regimes above overlap $~~$ & $ ~\ed^2/2 \lesssim \ephi$  \\&\\ \hline &\\
$~~$$s$-wave dominates, simple matching (Appendix \ref{sec:rev}) $~~$ & $ \epsilon_v \lesssim \ephi$ \\&\\ \hline &\\
Scattering cross section enhanced by dark force & $\epsilon_v , \ephi, \ed \lesssim1$ \\ &\\ \hline 
\end{tabular}}
\caption{A summary of the various approximations and assumptions used for deriving the wavefunction in different regimes. The top four rows characterize the different regimes in $r$ where various techniques can be applied to approximate the wavefunction in the vacinity of a Yukawa potential; the lower three rows describe constraints on the parameters which allow the approximation to hold and cause the scattering cross section to be enhanced. Despite these restrictions, the various regimes of validity overlap substantially, making these approximations useful in large swatches of parameter space.} \label{table:regime}
\end{table}

The full procedure for determining the approximate wavefunction $\psi(r)$ can be summarized as:
\begin{itemize}
\item For $r \lesssim 1$, we approximate the Yukawa potential as $V(r) \sim 1/r$, assume the potential term dominates (since $\ev^2$ and $\ed^2$ are assumed to be small), and solve the resulting Schr{\" o}dinger equation exactly.
\item For $r \gtrsim 1$, we propagate the small-$r$ solution outward using a WKB approximation. The validity of the WKB approximation in the potential-dominated regime requires that $\abs{\sqrt{V'(r)}/V(r)} = \frac{1}{2} \sqrt{r} \, e^{\ephi r /2} (\ephi + 1/r) \ll 1$, which is just equivalent to requiring that the spatial variation of the local de Broglie wavelength is sufficiently gradual. For $r >1$, the condition for validity of the approximation is that $\ephi /\sqrt{V(r)} \lesssim 1$, so the approximation breaks down when $V(r) \sim \ephi^2$. 
\item Where the WKB approximation breaks down (at $V(r)\lesssim \ephi^2$) or where the diagonalization approximation fails (at $V(r) \lesssim \ed^2/2$), we must match the WKB solution onto a large-$r$ solution. Therefore, in this work, we will choose the matching radius $r_M$ such that $\frac{e^{-\ephi r_M}}{r_M} = \max\left(\frac{\ed^2}{2}, \ephi^2\right)$. For large $r$ ($r \gtrsim 1/\ephi$, so the exponential behavior dominates) we can approximate the Yukawa potential as an exponential potential $V(r) \sim V_0 e^{-\mu r}$, which leads to a  Schr{\" o}dinger equation that can be solved exactly even in the two-state case \cite{PhysRevA.49.265}.
\end{itemize}

\begin{figure}[htb]
\includegraphics[width=\textwidth]{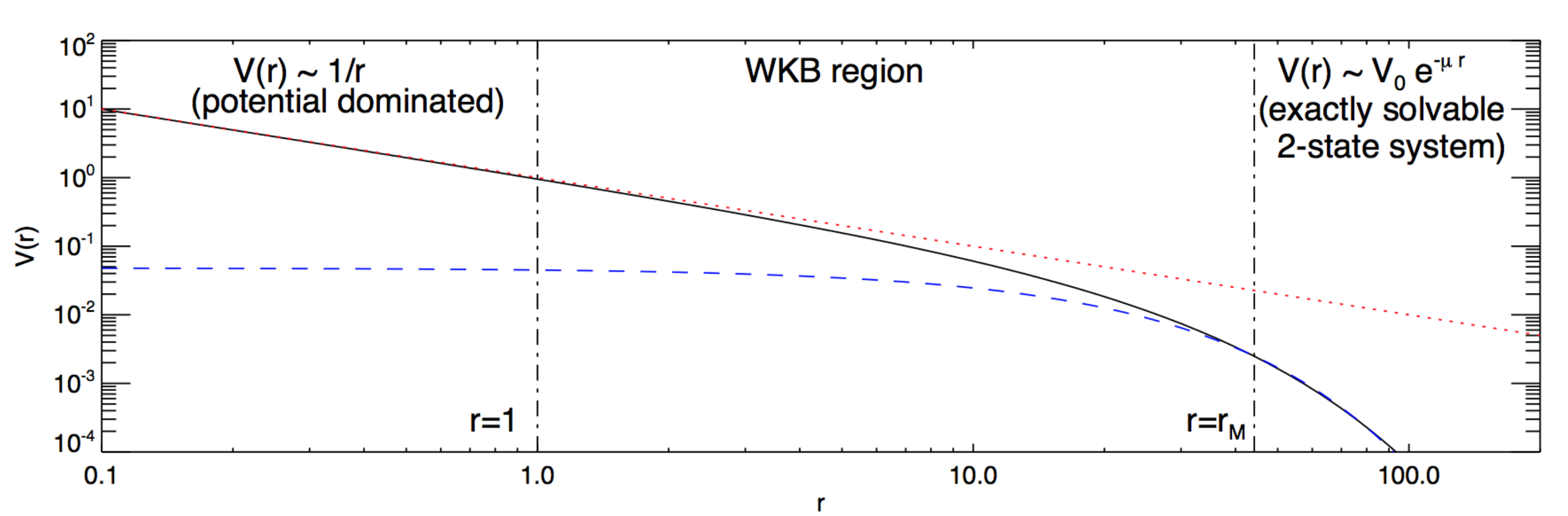}
\caption{Example of the different $r$-regimes and matching points for a sample parameter set ($\epsilon_v = 0.1, \, \epsilon_\delta = 0.02, \, \epsilon_\phi = 0.05$), following \cite{Slatyer:2009vg}. The plot shows the exact Yukawa potential (\emph{solid black line}) and the approximate potentials we employ, in their regimes of validity. In the $r \lesssim r_M$ region where the eigenstates are decoupled, for $r \lesssim 1$ the potential dominates the kinetic energy and mass splitting and is well approximated by $V(r) \approx 1/r$ (\emph{red dotted line}), whereas for $1 \lesssim r \lesssim r_M$ the WKB approximation is employed to obtain an approximate wavefunction. At $r \gtrsim r_M$ the WKB approximation may break down, but there $V(r) \approx V_0 e^{-\mu r}$ (\emph{dashed blue line}).}
\label{fig:matching}
\end{figure}

We summarize the different regimes in Figure \ref{fig:matching}. The large-$r$ solution involves hypergeometric functions which can only be analytically matched at $r_M$ by using an asymptotic expansion. As alluded to in Table \ref{table:regime} (noting that $\mu \sim \ephi$), this matching is simplest if $\ev \lesssim \mu$; if $\ev \gtrsim \mu$,  there is a term in the expansion that appears exponentially suppressed but can acquire an exponentially large prefactor. The mathematical condition that $\ev \lesssim \mu$ is physically equivalent to ignoring the contribution from the small-$r$ eigenstates of the $1/r$ potential that experience a repulsive interaction. (For further discussion, see Appendix \ref{sec:rev}. The breakdown of the approximation is not inevitable in the case where the mass splitting is substantial, as discussed in Appendix \ref{beyond}).
Additionally, for $\ev \gtrsim \mu$, we expect the $s$-wave term we compute here to be subdominant \cite{Tulin:2013teo} (see also Appendix \ref{beyond})\footnote{Numerical studies of inelastic DM scattering in this regime confirm that the higher partial waves provide large contributions \cite{Morris:2011dj}.}. The method developed here does not generalize straightforwardly to higher partial waves \cite{Slatyer:2009vg} --- it is useful for the ``resonant'' regime described by \cite{Tulin:2013teo}, not the ``classical'' regime, which even in the elastic-scattering case demands a different approach. Consequently we focus on the part of parameter space where $\ev \lesssim \mu \sim \ephi$.

We also require that $\ev$, $\ed$, and $\ephi$ are all less than 1 in order to see substantial enhancement to the $s$-wave cross section, beyond the geometric cross section associated with the mass of the DM particle (not the force carrier). If $\ev \gtrsim 1$, the kinetic energy is large compared to the potential energy; if $\ephi \gtrsim 1$, the range of the interaction is short; in both cases the presence of the potential does not significantly deform the wave-function. If $\ed \gtrsim 1$, the mass splitting is large compared to the Bohr potential energy ($\sim \alpha^2 m_\chi$, leading to a suppression of virtual excitations.) Since the potential is purely off-diagonal, this suppresses the elastic scattering cross section as well.

For completeness, we include a full description of the approximate wavefunctions in Appendix~\ref{sec:rev}, including the WKB matching between the small-$r$ and large-$r$ wavefunctions. A more in-depth derivation of these wavefunctions  (including more extensive discussion of the regimes of validity) can be found in \cite{Slatyer:2009vg}, albeit with different boundary conditions.

In this case, we use the regular boundary conditions, $\phi_+(0) = \phi_-(0)=0$. This corresponds to setting $\psi(0)=0$, which is equivalent to requiring that the physical wavefunction $\Psi(0)$ is finite at the origin. (In \cite{Slatyer:2009vg}, the Sommerfeld enhancement to annihilation was extracted from irregular solutions with $\phi_\pm(0) \neq 0$.)

Additionally we will impose one of two sets of boundary conditions: the radially ingoing particles will either be purely in the ground state or purely in the excited state (since the case of a ground state particle interacting directly with an excited state particle can be treated using existing methods in the literature \cite{Tulin:2013teo}.) We include a derivation of the dimensionless transfer cross sections in Appendix~\ref{sec:scattering} --- these must be multiplied by $\hbar^2/(c^2 \alpha^2 m_\chi^2)$ to obtain the physical cross sections, since we initially rescaled $r$ by $\alpha m_\chi c/\hbar$. 

\section{The Scattering Cross Sections}
\label{sec:scat}
\subsection{Analytic Results}

\begin{figure}[htb]
\center{\includegraphics[width=0.325\textwidth]{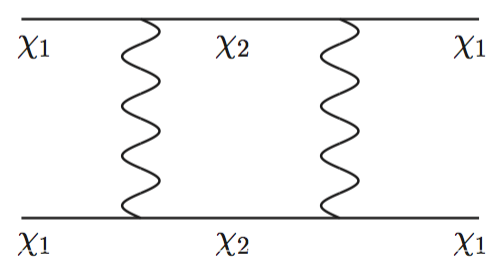}\hspace{0.6cm}
\includegraphics[width=0.36\textwidth]{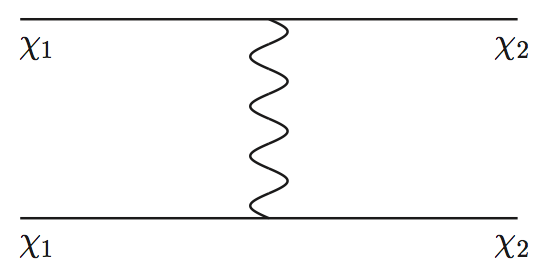}}
\caption{Schematic Feynman diagrams at lowest order for (left panel) ``ground-to-ground'' scattering (i.e. $\chi_1 \chi_1 \rightarrow \chi_1 \chi_1$) and (right panel) ``ground-to-excited'' scattering (i.e. $\chi_1 \chi_1 \rightarrow \chi_2 \chi_2$): swapping the fermion labels $1\leftrightarrow 2$ gives the ``excited-to-excited'' and ``excited-to-ground'' diagrams. The $u$-channel diagrams also contribute.}
\label{fig:feynman}
\end{figure}

While the interaction is purely off-diagonal, it is still possible for particles to scatter ``elastically'', i.e. two particles incoming in the ground (excited) state may scatter off each other while remaining in the ground (excited) state. This process does not occur at tree level, in the perturbative regime (see Appendix \ref{app:born}), but beyond the perturbative regime it is not suppressed. Figure \ref{fig:feynman} shows schematic perturbative Feynman diagrams for elastic and inelastic scattering; the non-perturbative regime requires resumming such ladder diagrams with arbitrary numbers of vector boson exchanges. Thus, below we present results for ground-to-ground and excited-to-excited elastic scattering, as well as the upscattering and downscattering rates.

Denoting the initial and final states as ``gr'' for ``ground'' and ``ex'' for ``excited'', we obtain the following dimensionless scattering cross sections:\\
\begin{equation}
 \sigma_{\text{gr}\rightarrow \text{gr}} = \frac{\pi}{\ev^2}\, \abs{ 1 + \left(\frac{V_0}{4\mu^2}\right)^{-\frac{2 i\epsilon_v}{\mu}} \left(\frac{\Gamma_v}{\Gamma_v^*} \right) 
 \left[\frac{\cosh\left(\frac{\pi (\epsilon_\Delta + \epsilon_v)}{2 \mu} \right) \sinh\left(\frac{\pi (\epsilon_v - \epsilon_\Delta)}{2\mu} + i \varphi \right)}{\cosh\left( \frac{\pi (\epsilon_\Delta - \epsilon_v)}{2 \mu}\right) \sinh \left(\frac{\pi (\epsilon_\Delta + \epsilon_v)}{2 \mu} - i \varphi \right)} \right]}^2 \label{eq:gg} \end{equation}
 \begin{equation}\sigma_{\text{ex}\rightarrow \text{ex}} = \frac{\pi}{\eD^2} \left|1 + \left(\frac{V_0}{4\mu^2}\right)^{-\frac{2 i\epsilon_\Delta}{\mu}} \left(\frac{ \Gamma_\Delta}{ \Gamma_\Delta^*} \right)   \left[\frac{\cosh\left(\frac{\pi (\epsilon_\Delta + \epsilon_v)}{2 \mu} \right) \sinh\left(\frac{\pi (\epsilon_\Delta - \epsilon_v)}{2\mu} + i \varphi \right)}{\cosh\left( \frac{\pi (\epsilon_\Delta - \epsilon_v)}{2 \mu}\right) \sinh \left(\frac{\pi (\epsilon_\Delta + \epsilon_v)}{2 \mu} - i \varphi \right)} \right] \right|^2\\\\ \label{eq:ee}
\end{equation}\\
\begin{equation} \sigma_{\text{gr}\rightarrow \text{ex}} =
 \frac{2\pi\, \text{cos}^2\varphi ~ \text{sinh} \left( \frac{\pi \epsilon_v}{\mu}\right) \text{sinh} \left( \frac{\pi \epsilon_\Delta}{\mu}\right)}{ \ev^2  \cosh^2 \left( \frac{\pi (\epsilon_\Delta - \epsilon_v)}{2 \mu}\right) \left(\cosh \left(\frac{\pi (\epsilon_v + \epsilon_\Delta)}{\mu}\right) - \cos(2 \varphi)\right)} \label{eq:ge}\end{equation}
\begin{equation} 
\sigma_{\text{ex}\rightarrow \text{gr}} = \frac{2\pi\, \text{cos}^2\varphi ~ \text{sinh} \left( \frac{\pi \epsilon_v}{\mu}\right) \text{sinh} \left( \frac{\pi \epsilon_\Delta}{\mu}\right)}{ \eD^2 \cosh^2 \left( \frac{\pi (\epsilon_\Delta - \epsilon_v)}{2 \mu}\right) \left(\cosh \left(\frac{\pi (\epsilon_v + \epsilon_\Delta)}{\mu}\right) - \cos(2 \varphi)\right)} \label{eq:eg}\end{equation}
where we have defined $\eD \equiv \sqrt{\ev^2 - \ed^2}$, and $\mu$ and $V_0$ are the defining parameters for the exponential potential $V_0 e^{-\mu r}$, given by: \begin{equation}
\begin{aligned}
&\mu = \ephi \left(\frac{1}{2} + \frac{1}{2} \sqrt{1 + \frac{4 }{\ephi r_M}}~ \right), & & V_0= \frac{ e^{\ephi r_M \big(-\frac{1}{2} + \frac{1}{2} \sqrt{1 + \frac{4 }{\ephi r_M}}\, \big) }}{r_M}, \end{aligned}\end{equation} with $r_M$ chosen so that $\frac{e^{-\ephi r_M}}{r_M} = \max\left(\frac{\ed^2}{2}, \ephi^2\right)$.
The terms $\Gamma_v$ and $\Gamma_\Delta$ come from matching the WKB wavefunction onto the wavefunction for the exponential potential, and are defined by,
\begin{equation}\begin{aligned}
&\Gamma_v \equiv \Gamma \left(1 + \frac{i \epsilon_v}{\mu}\right) \Gamma \left( \frac{i \epsilon_v - i \epsilon_\Delta}{2 \mu} + \frac{1}{2}\right) \Gamma \left( \frac{i \epsilon_v + i \epsilon_\Delta}{2 \mu} + \frac{1}{2}\right) \\ 
& \Gamma_\Delta  \equiv   \Gamma \left(1 + \frac{i \epsilon_\Delta}{\mu}\right) \Gamma \left( \frac{i \epsilon_\Delta - i \epsilon_v}{2 \mu} + \frac{1}{2}\right) \Gamma \left( \frac{i \epsilon_v + i \epsilon_\Delta}{2 \mu} + \frac{1}{2}\right), 
\end{aligned}
\end{equation} where $\Gamma$ denotes the gamma function. 
Finally, $\varphi$ is a phase that comes from extending the WKB solution to the matching region. Its full form is rather complicated, but where $\ed \lesssim \ephi$ and our other approximations hold (i.e. $\ephi \lesssim 1$, $\ev \lesssim \mu$), the phase $\varphi$ can be accurately approximated by $\varphi = \sqrt{2\pi/\ephi}$ (as noted in \cite{Slatyer:2009vg}). More generally, it is given by:\begin{equation}
i \varphi \equiv \int_0^{r_M} \sqrt{\lambda_-} dr' + \int_{r_M}^{r_s} \sqrt{\tilde{\lambda}_-} dr' + \frac{2 i\,\sqrt{ V_0\, }}{\mu}\,e^{- \mu r_s/2}.
\end{equation} 
Here $r_s$ must be chosen such that $V_0\,e^{-\mu r_s} \gg \mu^2, \ev^2,\, \ed^2$ (see Appendix \ref{sec:WKB}), but formally can (and should) be taken to $-\infty$. To see this, note that the phase can be rewritten in the form:
\begin{align} \varphi & = -i  \int_0^{r_M} \sqrt{\lambda_-(r')} dr' -  \int^{z_s}_{\max(\epsilon_\phi^4, \epsilon_\delta^4/4)/(16\mu^4)} \frac{1}{2\mu z} \left( \sqrt{\epsilon_v^2 - \frac{\epsilon_\delta^2}{2} + \sqrt{\left(\frac{\epsilon_\delta^2}{2}\right)^2 + 16 \mu^4 z}} - 2 \mu z^{1/4} \right) dz \nonumber \\
& + \left(\frac{\max(16\epsilon_\phi^4, 4 \epsilon_\delta^4)}{\mu^4}\right)^{1/4} \end{align}
where we have defined $z_s = V_0^2 e^{-2\mu r_s}{16\mu^4}$. We can now simply take $z_s \rightarrow \infty$. Furthermore, the second integral is analytically tractable, yielding (noting that $\lambda_-$ is always negative):
\begin{align} \varphi & =  \int_0^{r_M} \sqrt{|\lambda_-(r')|} dr' - \frac{1}{\mu} \left(-  2 \sqrt{\frac{\epsilon_\Delta^2 + \epsilon_v^2}{2} + \sqrt{\frac{\epsilon_\delta^4}{4} + \max(\epsilon_\phi^4, \epsilon_\delta^4/4)}} + \frac{i \pi}{2} ( \epsilon_v + \epsilon_\Delta) \right.  \nonumber \\
& \left.  +  \epsilon_\Delta \mathrm{arctanh} \left[ \frac{\sqrt{\frac{\epsilon_v^2 + \epsilon_\Delta^2}{2} + \sqrt{\frac{\epsilon_\delta^4}{4} +  \max(\epsilon_\phi^4, \epsilon_\delta^4/4)}}}{ \epsilon_\Delta} \right] +  \epsilon_v  \mathrm{arctanh} \left[ \frac{\sqrt{\frac{\epsilon_v^2 + \epsilon_\Delta^2}{2} + \sqrt{\frac{\epsilon_\delta^4}{4} +  \max(\epsilon_\phi^4, \epsilon_\delta^4/4)}}}{ \epsilon_v } \right]   \right), \end{align}
where we have chosen the convention for the branch cut such that $\mathrm{arctanh}(x) \rightarrow -i \pi/2$ as $x \rightarrow \infty$, for $x$ on the positive real line. There is still one numerical integral to perform, but it is fast and stable.

We emphasize again that these cross sections only hold in the regimes described in the lower half of Table I. Since we are only computing the $s$-wave piece of the scattering amplitude, which is angle-independent, the viscosity and transfer cross sections are trivially related to $\sigma$ (see \cite{Tulin:2013teo} for a discussion of the different cross sections and their relevance for the problem at hand).

The astute reader will notice the following salient feature of the scattering cross sections: the elastic and inelastic cross sections are the same whether the system starts in the ground state or the excited state, modulo a swap of $\ev$ with $\eD$ (assuming that $\eD$ is the same in both cases, which requires the system to be above threshold.) The scattering \emph{amplitudes} are identical. This reflects the identical interactions of the ground and excited states: swapping $\ev \leftrightarrow \eD$ simply corresponds to relabeling the states. The result also agrees with our intuition from quantum mechanical scattering off a 1D step potential: the transmission and reflection probabilities are the same for ``downhill'' and ``uphill'' scattering when the particle's energy is greater than the potential barrier, and the same is true in this system above the mass-splitting threshold. 


\subsection{Numerical Cross-Checks}

\begin{figure}[h!]
\center{\includegraphics[width=0.325\textwidth]{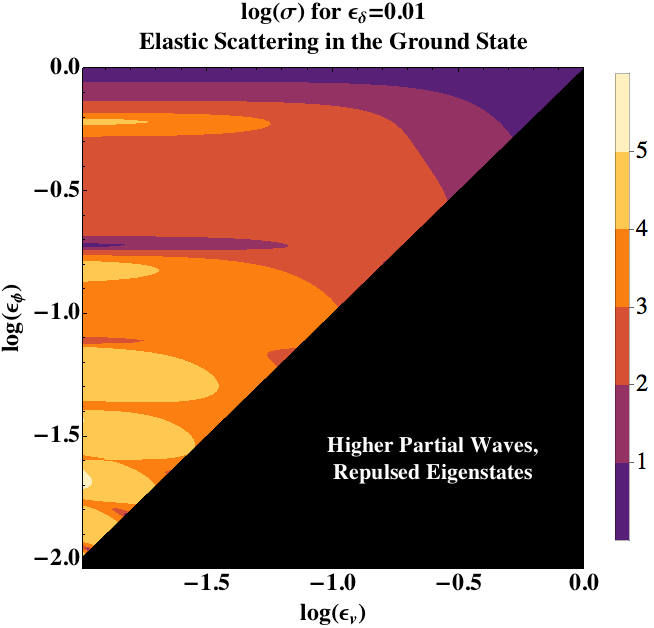}\hspace{0.5cm}\includegraphics[width=0.3\textwidth]{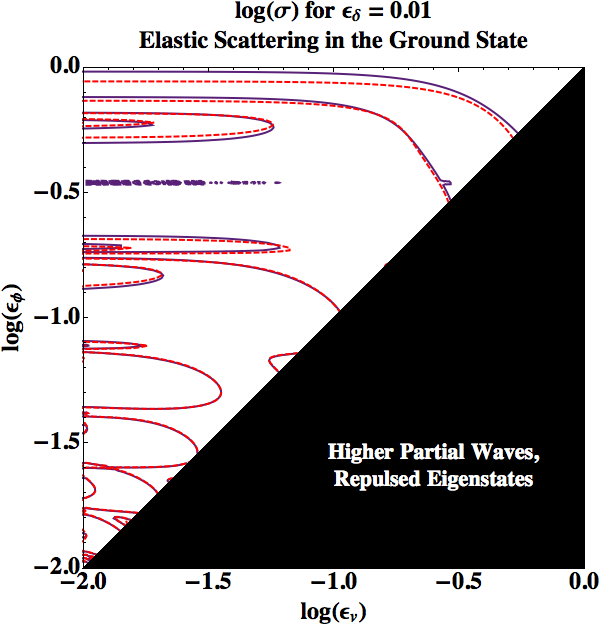}\vspace{0.1cm}\\
\includegraphics[width=0.325\textwidth]{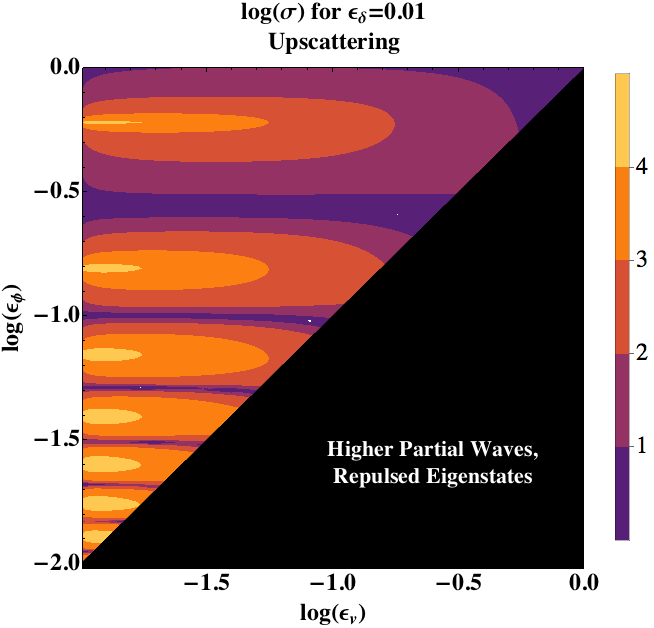}\hspace{0.5cm}\includegraphics[width=0.3\textwidth]{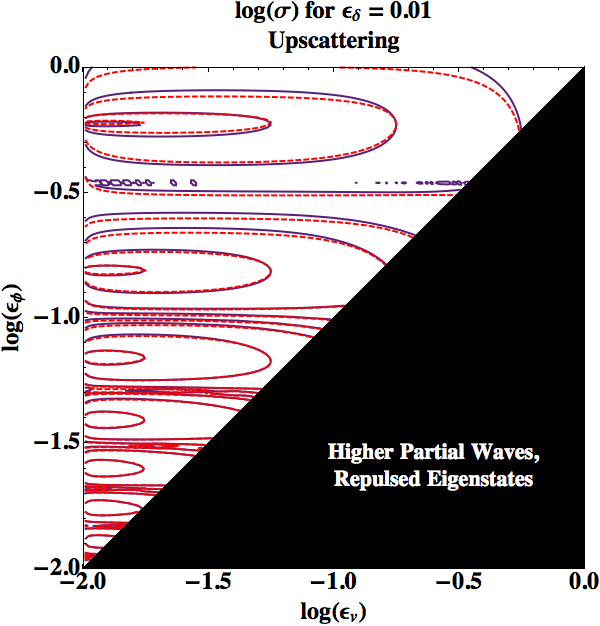}\vspace{0.1cm}\\
\includegraphics[width=0.325\textwidth]{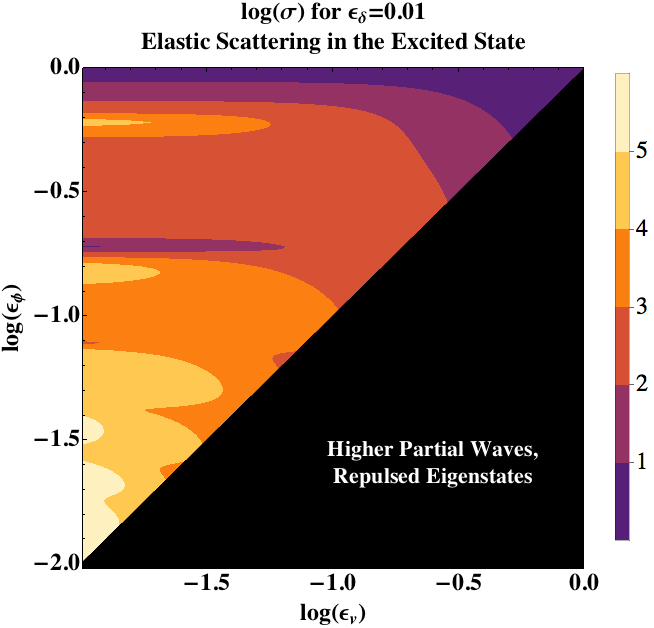}\hspace{0.5cm}\includegraphics[width=0.3\textwidth]{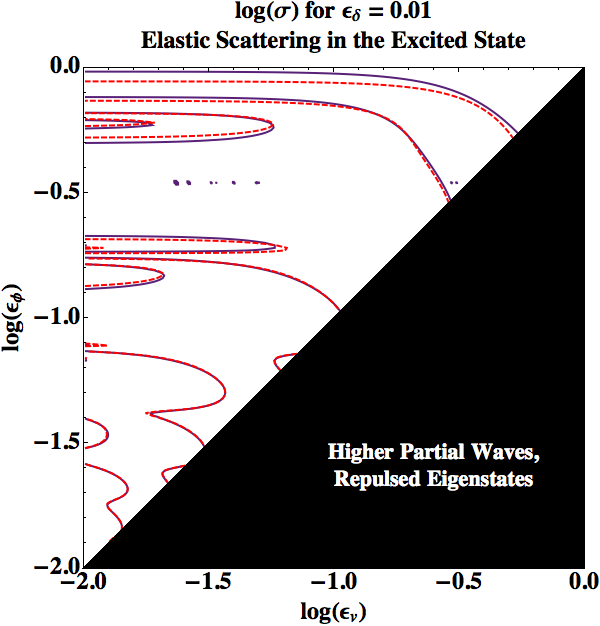}\vspace{0.1cm}\\
\includegraphics[width=0.325\textwidth]{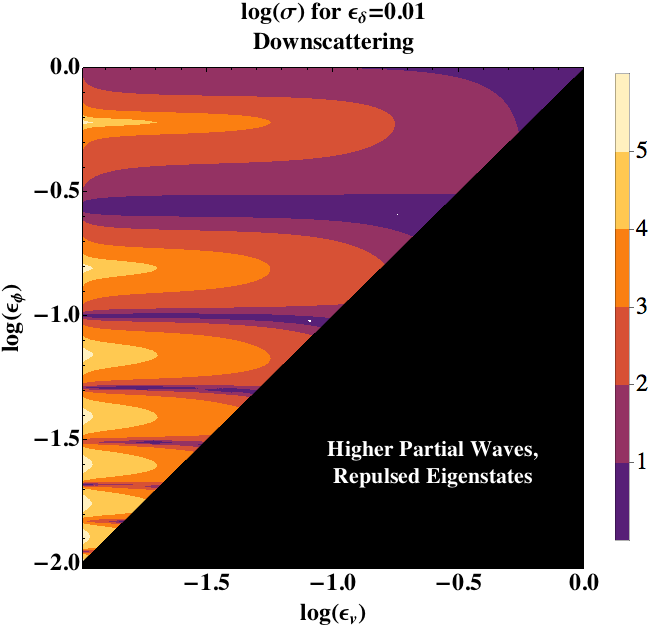}\hspace{0.5cm}\includegraphics[width=0.3\textwidth]{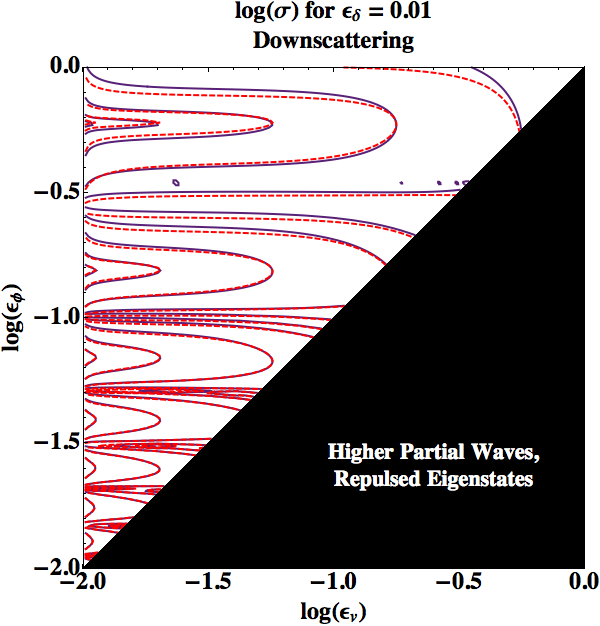}\\}
\caption{The left panels show the dimensionless cross sections computed using our approximations, for $\ed=0.01$. The right panels compare these results (red dashed lines) with the exact numerical results (solid lines). Note that ``log'' indicates base 10.}
\label{fig:comp1}
\end{figure}

\begin{figure}[h!]
\center{\includegraphics[width=0.325\textwidth]{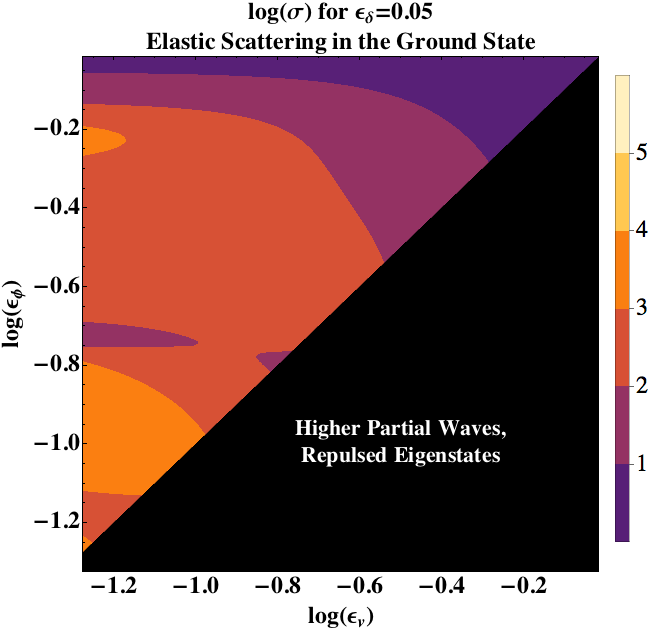}\hspace{0.5cm}\includegraphics[width=0.3\textwidth]{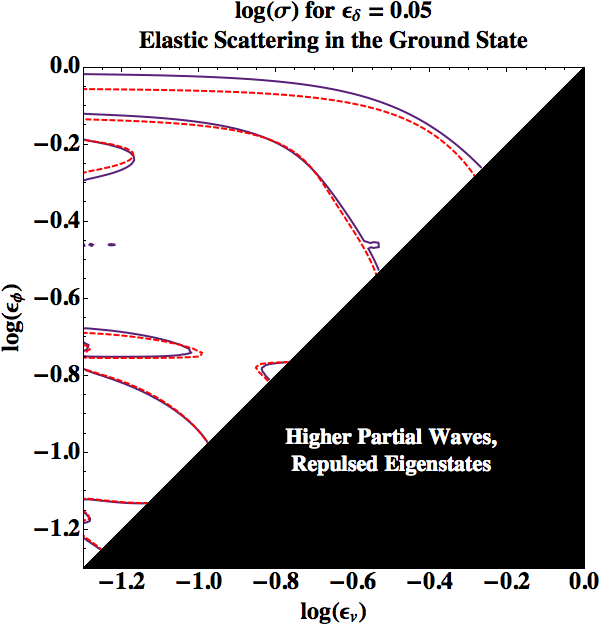}\vspace{0.3cm}\\
\includegraphics[width=0.325\textwidth]{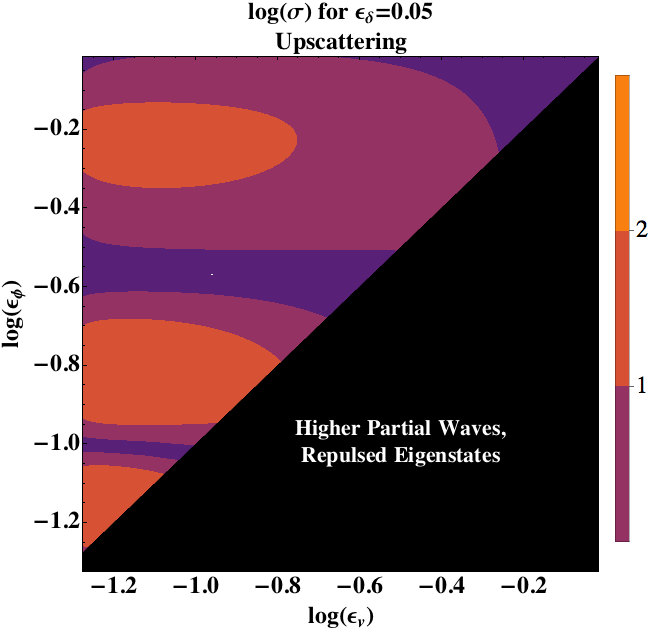}\hspace{0.5cm}\includegraphics[width=0.3\textwidth]{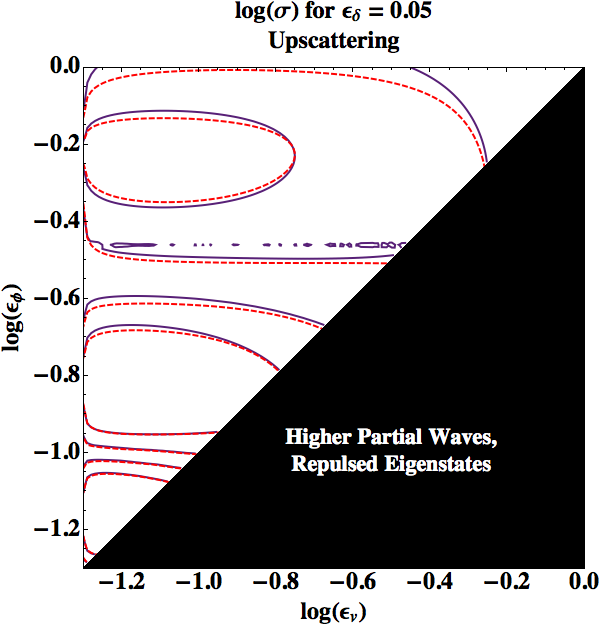}\vspace{0.3cm}\\
\includegraphics[width=0.325\textwidth]{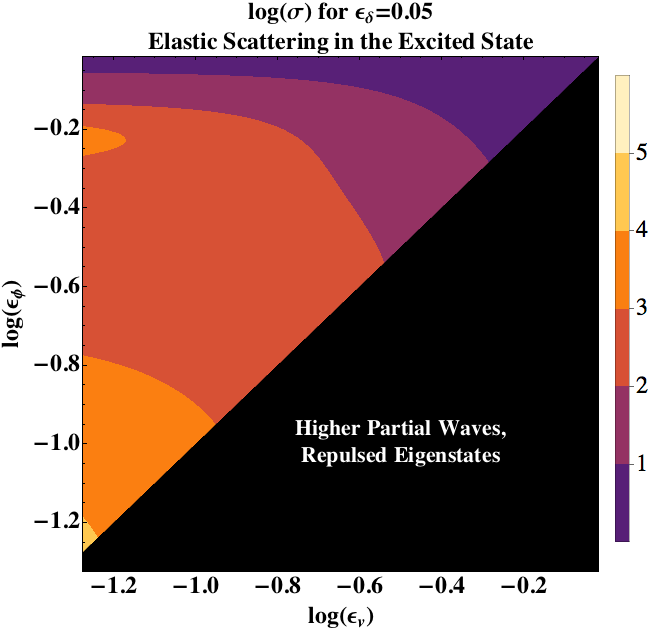}\hspace{0.5cm}\includegraphics[width=0.3\textwidth]{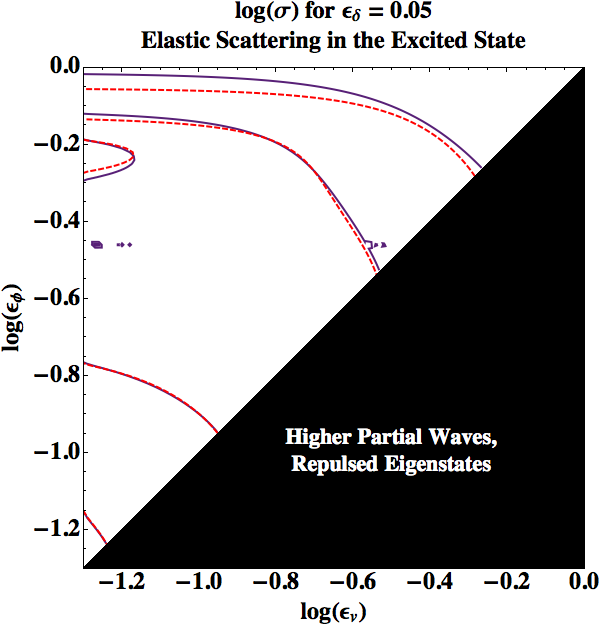}\vspace{0.3cm}\\
\includegraphics[width=0.325\textwidth]{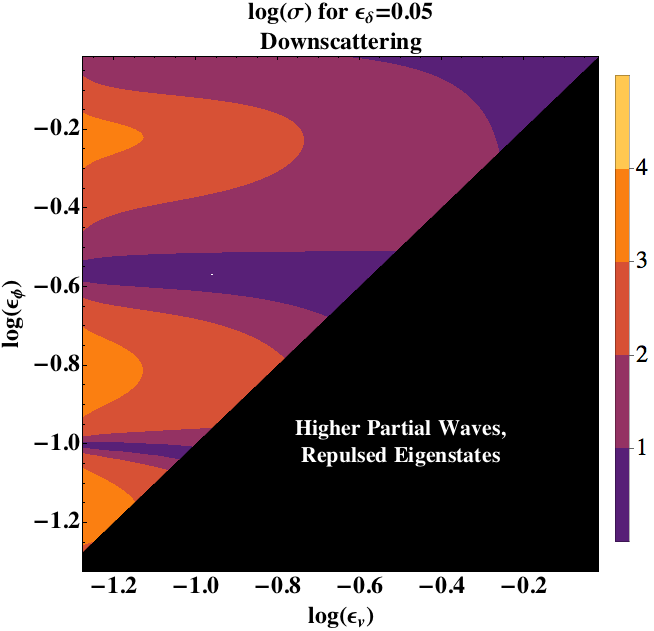}\hspace{0.5cm}\includegraphics[width=0.3\textwidth]{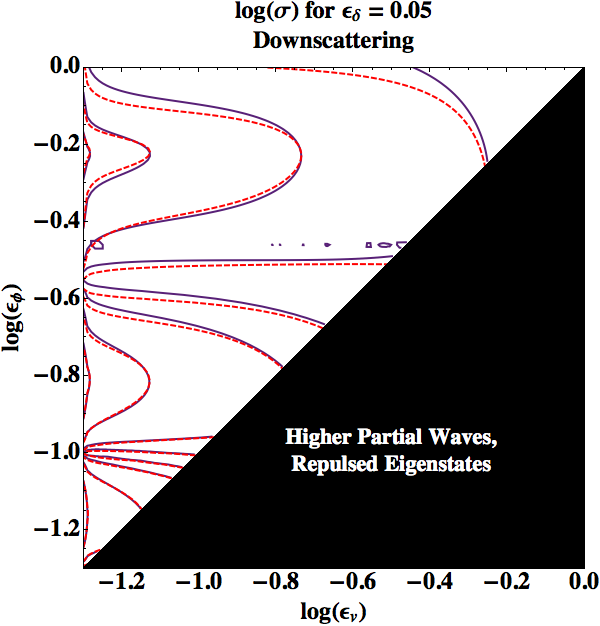}\\}
\caption{As Figure \ref{fig:comp1}, but with $\ed = 0.05$.}
\label{fig:comp2}
\end{figure}

Here we compare the results of our analytic approximation to the exact $s$-wave cross sections, obtained by numerically solving the matrix Schr{\"o}dinger equation using \texttt{Mathematica}. Numerically solving for the scattering amplitudes proved computationally expensive and/or unstable in certain regions of parameter space, and was in all cases several orders of magnitude slower than computing the semi-analytic results, further motivating the use of these approximations.  The results are shown in Figures \ref{fig:comp1}-\ref{fig:comp2}, for two sample choices of the mass splitting parameter $\ed = 0.01, 0.05$,
 and for elastic scattering in the ground and excited states (i.e. ``gr$\rightarrow$gr'' and ``ex$\rightarrow$ex''  respectively), upscattering (``gr$\rightarrow$ex'') and downscattering (``ex$\rightarrow$gr''). We show only results for $\ev > \ed$ in this figure, since if this is not the case only ground-to-ground scattering is possible, and also restrict $\ev < \ephi$ since otherwise we expect both that our approximation may break down and that higher partial waves will become important. We note that in all cases, notable resonances and antiresonances develop at particular values of $\ephi$.

We find that our approximations agree with the numerics to within 10\% away from resonances (at resonances, minor shifts can cause huge fractional disagreement in spite of the fact that the approximation actually does capture the behavior quite well), and correctly describe the resonance positions. (Note that the minor numerical artifacts in these plots reflect the instability of the numerical calculation, and should be ignored for purposes of comparison.)

\begin{figure}[h!]
\includegraphics[width=0.49\textwidth]{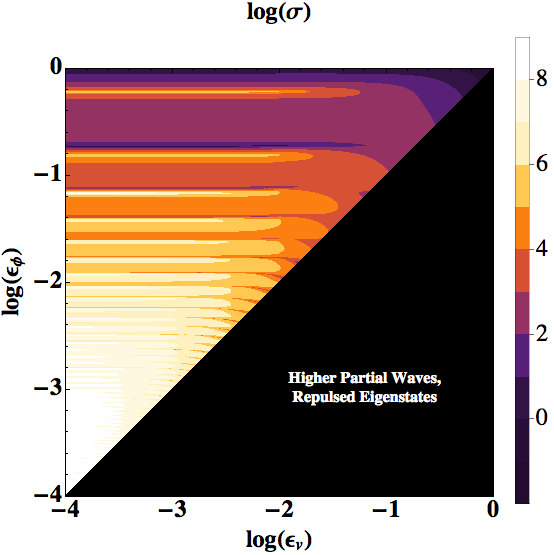}
\includegraphics[width=0.49\textwidth]{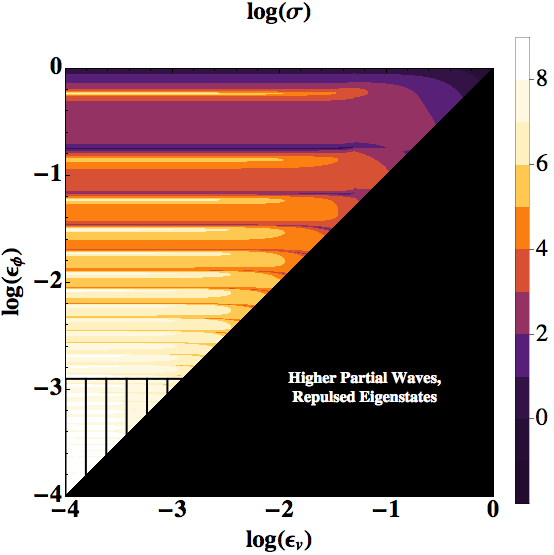}
\caption{An analytic calculation of the elastic ground$\rightarrow$ground dimensionless cross section in the $\ephi$ vs. $\ev$ plane, with $\ed = 0.01$ (\emph{left}) and $\ed = 0.05$ (\emph{right}). The resonance structures in the upper left are very similar to those depicted in Figure 1 of \cite{Tulin:2013teo}. The black areas to the lower right of the diagonal on the plots indicates where our approximation is no longer valid because of the repulsed eigenstates and higher partial waves (i.e. when $\ev>\ephi$.) The black cross-hatching on the bottom of the $\ed = 0.05$ plot indicates where $\ephi < \ed^2/2$ which is a region where our approximation is less accurate (see Table \ref{table:regime}.) Note that the use of ``log'' indicates the base 10 logarithm.\vspace{-0.4cm}} \label{fig:2d}
\end{figure}

Figure \ref{fig:2d} shows the analytic calculation of the effect of a non-zero mass splitting on elastic scattering in the ground state, including for $\ev < \ed$, for the same two choices of $\ed$. For ground-state elastic scattering, the main effect of the mass splitting is to cause the positions of the resonances to shift. The effect is much more pronounced for small $\ephi$. We will understand this behavior in the following subsection, by studying analytically tractable limits of eqs. \ref{eq:gg}-\ref{eq:eg}. The numerical results again agree with the analytic results in this regime (we will show an explicit comparison in the case of small $\ev$ in Figure \ref{fig:elasticlowvlimit}).

\subsection{Features and Limits of the Scattering Cross Sections}

We will now study various limits of the scattering cross sections using our analytic approximation. First, however, let us emphasize a point regarding our definition of $\epsilon_v$. This quantity relates to the energy of the particle \emph{relative to the ground state}. A particle freely propagating in the excited state will \emph{always} have $\epsilon_v > \epsilon_\delta$. While, as noted above, the ``uphill'' and ``downhill'' scattering \emph{amplitudes} are identical above the kinematic threshold given by the mass splitting,  the cross section for downscattering diverges as $1/\eD$ close to threshold (except at anti-resonances) while the cross section for upscattering goes identically to zero. These different behaviors originate solely from the very different phase-space factors near threshold, and are demonstrated in Figures \ref{fig:comp1}-\ref{fig:comp2}. For larger $\ev$, upscattering and downscattering are equally likely, since there is relatively little energetic overhead to upscattering.  As well as Figures \ref{fig:comp1}-\ref{fig:comp2}, the threshold behavior is shown for a constant-$\ephi$ slice in Figure \ref{fig:inel}.
 
These cross sections, with the same $\ev$, correspond to different physical scenarios in the context of a virialized DM halo --- for instance, an \emph{excited} state particle in a virialized halo will give a larger ground state velocity and thus correspond to a larger resulting $\ev$ than for virialized \emph{ground} state particles in the halo. We further discuss the astrophysical relevance of these scenarios in Section \ref{astropheno}. 

\subsubsection{The Degenerate Limit} 
In the limit where $\delta \rightarrow 0$, analytic expressions for the scattering cross section in a repulsive or attractive potential have been previously derived \cite{Tulin:2013teo}. In this limit, our ``elastic'' and ``inelastic'' cross sections refer to scatterings between particular linear combinations of the attracted and repulsed two-body eigenstates, and it is natural to switch to the basis of (dark) charge eigenstates accordingly (which in this limit are also mass eigenstates). In Appendix \ref{app:degeneratelimit} we show in detail how this conversion is done, and find agreement with previous results \cite{Tulin:2013teo}. In particular, we find that the resonance positions occur when $\varphi = n\pi$ and $n$ is an integer; using the approximation $\varphi \approx \sqrt{2\pi/\ephi}$ (valid for $\epsilon_\delta \lesssim \ephi$), meaning the resonances occur at $\ephi \approx \frac{2}{\pi} \frac{1}{n^2}$. In the analysis of \cite{Tulin:2013teo} the resonances occur at $\ephi = \frac{1}{\kappa}\frac{1}{n^2}$ where the parameter $\kappa$ is chosen to be 1.6, in close agreement.

For completeness, we include the $\ed \rightarrow 0$ expansion here, as well as in Appendix \ref{app:degeneratelimit}.
The ``elastic'' scattering cross section to first order in $\ed$ (for both the ground and excited state, since they are now degenerate) becomes:
\begin{equation}
\sigma_\text{``elastic''}= \frac{\pi}{\ev^2}\abs{ 1 + \frac{\Gamma\left(1 + \frac{i \epsilon_v}{\mu} \right)}{\Gamma\left(1 - \frac{i \epsilon_v}{\mu} \right)} \left(\frac{i \sin \varphi \cosh\left(\frac{\pi \epsilon_v}{\mu}\right)}{\sinh\left(\frac{\pi \epsilon_v}{ \mu} - i \varphi \right)}\right) \left(\frac{V_0}{\mu^2} \right)^{-2 i\epsilon_v/\mu} }^2.
\end{equation}
Off-resonance, in the limit as $\epsilon_v \rightarrow 0$, $\sinh\left(\frac{\pi \epsilon_v}{ \mu} - i \varphi \right) \rightarrow - i \sin \varphi$, and the cross section scales as $1/\mu^2$. More precisely, a Taylor expansion yields:
\begin{equation}\label{lowdeqn} \sigma_\text{``elastic''} \rightarrow \frac{\pi}{\mu^2} \left[  \pi \cot \varphi + 2 \ln\left(\frac{V_0}{\mu^2} \right) + 2\gamma \right]^2,\end{equation}
where $\gamma$ is the Euler-Mascheroni constant.
On-resonance, where $\sin \varphi = 0$, the ``elastic'' scattering amplitude is simply 1, and the dimensionless cross section is accordingly $\sigma_\text{``elastic''} = \pi/\epsilon_v^2$. 
\begin{figure}[htb] 
\includegraphics[width = 0.49\textwidth]{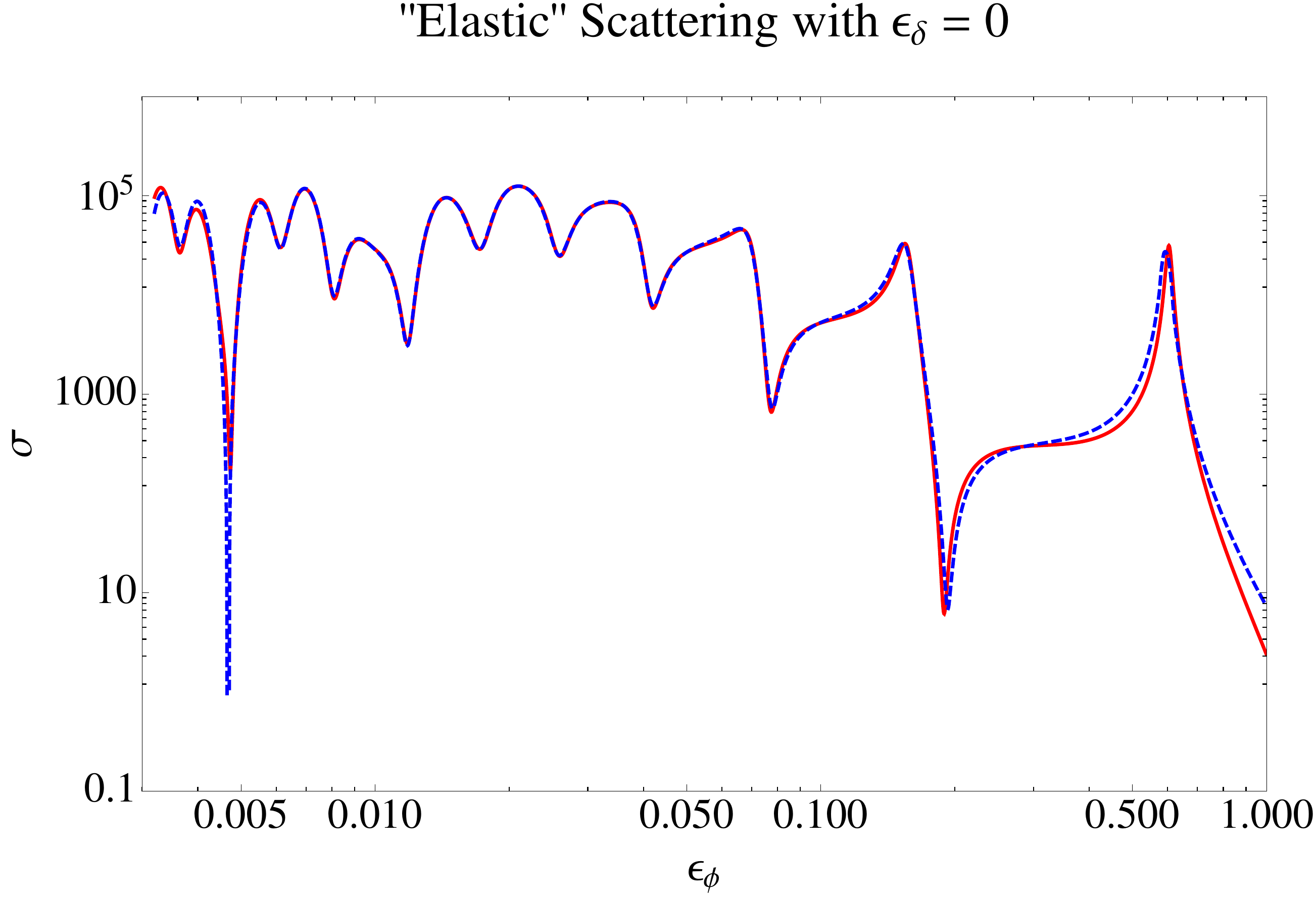}
\includegraphics[width = 0.49\textwidth]{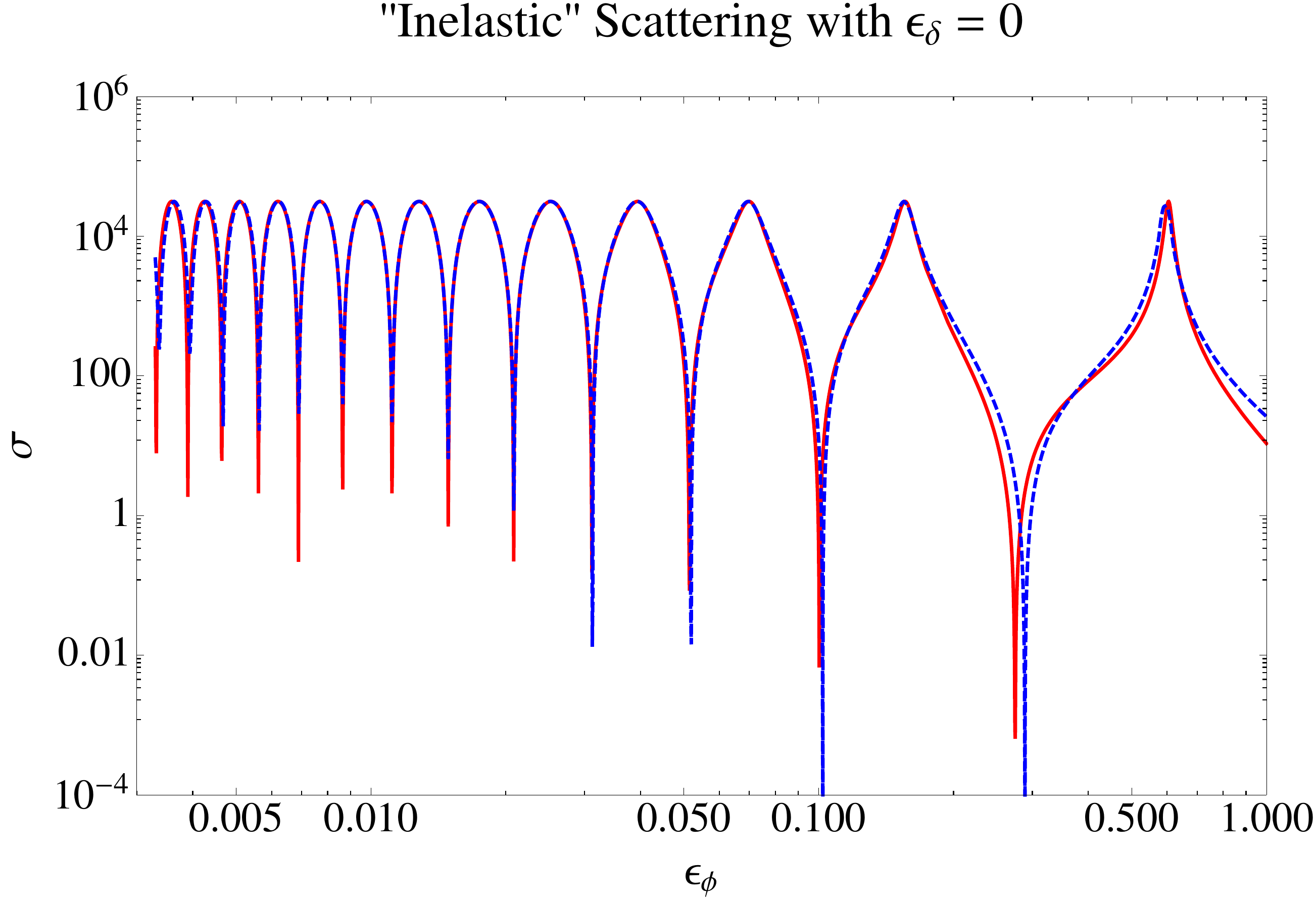}
\caption{Scattering with $\ed=0$ and $\ev =0.01$, where the red solid curve comes from our analytic approximation and the blue dashed curve comes from numerically solving the Schr\"odinger equation. The quotation marks in the plot titles serve as a reminder that with $\ed=0$ there is no inelastic scattering because the states are degenerate. Resonances occur as predicted by our Taylor expansions.} \label{fig:delta0limit}
\end{figure}

Meanwhile, for the ``inelastic'' case, setting  $\ev = \eD$ yields
\begin{equation}
\sigma_\text{``inelastic''}= \frac{2 \pi\, \text{cos}^2\varphi ~ \text{sinh}^2 \left( \frac{\pi \epsilon_v}{\mu}\right) }{\ev^2\left( \cosh \left(\frac{2 \pi \epsilon_v }{\mu}\right) - \cos(2 \varphi)\right)}.
 \end{equation}
 Off-resonance, as $\epsilon_v \rightarrow 0$, this cross section approaches \begin{equation}\sigma_\text{``inelastic''} \rightarrow \pi \left( \frac{\pi  \cot\varphi}{\mu} \right)^2.\end{equation} On-resonance, where $\cos(2\varphi) = 1$, the ``inelastic'' scattering amplitude approaches 1, and again $\sigma_\text{``inelastic''} \rightarrow \pi/\ev^2$. 
 
We plot our results for the cross section in this limit in Figure \ref{fig:delta0limit}, again finding good agreement between our full semi-analytic approximation and the numerical results for $\ephi < 1$. These results can be used more generally for elastic and inelastic scattering where $\ed$ is small relative to all the other parameters in the problem.

\subsubsection{The Low-Velocity Limit}
\label{sec:lowvelocitylimit}
Now let us consider the case where $\ev \rightarrow 0$, \emph{without} first setting $\ed \rightarrow 0$. In this limit, scattering into the excited state is forbidden, so we will only examine elastic scattering in the ground state. Expanding $\sigma_{\text{gr}\rightarrow \text{gr}}$ to first order in $\ev$ yields 
\begin{equation}
\label{lowveqn}
\sigma_{\text{gr}\rightarrow \text{ gr}} \approx \frac{\pi}{\mu^2} \left( \pi \cot\left(\varphi - \frac{\pi \ed}{2\mu}\right) + 2 \ln\left(\frac{V_0}{\mu^2}\right) + 2 \gamma - 4\ln 2 - 2 \psi_0 \left( \frac{\ed}{2 \mu} + \frac{1}{2}\right)  \right)^2
 \end{equation}
 where $\psi_0$ is the digamma function and $\gamma$ is the Euler-Mascheroni constant.
 \begin{figure}[htb]
\center{ \includegraphics[width=0.48\textwidth]{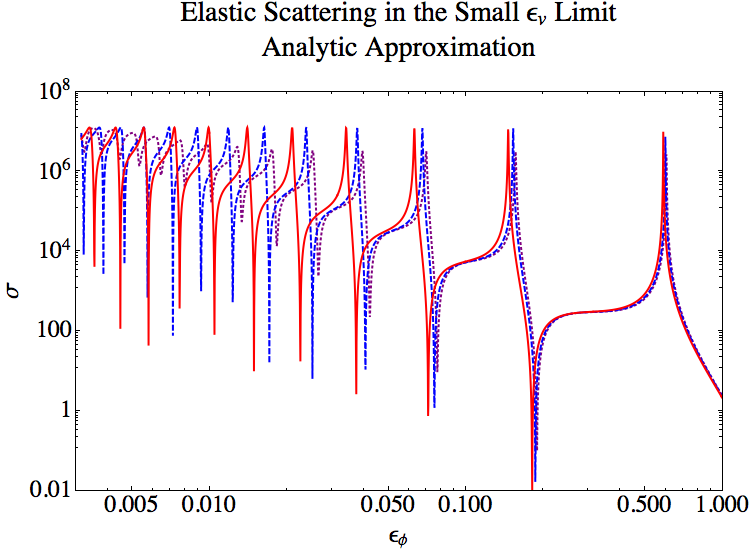}\includegraphics[width=0.48\textwidth]{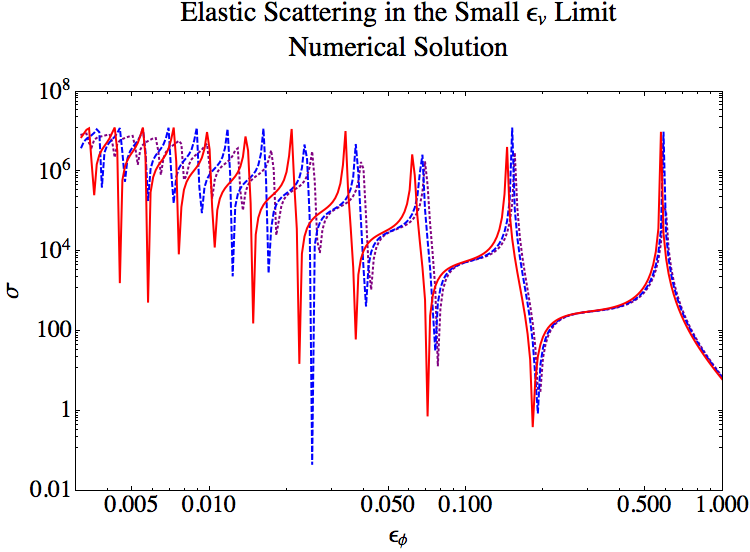}\\
\includegraphics[width=0.5\textwidth]{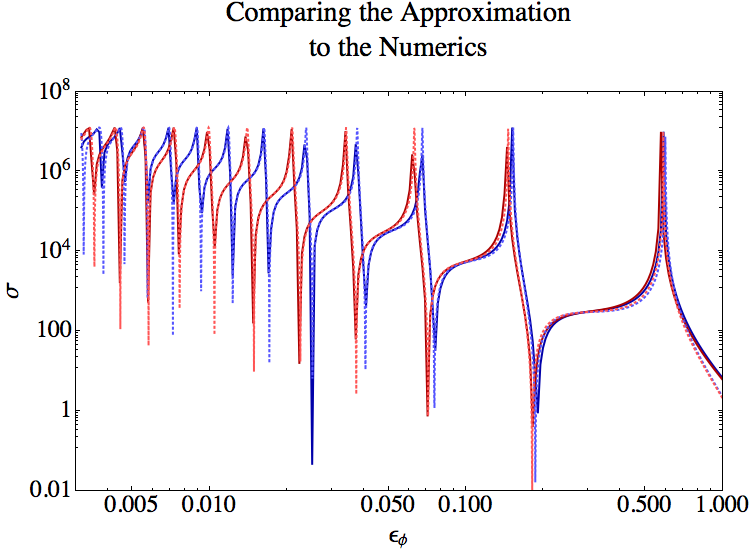}}
 \caption{Scattering from the ground state to the ground state with $\ev= 0.001$. Shown here are different scenarios with $\ed=0$ (purple, dotted), $\ed=0.01$ (dashed, blue), and $\ed = 0.03$ (solid, red). The upper left panel shows our analytic approximation and the upper right panel shows the numerical solution, which are in close agreement. To show the agreement between the two calculations explicitly, in the lower panel we have overlaid the analytic and numerical results for the $\ed=0.01$ and $\ed=0.03$ cases, with the solid dark lines representing the numerical computation and the lighter dashed lines representing our approximation. The analytic result accurately reproduces the numerics, and we can see that the presence of a mass splitting --- even a small one--- shifts the resonance positions, most noticeably for small $\ephi$.} \label{fig:elasticlowvlimit}
 \end{figure}

We see that the cross section does not vary with $\ev$ in this low-velocity limit, with the cross section approaching the expected geometric size of $\sim \pi/m_\phi^2$, up to constant prefactors, once we convert to dimensionful parameters (making the approximation $\mu \sim \ephi = m_\phi / (\alpha m_\chi)$, and then multiplying the dimensionless cross section by $1/(\alpha m_\chi)^2$ as usual). Resonances occur when $\varphi = (n + \epsilon_\delta/2 \mu) \pi$, and in the case where $\delta=0$ (as mentioned previously) the resonance positions are $\varphi = n\pi$. Thus the presence of a mass splitting induces a shift to the resonance positions at velocities below the threshold, to smaller $\ephi$: Figure \ref{fig:elasticlowvlimit} shows the impact of the mass splitting on the resonances at low velocity for several choices of $\delta$. The effect on the resonances is the same as found for the case of annihilation \cite{Slatyer:2009vg}. We also demonstrate in Figure \ref{fig:elasticlowvlimit}  that our analytic approximation accurately reproduces the numerical results for low-$\ev$ scattering below threshold.

Except for the shift in resonance positions, this cross section is very similar in form to \eqref{lowdeqn}; in the limit as $\ed \rightarrow 0$ (but $\ev < \ed$) they are identical, except that \eqref{lowveqn} has a 4$\gamma$ term rather than $2\gamma$ (two extra $\gamma$'s come from $\psi_0(\frac{1}{2})$). This is a subdominant correction; generally larger contributions will arise from the $\cot$ and log terms. So we see that for elastic scattering in the ground state, the effect of the mass splitting is primarily just to shift the resonance positions; this contrasts with the case of annihilation where switching on the mass splitting can lead to a generic enhancement of the cross section by a factor of 2-4 at low velocities \cite{Slatyer:2009vg}.

\subsubsection{The Threshold ($\ev = \ed$) Limit}
\label{sec:thresholdlimit}
Scattering amplitudes involving the excited state will be suppressed by $\eD$ as $\eD$ approaches zero from above, but the corresponding cross sections need not vanish. The case where $\ev \approx \ed$ corresponds, for particles initially in the excited state, to very low physical velocities. We perform a Taylor expansion in small (but real and positive) $\eD$, finding for the cross sections:
\beq 
\sigma_{\text{gr} \rightarrow \, \text{ex}} \approx \frac{4 \pi^2 \eD \cos^2 \varphi \tanh \left( \frac{\pi \ed}{2\mu}\right)}{\ed^2\, \mu \left( \cosh\left( \frac{\pi \ed}{\mu}\right) - \cos( 2 \varphi)\right)} 
\eeq
\beq 
\sigma_{\text{ex} \rightarrow \, \text{gr}} \approx \frac{4 \pi^2  \cos^2 \varphi \tanh \left( \frac{\pi \ed}{2\mu}\right)}{\eD\, \mu \left( \cosh\left( \frac{\pi \ed}{\mu}\right) - \cos( 2 \varphi)\right)} 
\eeq
\beq 
\sigma_{\text{ex} \rightarrow \, \text{ex}} \approx \frac{1}{\mu^2} \left(\zeta^2 + \frac{2 \pi \cos \varphi}{\cosh\left(\frac{\pi \ed}{\mu}\right) - \cos(2 \varphi)}  \left(2\, \zeta \sin\varphi + \cos\varphi \sech^2 \left(\frac{\pi \ed}{2\mu}\right) \right)\right)^2
\eeq
where for convenience, we have defined the subdominant $\mathcal{O}(1)$ term \beq
\zeta \equiv 2 \gamma - 2\psi_0\left(\frac{i \ed}{2 \mu} + \frac{1}{2}\right) + i \pi \tanh\left(\frac{\pi \ed}{2\mu}\right) + \ln \left(\frac{V_0^2 }{16 \mu^4}\right), \eeq which is a real quantity because the term with the hyperbolic tangent cancels out the imaginary part of the digamma function. 

We see that in all cases there is a potentially large enhancement corresponding to the zero-$\delta$ resonances, $\varphi = n\pi$ so $\cos(2\phi) =1$. The cross section does not actually diverge at these pseudo-resonances, but scales as $1/(\cosh(\pi \ed/\mu) - 1)$, and so is large when $\ed \ll \mu$.

The upscattering cross section vanishes as $\eD \rightarrow 0$, as expected, as the phase space for newly-excited particles shrinks to zero. The elastic scattering cross section for the particles in the excited state scales parametrically as $1/\mu^2$, except close to the resonances, where it instead scales as $1/\ed^2$ if $\ed \ll \mu$. Both these behaviors correspond to geometric cross sections, one governed by the range of the force and one by the momentum transfer associated with virtual de-excitation to the ground state. The downscattering cross section --- which is perhaps most interesting for scenarios where an abundant relic population of dark matter in the halo exists in the excited state --- diverges as $1/\epsilon_\Delta$, meaning that if $v_\mathrm{ex} = \alpha \epsilon_\Delta$ is the physical velocity of the incoming particles in the excited state, $\sigma v_\mathrm{ex}$ will approach a constant value at low velocities. For $\ed \ll \mu$, the cross section scales as $\ed/(\eD \mu^2)$ away from the resonances, and $1/(\eD \ed)$ close to the resonances. Inserting the dimensionful prefactors, the physical cross sections for downscattering and elastic scattering in the excited state have the following scaling behavior:
\beq 
\sigma_{\text{ex} \rightarrow \, \text{gr}}~ v_\mathrm{ex} \propto \frac{1}{m_\phi^2}  \sqrt{\frac{\delta}{m_\chi}}   \, \text{off-resonance}, \quad 
\frac{1}{m_\chi \delta} \sqrt{\frac{\delta}{m_\chi}}  \, \text{near-resonance}
\label{eq:lowvdownscattering}
\eeq
\beq 
\sigma_{\text{ex} \rightarrow \, \text{ex}} \propto \frac{1}{m_\phi^2} \, \text{off-resonance}, \quad \frac{1}{m_\chi \delta} \, \text{near-resonance}.
\eeq

We note that for slow-moving particles initially in the excited state, inelastic downscattering will generally dominate over elastic scattering (due to the $1/v$ scaling). The constant $\sigma v$ for downscattering implies that the argument given in \cite{Loeb:2010gj} (which predicts a constant density core in dwarf galaxies as a direct result of a constant $\sigma v$ for exothermic interactions) holds even at low velocities where the perturbative approach used in that work is not valid. However, the scaling of the constant $\sigma v$ with the parameters of the model is quite different to the perturbative case. Note in particular that in regions of parameter space close to a resonance, large scattering cross sections can be achieved even for large $m_\phi$ (provided $m_\phi \lesssim \alpha m_\chi$ so our approximation holds), depending only on the mass splitting and the dark matter mass rather than the mediator mass.

The cross section for elastic scattering in the ground state does not have a simple behavior close to threshold, since there is nothing special about $\ev \approx \ed$ from the perspective of the ground state.  Setting $\eD=0$ we obtain:
\begin{equation}
\Gamma_v \rightarrow \Gamma \left(1 + \frac{i \epsilon_v}{\mu}\right) \Gamma^2\left( \frac{i \epsilon_v }{2 \mu} + \frac{1}{2}\right),
\end{equation}
\begin{equation}
 \sigma_{\text{gr}\rightarrow \text{gr}} = \frac{\pi}{\ev^2}\, \abs{ 1 + \left(\frac{V_0}{4\mu^2}\right)^{-\frac{2 i\epsilon_v}{\mu}} \left(\frac{\Gamma_v}{\Gamma_v^*} \right) 
 \left[\frac{ \sinh\left(\frac{\pi \epsilon_v }{2\mu} + i \varphi \right)}{ \sinh \left(\frac{\pi \epsilon_v}{2 \mu} - i \varphi \right)} \right]}^2.
\end{equation}
\hspace{-0.1cm}The term involving $\sinh$'s approaches 1 when $\varphi \rightarrow n \pi$ and -1 when $\varphi \rightarrow (n + 1) \pi/2$, which gives rise to the characteristic resonances and anti-resonances in the low-$v$ limit. More explicitly, we can perform a Taylor expansion in the low-velocity limit $\ev \ll \mu$ (here having already set $\ev = \ed$), obtaining:
 \begin{equation}
 \sigma_{\text{gr}\rightarrow \text{gr}}\, \approx \frac{\pi}{\mu^2} \left(\pi \cot(\varphi) + 2 \ln\left(\frac{V_0}{\mu^2}\right) + 4 \gamma \right)^2. \end{equation}
We see that this cross section has the same form as the other low-velocity and low-mass-splitting limits we have studied; it is identical to the expression obtained by first taking $\ev \rightarrow 0$ and then $\ed \rightarrow 0$. 
\section{Applications to Dark Matter Halos}
\label{astropheno}

\subsection{Regimes of Interest for Modification of DM Halos}

\subsubsection{Parameter Choices for Inelastic Scattering}

On dwarf galaxy scales, an elastic scattering cross section of roughly $\sigma/m_\chi \gtrsim 0.1$ cm$^2$/g is required in order for dark matter self-scattering to have a significant impact on the internal structure \cite{Zavala:2012us}. This corresponds to particles in the core interacting once on average over the age of the universe \cite{Loeb:2010gj}, and so is likely also a necessary condition for exothermic downscattering to be relevant. We will thus use this cross section as a benchmark.

As discussed in \cite{Loeb:2010gj}, requiring a significant relic population of particles in the excited state at late times (that was not depleted by scatterings in the early universe) requires $m_\chi$ at the MeV scale or lighter. However, the excited state might be populated non-thermally, in which case much heavier DM masses might also be viable.

For the non-degeneracy of the excited state to have a significant impact on scattering in dwarf galaxies, the mass splitting should be significant compared to the typical kinetic energy of the dark matter particles. Taking the typical velocity in dwarf galaxies to be 10 km/s$ \sim 3 \times 10^{-5} c$ \cite{Walker:2009zp}, this implies $\delta \gtrsim 10^{-9} m_\chi$ in order to see differences from purely elastic scattering. Our requirements that $\epsilon_\delta \lesssim 1, \, \epsilon_\phi \lesssim 1$ impose that $\delta \lesssim \alpha^2 m_\chi$ and $\alpha m_\phi \lesssim \alpha^2 m_\chi$\footnote{Natural scales for the splitting can include $\alpha^2 m_\chi$ (analogous to the splitting between atomic energy levels), $\alpha m_\phi$ or $\alpha^2 m_\phi$ (if the splitting is generated by loops of the mediator) \cite{ArkaniHamed:2008qn}; if the mass splitting is generated by some higher-dimension operator as in \cite{Finkbeiner:2010sm}, then its size depends on the heavy mass scale.}. So in order for our approximation to be valid and the mass splitting to be interestingly large, we will focus on the range $\alpha \gtrsim 10^{-4}$ (or higher for larger $\delta$: $\alpha \gtrsim \sqrt{\delta/m_\chi}$), which will also guarantee $\epsilon_v = v/\alpha \lesssim 1$ as required. For a vector mediator, this is in agreement with broad expectations from the Standard Model if the coupling is not fine-tuned to be small.

In general, we will treat $\alpha$ as a free parameter within the range $10^{-4} \lesssim \alpha \lesssim 1$, since the constraints on it are rather model-dependent. In particular, we do \emph{not} impose the constraint that annihilations of the DM to the force carrier should generate the correct relic density. The self-interacting DM could be non-thermal or a sub-dominant component of the total DM if $\alpha$ is higher than the value expected for a thermal relic, or annihilation channels not involved in the self-scattering could prevent the overclosure of the universe if $\alpha$ is too small. However, for calibration, \cite{Finkbeiner:2010sm} found typical values for $\alpha$ (yielding the correct relic density) of a few times $10^{-2}$ for TeV-scale DM in a model with the same potential as the one employed in this work, and the annihilation of the dark matter to the force carriers has a cross section scaling as $\alpha^2/m_\chi^2$ at freezeout, so lighter DM will generally imply a smaller value of $\alpha$ if the DM is indeed a thermal relic.

There are few model-independent constraints on the mediator mass $m_\phi$; the coupling of the force carrier to Standard Model particles is independent of its role here of mediating dark matter scattering. For significant scattering we require that $m_\phi \lesssim \alpha m_\chi$, and for $s$-wave scattering to dominate and our approximation to be valid we will generally require that $\epsilon_v \lesssim \epsilon_\phi$ in dwarf galaxies, i.e. $m_\phi \gtrsim m_\chi v \sim 3 \times 10^{-5} m_\chi$.

When we consider the exothermic scenario with a significant population of particles initially in the excited state, their scatterings have $\epsilon_v \sim \epsilon_\delta$ in our notation, assuming the kinetic energy of the excited-state particles (limited by the escape velocity of the dwarf) is small compared to the mass splitting. Thus for this scenario we will also require $\epsilon_\delta \lesssim \epsilon_\phi$, i.e. $2 \delta / \alpha^2 m_\chi \lesssim m_\phi^2 / \alpha^2 m_\chi^2 \Rightarrow m_\phi \gtrsim \sqrt{\delta m_\chi}$. The requirement that $m_\phi \lesssim \alpha m_\chi$ means that the higher $\alpha$ is above its lower bound of $\sqrt{\delta/m_\chi}$, the more valid parameter space there will be for $m_\phi$ (although raising $m_\phi$ above the upper bound simply means we should use the Born approximation, as in Appendix \ref{app:born}, rather than the approximations derived in this article.)

\subsubsection{Results}

With the above reasoning as our guide, we performed several cuts through parameter space to estimate phenomenologically interesting regions in DM mass, mediator mass, mass-splitting, and coupling. We selected a few fiducial velocity scales which correspond to the virial velocities of dwarf galaxies, galaxies the size of the Milky Way (MW), and clusters. Practically speaking, the DM scattering rate within a halo is not determined by one fixed velocity, but rather some distribution of velocities as DM particles pass through different parts of the halo. In order to understand these dynamics in detail, one must perform numerical simulations which lie outside the scope of this work. However, the simple estimates presented here can serve as benchmarks for simulations of the dynamical behavior of inelastically scattering DM in an actual halo.

\begin{figure}[h!]
\includegraphics[width=0.32\textwidth]{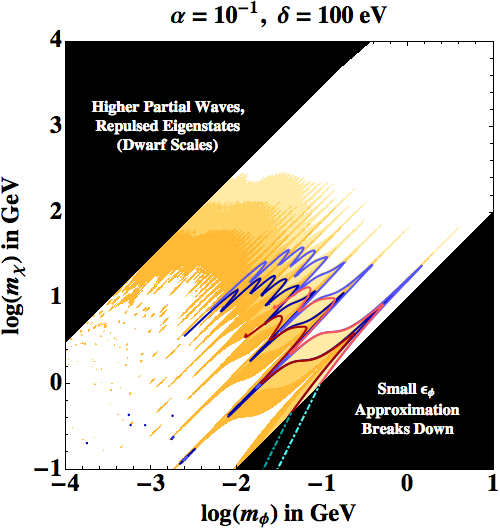}
\includegraphics[width=0.32\textwidth]{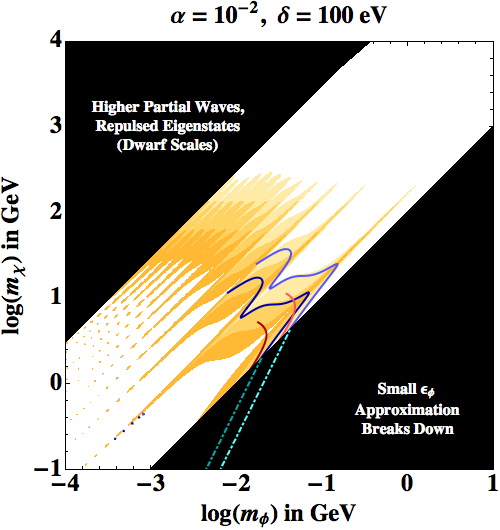}
\includegraphics[width=0.32\textwidth]{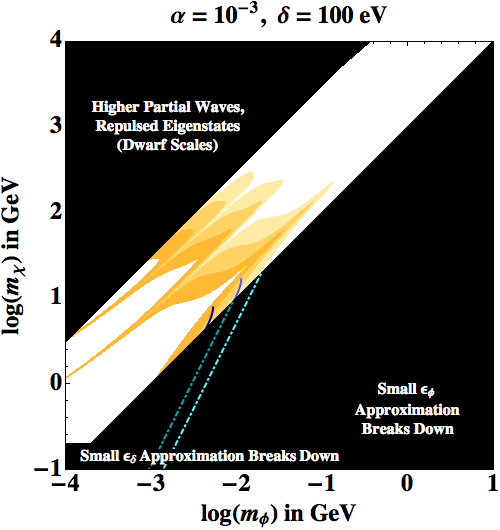}\\\\
\includegraphics[width=0.32\textwidth]{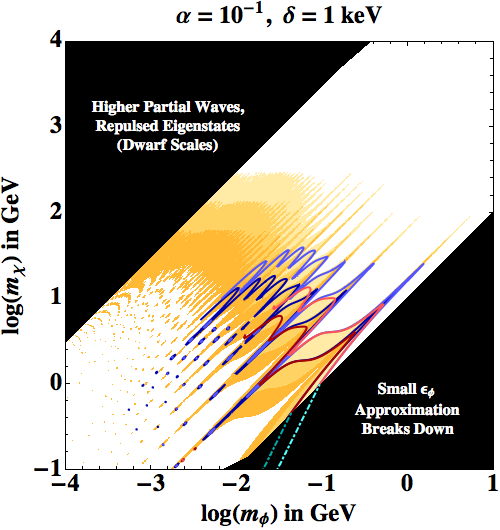}
\includegraphics[width=0.32\textwidth]{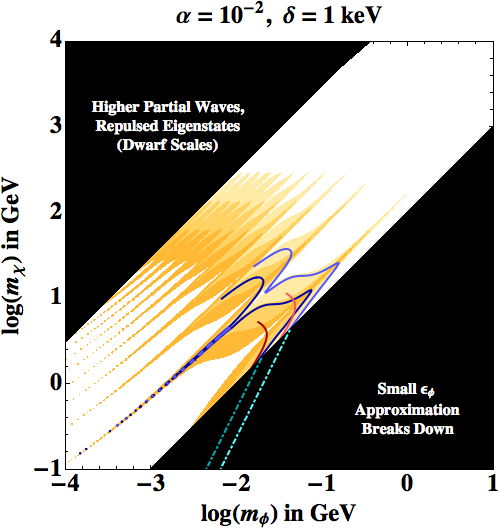}
\includegraphics[width=0.32\textwidth]{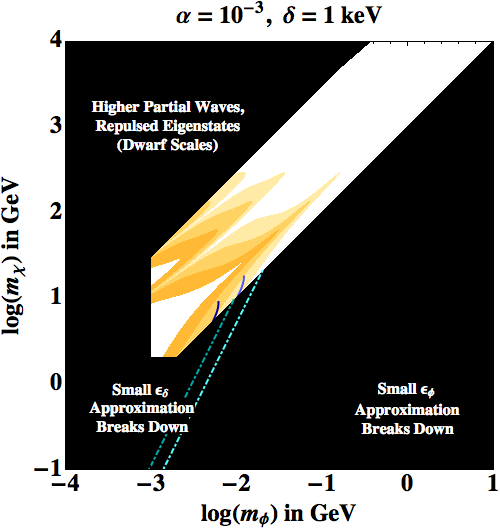}\\\\
\includegraphics[width=0.32\textwidth]{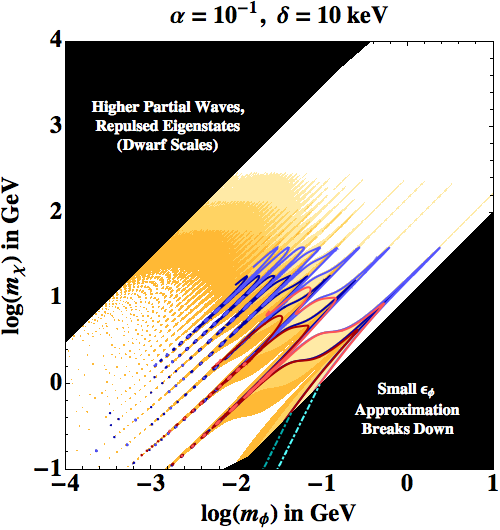}
\includegraphics[width=0.32\textwidth]{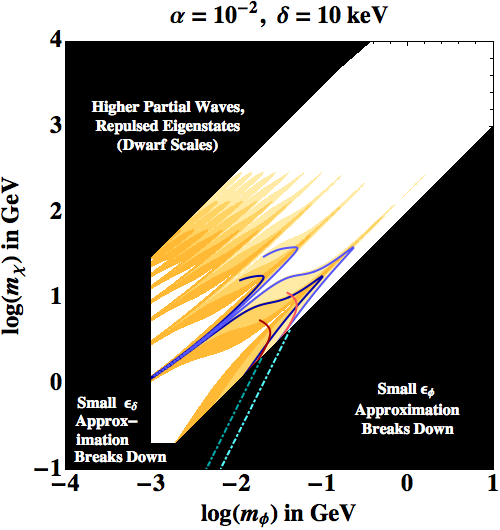}
\includegraphics[width=0.32\textwidth]{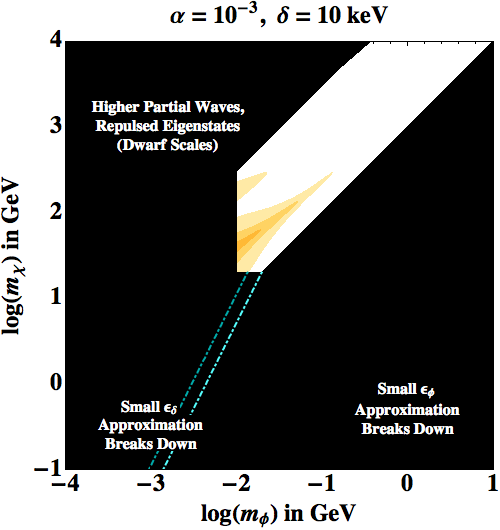}\\\\
\centerline{\includegraphics[height=0.5cm]{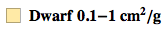}
\includegraphics[height=0.5cm]{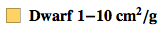}
\includegraphics[height=0.5cm]{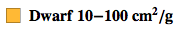}}\\
\centerline{\includegraphics[height=0.6cm]{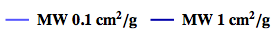}
\includegraphics[height=0.6cm]{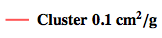}
\includegraphics[height=0.6cm]{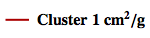}
\includegraphics[height=0.6cm]{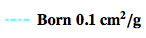}
\includegraphics[height=0.6cm]{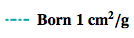}}
\caption{DM and mediator masses with ground$\rightarrow$ground elastic scattering cross sections $\sigma/m_\chi$ in the 0.1-10 cm$^2$/g range, for couplings and mass splittings as labeled. Note that the use of ``log'' indicates the base 10 logarithm.}
\label{fig:g2g}
\end{figure}

\begin{figure}[h!]
\includegraphics[width=0.32\textwidth]{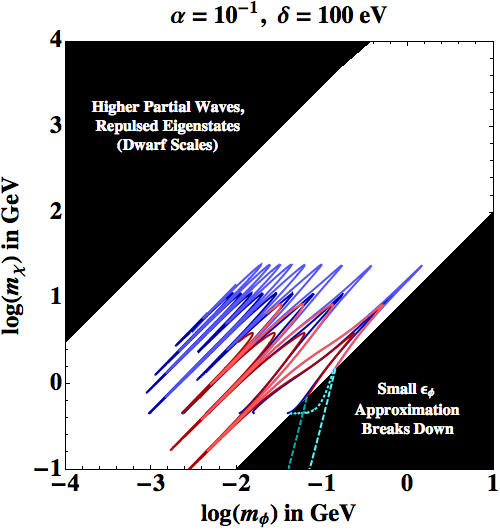}
\includegraphics[width=0.32\textwidth]{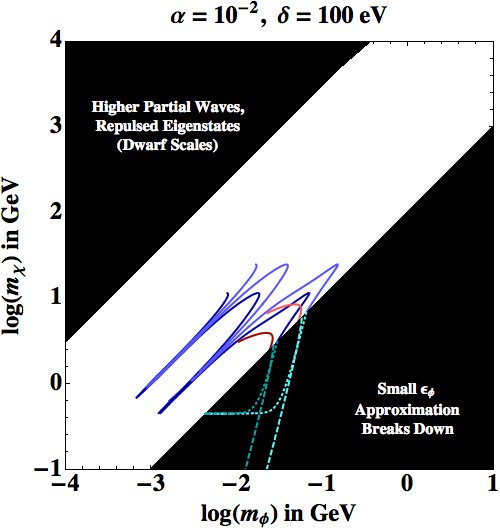}
\includegraphics[width=0.32\textwidth]{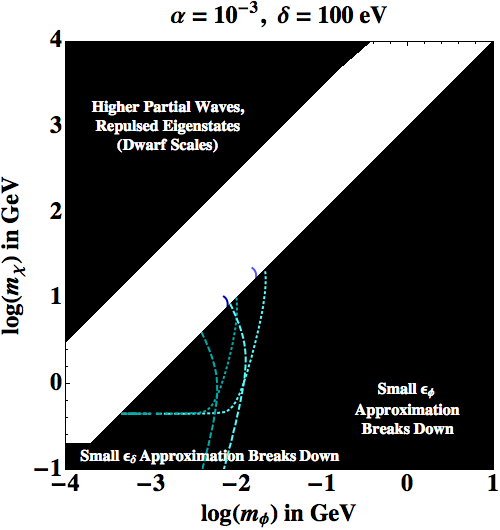}\\\\
\includegraphics[width=0.32\textwidth]{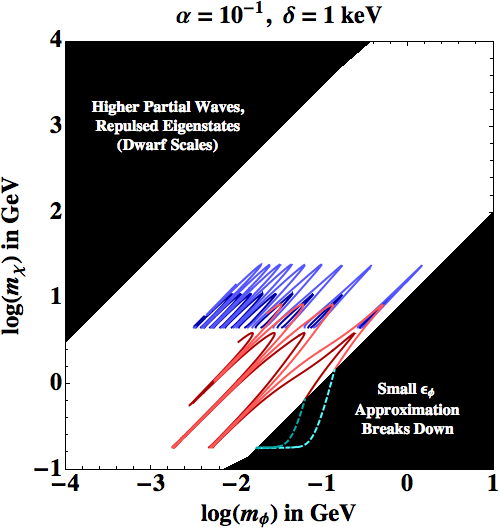}
\includegraphics[width=0.32\textwidth]{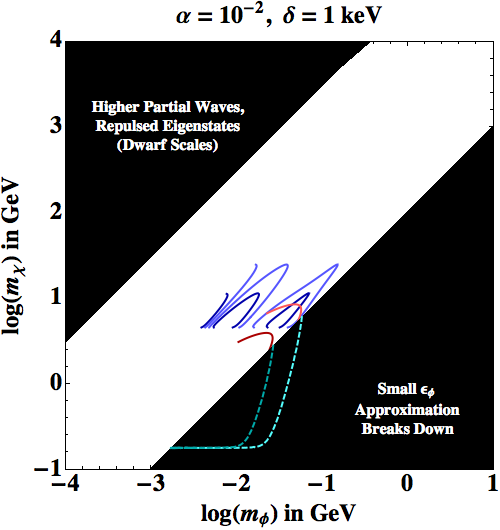}
\includegraphics[width=0.32\textwidth]{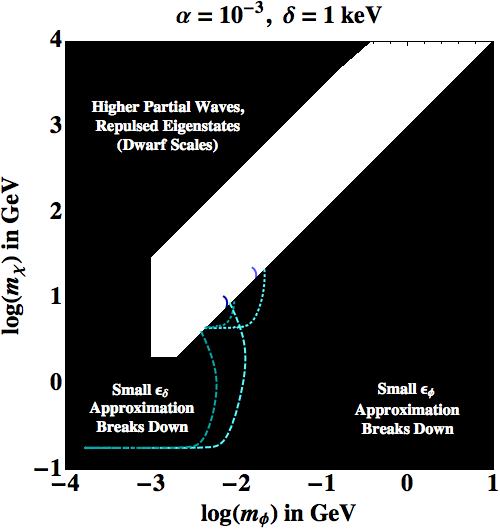}\\\\
\includegraphics[width=0.32\textwidth]{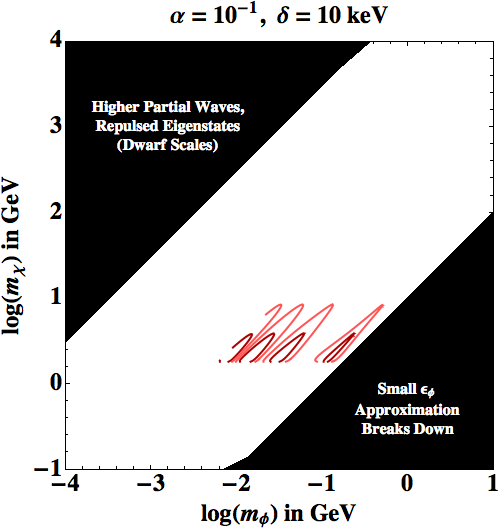}
\includegraphics[width=0.32\textwidth]{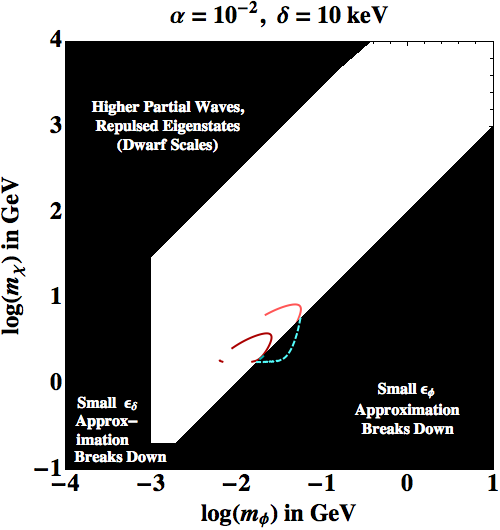}
\includegraphics[width=0.32\textwidth]{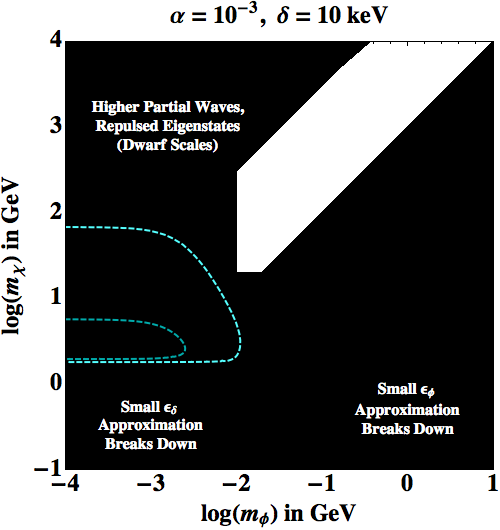}\\\\
\centerline{\includegraphics[height=0.5cm]{dwarf01.png}
\includegraphics[height=0.5cm]{dwarf1.png}
\includegraphics[height=0.5cm]{dwarf10.png}}\\
\centerline{\includegraphics[height=0.6cm]{mw.png}
\includegraphics[height=0.6cm]{cluster01.png}
\includegraphics[height=0.6cm]{cluster1.png}}\\
\centerline{\includegraphics[height=0.6cm]{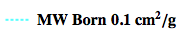}
\includegraphics[height=0.6cm]{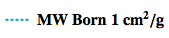}
\includegraphics[height=0.6cm]{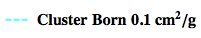}
\includegraphics[height=0.6cm]{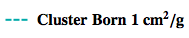}}
\caption{As Figure \ref{fig:g2g}, except for inelastic scattering from the ground state to the excited state (upscattering). Upscattering in dwarf-sized halos, even where kinematically allowed, was never significant for the parameters we sampled (i.e. $\sigma/m_\chi \lesssim 0.01$ cm$^2$/g for all of the parameter space.) Note that the horizontal cutoffs of the contours come from the mass splitting threshold, as upscattering does not occur below threshold.}
\label{fig:g2e}
\end{figure}

\begin{figure}[h!]
\includegraphics[width=0.32\textwidth]{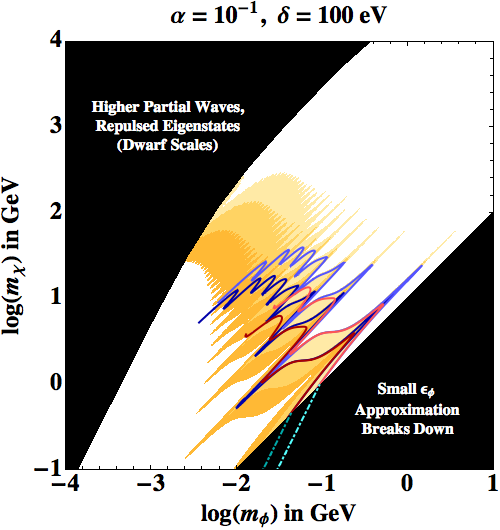}
\includegraphics[width=0.32\textwidth]{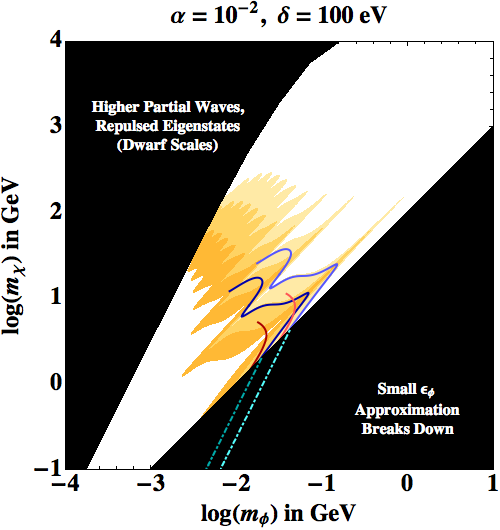}
\includegraphics[width=0.32\textwidth]{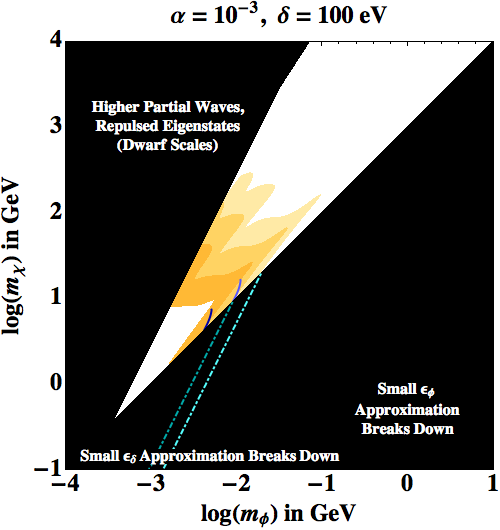}\\\\
\includegraphics[width=0.32\textwidth]{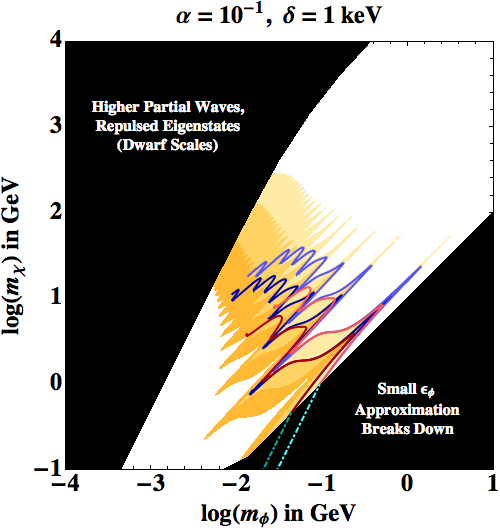}
\includegraphics[width=0.32\textwidth]{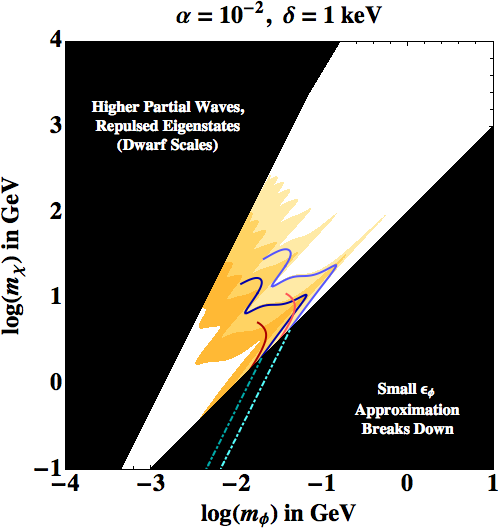}
\includegraphics[width=0.32\textwidth]{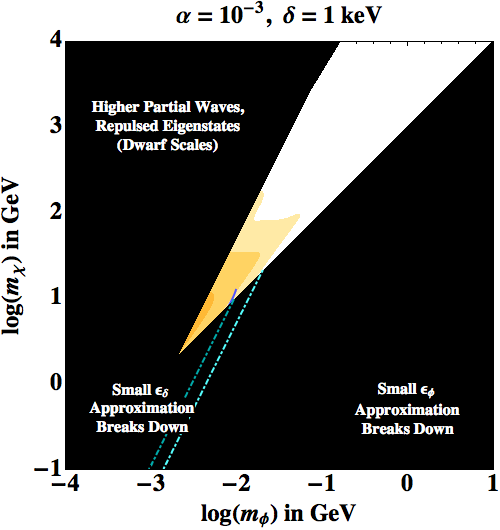}\\\\
\includegraphics[width=0.32\textwidth]{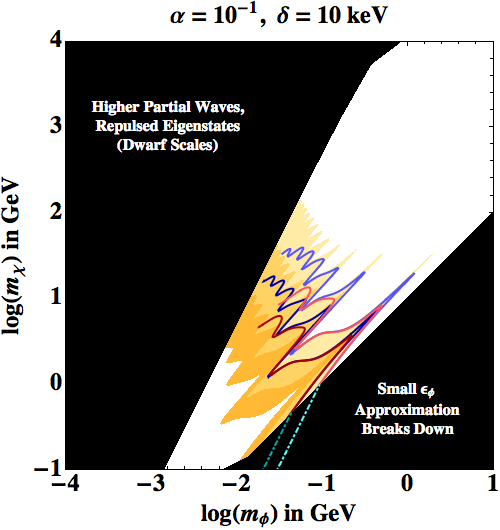}
\includegraphics[width=0.32\textwidth]{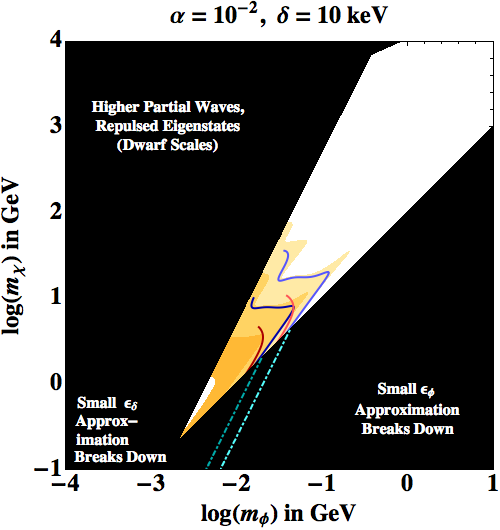}
\includegraphics[width=0.32\textwidth]{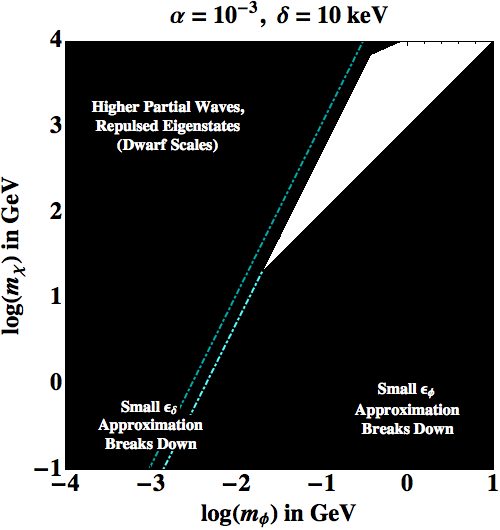}\\\\
\centerline{\includegraphics[height=0.5cm]{dwarf01.png}
\includegraphics[height=0.5cm]{dwarf1.png}
\includegraphics[height=0.5cm]{dwarf10.png}}\\
\centerline{\includegraphics[height=0.6cm]{mw.png}
\includegraphics[height=0.6cm]{cluster01.png}
\includegraphics[height=0.6cm]{cluster1.png}
\includegraphics[height=0.6cm]{born01.png}
\includegraphics[height=0.6cm]{born1.png}}
\caption{As Figure \ref{fig:g2g}, except for elastic scattering from the excited state to the excited state.}
\label{fig:e2e}
\end{figure}

\begin{figure}[h!]
\includegraphics[width=0.32\textwidth]{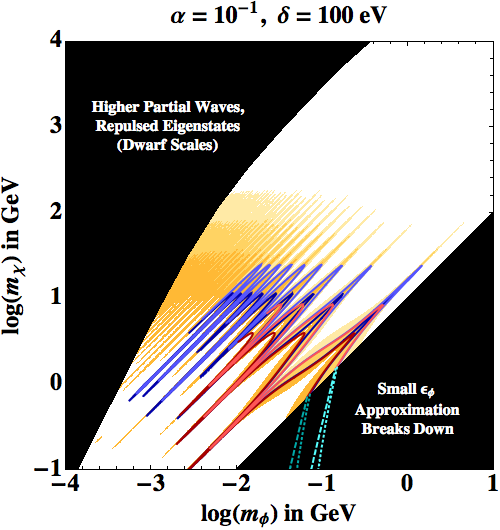}
\includegraphics[width=0.32\textwidth]{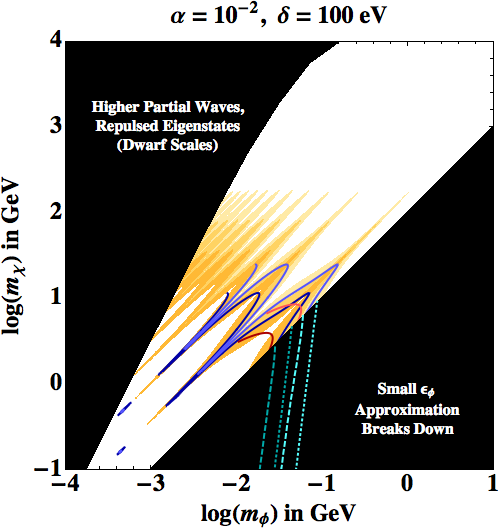}
\includegraphics[width=0.32\textwidth]{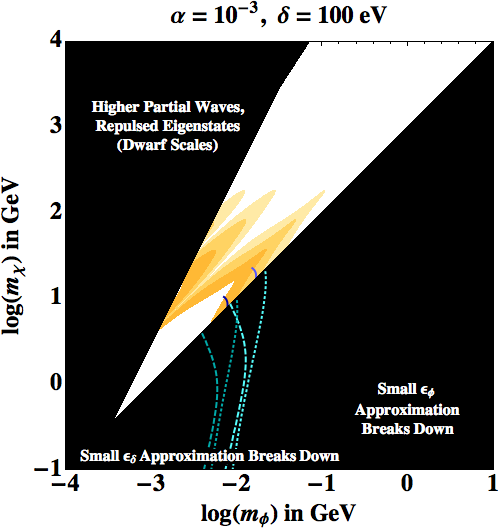}\\\\
\includegraphics[width=0.32\textwidth]{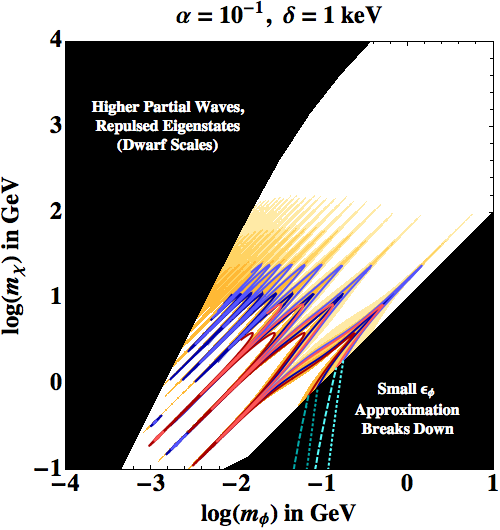}
\includegraphics[width=0.32\textwidth]{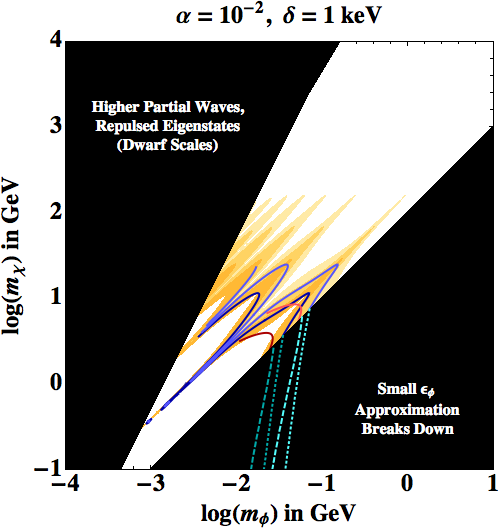}
\includegraphics[width=0.32\textwidth]{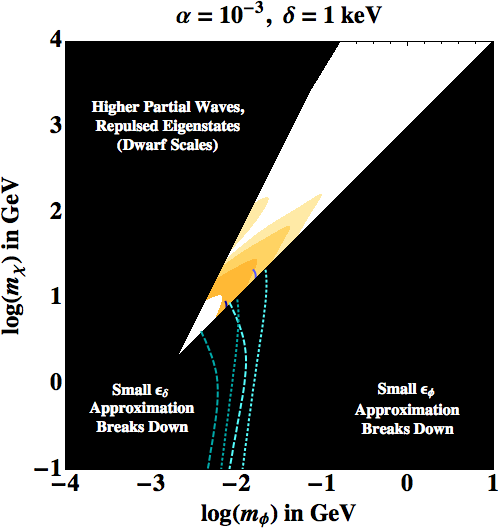}\\\\
\includegraphics[width=0.32\textwidth]{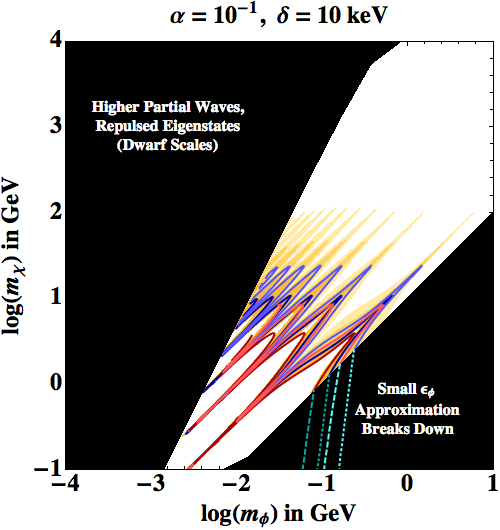}
\includegraphics[width=0.32\textwidth]{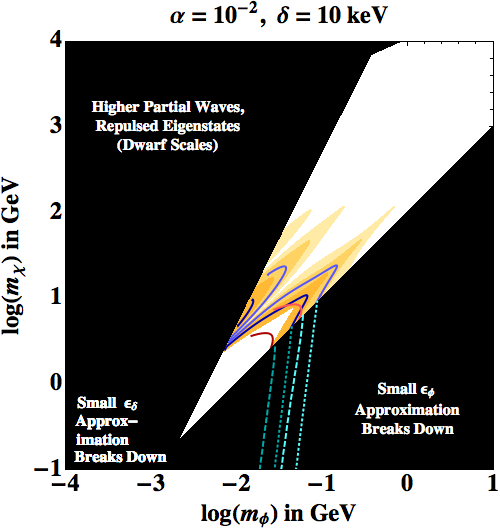}
\includegraphics[width=0.32\textwidth]{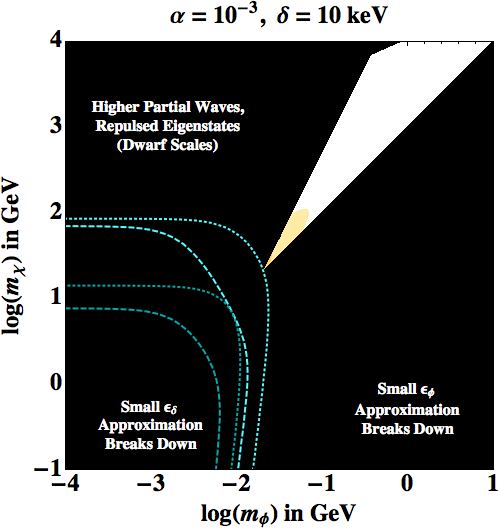}\\\\
\centerline{\includegraphics[height=0.5cm]{dwarf01.png}
\includegraphics[height=0.5cm]{dwarf1.png}
\includegraphics[height=0.5cm]{dwarf10.png}}\\
\centerline{\includegraphics[height=0.6cm]{mw.png}
\includegraphics[height=0.6cm]{cluster01.png}
\includegraphics[height=0.6cm]{cluster1.png}}\\
\centerline{\includegraphics[height=0.6cm]{mwborn01.png}
\includegraphics[height=0.6cm]{mwborn1.png}
\includegraphics[height=0.6cm]{clusterborn01.png}
\includegraphics[height=0.6cm]{clusterborn1.png}}
\caption{As Figure \ref{fig:g2g}, except for inelastic scattering from the excited state to the ground state (downscattering). There are significant portions of parameter space where downscattering could be rapid and where the velocity ``kick'' to DM particles exceeds the halo escape velocity. Such downscattering could potentially dissipate halos and help solve the missing satellites problem.}
\label{fig:e2g}
\end{figure}

The results of our parameter sweeps for four different scattering scenarios (ground to ground, ground to excited, excited to excited, and excited to ground) can be found in Figures \ref{fig:g2g}-\ref{fig:e2g}, respectively. All plots depict $\sigma_T / m_\chi$ in nine slices of the $m_\chi$ vs. $m_\phi$ plane with $\alpha$ and $\delta$ held fixed (note we label the cross section as the transfer cross section $\sigma_T$ for comparison to the literature, but in our $s$-wave approximation it is identical to the cross section we have discussed so far; see Appendix \ref{app:transferxsec} for a discussion). Each slice has different values of $\alpha$ and $\delta$, with $\alpha$ decreasing from left to right and $\delta$ increasing from top to bottom (with each subplot labelled accordingly.) We also include three velocity scales: 10 km/s for dwarf galaxies, 200 km/s for MW-sized galaxies, and 1000 km/s for clusters. In terms of our model, these velocities correspond to $\ev$ for situations where the particles are initially in the ground state and $\eD$ for situations where the particles are initially in the excited state. Additional features of the plots are listed as follows:
\begin{itemize}
\item Black regions show where our approximation breaks down. Since we know that our approximations will give wrong or misleading cross sections in these regions, we cover them entirely. Black, triangular regions in the lower right corner show when $\ephi >1$. Black, horizontal strips across the bottom of the plots (when present) show when $\ed >1$. Black, vertical strips across the left side of the plots (when present) show when $\ed^2/2 > \ephi$. Black regions in the upper left show when higher partial waves and repulsed eigenstates can no longer be ignored, which happens when $\ev> \ephi$. Note that we chose to black out the region with $\ev$ corresponding to dwarf scales, e.g. the blacked out regions have $\ephi < 3 \times 10^{-5} / \alpha$ for scenarios where the particles are incoming in the ground state. Since this criterion is velocity-dependent, we only plot cross sections for DM in the MW and clusters when our approximations are valid at those velocity scales (rather than blacking out regions where the approximation is not valid for those scales.)
\item Yellow regions show $0.1~ \text{cm}^2/\text{g} < \sigma_T/m_\chi <1~ \text{cm}^2/\text{g}$ (light), $1~ \text{cm}^2/\text{g} < \sigma_T/m_\chi <10~ \text{cm}^2/\text{g}$ (medium), and $10~ \text{cm}^2/\text{g} < \sigma_T/m_\chi <100~ \text{cm}^2/\text{g}$ (dark) on dwarf velocity scales with $v \sim$ 10 km/s.
\item Blue contours represent $\sigma_T/m_\chi$ = 0.1 cm$^2/$g (light) and $\sigma_T/m_\chi$ = 1 cm$^2/$g (dark) on MW velocity scales with $v \sim$ 200 km/s. We emphasize that the contours are only plotted when $\ev<\ephi$ on these scales, which ensures that the repulsed eigenstates and higher partial waves are safely neglected.
\item Red contours represent $\sigma_T/m_\chi$ = 0.1 cm$^2/$g (light) and $\sigma_T/m_\chi$ = 1 cm$^2/$g (dark) on cluster velocity scales with $v \sim$ 1000 km/s. We emphasize that the contours are only plotted when $\ev<\ephi$ on these scales, which ensures that the repulsed eigenstates and higher partial waves are safely neglected. The strongest robust constraints on DM self-interaction come from observations of merging clusters, with the upper bound being around $\sigma_T/m_\chi \sim$ 1 cm$^2/$g (e.g. \cite{Randall:2007ph}). Thus, parameter space to the lower left of the dark red contours is effectively ruled out for our model at the selected masses and couplings.
\item Teal contours represent the Born approximation with $\sigma_T/m_\chi$ = 0.1 cm$^2/$g (light) and $\sigma_T/m_\chi$ = 1 cm$^2/$g (dark). 
We present a full derivation of the Born cross sections in \ref{app:born} but report them here.
For elastic scattering in the limit of slow-moving initial particles, the cross sections for ground-state elastic scattering and excited state elastic scattering are respectively \begin{equation} \sigma = \frac{\pi \alpha^4 m_\chi^4}{m_\phi^2 (m_\phi + \sqrt{2 \delta m_\chi})^4}, \quad \sigma = \frac{\pi \alpha^4 m_\chi^4}{m_\phi^2 (m_\phi^2 + 2 \delta m_\chi)^2}.  \end{equation} The Born approximation for the elastic cases is shown in dot-dashed lines (we wish to emphasize that we have taken a low-velocity limit in computing the Born approximation, so the result does not actually depend on the velocity of the incoming particles; it is most likely to be valid for dwarfs where the virial velocities are very low).\\\\
Meanwhile, for the inelastic cases, the upscattering cross section is \begin{equation} \sigma =  \frac{4 \pi \alpha^2 m_\chi^2 \sqrt{1 - \frac{2\delta}{m_\chi v^2}}}{m_\phi^4 \left[\left(1 - 2 \delta m_\chi/m_\phi^2\right)^2 + 4 m_\chi^2 v^2/m_\phi^2 \right]}  \end{equation} and the downscattering cross section is \begin{equation} \sigma =  \frac{4 \pi \alpha^2 m_\chi^2}{ \sqrt{1 - \frac{2\delta}{m_\chi v^2}} m_\phi^4 \left[\left(1 - 2 \delta m_\chi/m_\phi^2\right)^2 + 4 m_\chi^2 v^2/m_\phi^2 \right]}.  \end{equation}  For the inelastic cases, the Born approximation is shown in dotted lines for MW velocities and dashed lines for cluster velocities.

%

\end{itemize} \newpage
\newpage
\newpage
\newpage
\subsubsection{Discussion}
The results in Figures \ref{fig:g2g}-\ref{fig:e2g} show that in the parameter region where our approximation holds,  parameters below the dark red line are definitively ruled out by cluster mergers while yellow regions are favored because they potentially solve the small-scale structure anomalies. Interestingly, tentative evidence from cluster mergers suggests that the dark matter self-interaction cross section may be nonzero, and the favored value of $\sigma/ m_\chi$ is 0.8 cm$^2$/g at around $1\sigma$ \cite{dawson}. Thus, the discovery of future mergers may further constrain self-interacting dark matter or even possibly pick out a strongly-preferred interaction cross section at that velocity scale.

The resonances occur in thin bands of roughly constant $\ephi$  from the lower left corner of the $m_\chi$ vs. $m_\phi$ plane up to the upper right corner. Decreasing $\alpha$ spaces the resonances further apart and broadens them due to the weak dependence of the ratio $\ed/\mu$ (which appears in our expressions for predicted resonance positions) on $\alpha$. Increasing $\delta$ shifts the resonance positions slightly. For elastic scattering in the ground state, increasing $\delta$ elongates the resonance bands and makes them slightly more curved; the most pronounced curvature develops at small $m_\phi$, as expected from the discussion in Section \ref{sec:lowvelocitylimit}. Increasing $\delta$ has the opposite effect on elastic scattering in the excited state, in that it dampens the resonances (as discussed in Section \ref{sec:thresholdlimit}) and so reduces their effect on the favored regions. Our approximations match smoothly onto the perturbative expressions as $\epsilon_\phi \rightarrow 1$, as expected. 

We see that for these mass splittings, in the region of greatest astrophysical interest the ground-ground elastic scattering cross section is generally similar to the case where the DM states are degenerate \cite{Tulin:2013teo}. Additionally, the favored regions are quite comparable for elastic scattering and downscattering, so in the context of this model it is possible for both effects to simultaneously contribute to the dynamics of DM within halos and help alleviate small-scale structure issues. This is a consequence of our choice of mass splittings fairly similar to the kinetic energy of virialized DM particles, since the (non-resonant) ratio of the scattering cross sections at low velocity scales as $\sqrt{\delta/(m_\chi v^2)}$ (see Eq. \ref{eq:lowvdownscattering}). However, for much higher mass splittings the scattering will most likely be in the classical high-velocity regime for most of the parameter space of interest, unless the mediator mass is also raised. However, as expected from our earlier discussion, at small $m_\phi$ the resonances occur for different parameters for downscattering, compared to elastic scattering in the ground state. Consequently, there are regions of parameter space where the ground-state elastic scattering cross section is large and the downscattering rate small, and vice versa.


Interestingly, for the case of upscattering there are no regions in the sampled parameter space where upscattering was significant in dwarf halos (i.e. there were no regions where upscattering exceeded 0.1 cm$^2/$g.) From the perspective of avoiding dark matter ``cooling" in dwarf halos (which could potentially worsen the core-cusp problem, etc.) the ease of suppressing upscattering is an appealing feature. However, there are substantial regions of parameter space where upscattering is significant for the MW and for clusters. It is possible that upscattering could contribute to \emph{steepening} of the density profile in the central parts of such large halos.


\subsection{Regimes of Interest for XrayDM}

There has been a great deal of recent interest in the detection of an apparent $\sim 3.5$ spectral keV line in radio observations of galaxies and galaxy clusters, as a potential signal from dark matter \cite{Bulbul:2014sua, Boyarsky:2014jta}. However, there appears to be some tension between the interpretation of this line as originating from the decays of keV-scale dark matter, and its non-detection in the Virgo cluster. It has been proposed that the signal could instead originate from the upscattering of (weak-scale) dark matter to an excited state $\sim 3.5$ keV heavier than the ground state, followed by decay back to the ground state with emission of a photon \cite{Finkbeiner:2014sja}. This scenario was termed ``XrayDM''. Since the upscattering process involves two particles, the rate of excitations (and hence decays) scales with the density \emph{squared} rather than the density (as would also be the case for annihilation, e.g. \cite{Dudas:2014ixa}), and also depends on the typical velocity of the DM particles: this can modify the relative strength of the signal in different regions\footnote{Recent studies have also claimed tension with or exclusion of the interpretation of the signal as DM decay, based on samples of dwarfs and galaxies, e.g. \cite{Malyshev:2014xqa, Anderson:2014tza}; it is not clear how those exclusions generalize to cases with a different density/velocity dependence than decay, although one would expect a suppression of upscattering in dwarfs due to their lower virial velocities.}. 

\begin{figure}
\center{\includegraphics[width=0.65\textwidth]{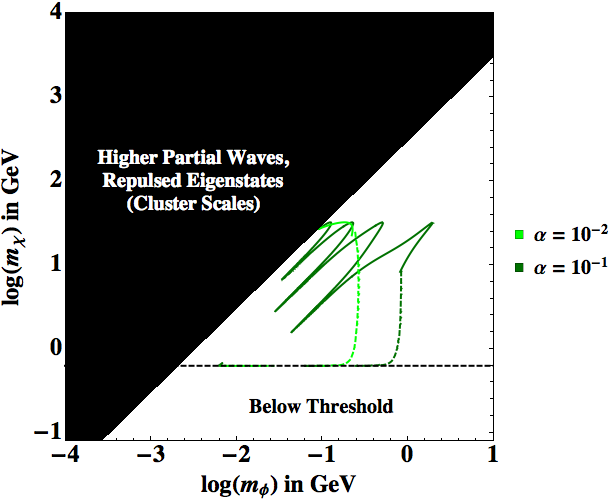}}
\caption{Contours satisfying \eqref{xraydmeqn} for the characteristic velocity scale of clusters, $v = 1000$ km/s. These lines serve as illustrative benchmarks -- in reality, there will of course be some distribution of velocities in any cluster. The region where our approximation breaks down because $\ev > \ephi$ is masked out, but it is likely that this part of parameter space can also furnish appropriate cross sections. The horizontal dashed line indicates the threshold for upscattering to be kinematically allowed, $\ev > \ed$.}
\label{fig:xraydm}
\end{figure}

The example model employed in the XrayDM scenario of \cite{Finkbeiner:2014sja} is \emph{exactly} the simple model studied here: accordingly, we can now use our approximation to calculate which regions of parameter space can give rise to a sufficiently large cross section to explain the 3.5 keV line. Due to the large virial velocities of clusters, for small $m_\phi$ and/or large $m_\chi$ our approximation becomes invalid (as higher partial waves become important), but there is a significant region of interesting parameter space where the $s$-wave contribution dominates, as shown in Figure \ref{fig:xraydm}. 
Here we have imposed the criterion that:
\begin{equation} \sigma_{\mathrm{gr}\rightarrow \mathrm{ex}}   = \left( \frac{10^{-19}~ \mathrm{cm}^3\mathrm{/s}}{1000~ \mathrm{km/s}} \right) \times \left(\frac{m_\chi}{10 \mathrm{GeV}}\right)^2, \label{xraydmeqn}\end{equation}
following \cite{Finkbeiner:2014sja}, with $v = 1000$ km/s and $\delta = 3.5$ keV. As previously, we only show the result for a single velocity, rather than integrating over a distribution, since our purpose here is to provide estimates rather than a detailed analysis of the allowed parameter space.

Within the regime of validity of our approximation, we find that DM masses in the range of a few GeV to a few tens of GeV, moderately large values of $\alpha$, and mediator masses in the range of 10 MeV -- 1 GeV can naturally produce the required flux. We expect that similar DM masses and lower mediator masses will also provide viable explanations, but higher partial waves become important in these cases. We do not show results for smaller $\alpha$ as in that case, for interesting cross sections, the perturbative regime transitions directly to the classical regime (where high partial waves are important) without an intermediate resonant regime -- this does not mean no parameter space is open for smaller $\alpha$, just that it does not require the use of our results.

A similar analysis could be performed for the original ``XDM'' scenario \cite{Finkbeiner:2007kk}; we do not carry it out here because in that case the mass splitting is quite large ($\sim 1$ MeV), meaning that only particles in the high-velocity tail of the velocity distribution (for DM in the Milky Way) are typically able to upscatter. The viability of this scenario thus depends critically on the details of the velocity distribution, and also typically requires a more complex model where both particles in a collision do not need to excite simultaneously (to reduce the energy requirement for upscattering).

\section{Conclusions}
\label{conclusions}
We have derived a semi-analytic approximation for the scattering cross sections for dark matter interacting via an off-diagonal dark Yukawa potential. This physical model demonstrates several new features relative to a system with purely elastic scattering, and we have presented simple analytic forms for the upscattering and downscattering rates in the nonperturbative resonant regime. Within the regime of validity of our approximations, we readily obtain cross sections that appear large enough to modify the inner regions of dwarf halos or give rise to interesting signatures in indirect DM searches.

Our approximations are valid when $v \lesssim m_\phi/m_\chi \lesssim \alpha$ and (for particles initially in the excited state) $\delta \lesssim m_\phi^2/m_\chi \lesssim \alpha^2 m_\chi$, where $m_\chi$ is the DM mass and $m_\phi$ the mass of the mediator. Within this regime, the scattering cross section exhibits resonant enhancement at particular values of the parameter $m_\phi/(\alpha m_\chi)$. For particles initially in the excited state, or scatterings of ground-state particles with energies above the excitation threshold, the resonance positions are very similar to those in the case with no mass splitting: however, for scatterings in the ground state below the excitation threshold, the resonances are shifted by the presence of the mass splitting when $\delta$ becomes comparable to $m_\phi^2/m_\chi$. Consequently, the relationship between the elastic and inelastic scattering cross sections can be quite complex. Away from the resonances, we recover the expected geometric cross sections at low velocities: $\sigma \propto 1/m_\phi^2$ for elastic scattering, $\sigma v \propto \sqrt{\delta/m_\chi} (1/m_\phi^2)$ for downscattering. Our approximations match smoothly onto the corresponding perturbative expressions when $m_\phi \approx \alpha m_\chi$.

We hope that in future work, the incorporation of these semi-analytic scattering cross sections into numerical simulations will allow the first detailed studies of halos containing inelastically scattering dark matter.

\acknowledgments
It is our pleasure to thank Doug Finkbeiner, Adrian Liu, Neal Weiner, and Kathryn Zurek for helpful discussions. We especially thank Hai-Bo Yu for the suggestion that led to this analysis. This work was supported by the U.S. Department of Energy under cooperative research agreement Contract Number DE-FG02-05ER41360. KS was supported in part by an undergraduate research fellowship from the Lord Foundation, a National Science Foundation Graduate Research Fellowship, and a Hertz Foundation Fellowship. 

\appendix
\section{A Brief Review of Previous Results}
\label{sec:rev}
\subsection{Approximate Small-$r$ Wavefunctions}
\label{sec:smallr}
For small $r$, we can approximate the Yukawa potential as $1/r$, which therefore dominates the eigenvalues at small $r$ so $\lambda_\pm \approx \pm 1/r$. The $s$-wave solutions to the rescaled Schr{\"o}dinger equation can be expressed in terms of Bessel functions:
\begin{equation}
\begin{aligned}
& \phi_-(r) = A_- \sqrt{r} J_1 (2 \sqrt{r}) - \pi \phi_-(0) \sqrt{r} Y_1 ( 2 \sqrt{r})  \\
& \phi_+(r) = A_+ \sqrt{r} I _1 (2 \sqrt{r}) + 2 \phi_+(0) \sqrt{r} K_1 (2 \sqrt{r}),
\end{aligned}
\end{equation}
where $A_+$ and $A_-$ are the coefficients of the repulsed and attracted eigenstates, respectively.
Moving radially outward (but still within the regime of validity for the $V \sim 1/r$ approximation), the large-$r$ asymptotics of the Bessel functions give
\begin{equation}
\label{smallwkb1}
\begin{aligned}
& \phi_+(r) = \frac{1}{\lambda_+^{1/4}} \left( \frac{A_+}{2 \sqrt{\pi}} e^{\int_0^r \sqrt{\lambda_+(r')}\, dr'} + \left(\sqrt{\pi} \phi_+(0) - \frac{i A_+}{2 \sqrt{\pi}}\right) e^{-\int_0^r \sqrt{\lambda_+(r')}\, dr'}\right)   \\
& \phi_-(r) = \frac{1}{\lambda_-^{1/4}} \left(-\frac{1}{2\sqrt{\pi}} \left(A_- + i \pi \phi_-(0)\right) e^{\int_0^r \sqrt{\lambda_-(r')}\, dr'} + \frac{i}{2 \sqrt{\pi}} \left(A_- - i \pi \phi_-(0)\right) e^{-\int_0^r \sqrt{\lambda_-(r')}\, dr'}\right).
\end{aligned}
\end{equation}
If the particle velocity is high enough (above threshold) such that there exists a radius $r^*$ in this regime where $V(r^*) = \epsilon_v \sqrt{\epsilon_v^2 - \epsilon_\delta^2}$ then $\lambda_+(r^*) =0$ and we must perform a WKB approximation about the turning point. Linearizing the potential and matching the wavefunctions on either side using the connection formulae yields:
\begin{equation}
\label{smallwkb2}
\begin{aligned}
\phi_+ = \,&\frac{1}{\abs{\lambda_+}^{1/4}} \Bigg[\left(\frac{-i A_+}{2 \sqrt{\pi}} + \frac{1}{2} \left(\sqrt{\pi} \phi_+(0) - \frac{i A_+}{2 \sqrt{\pi}}\right) e^{-2 \int_0^{r^*} \sqrt{\lambda_+(r')} \,dr'}\right) e^{\frac{i \pi}{4} + \int_0^r \sqrt{\lambda_+(r')} \, dr'} \\
&+ \left(\frac{i A_+}{2 \sqrt{\pi}} e^{2 \int_0^{r^*} \sqrt{\lambda_+(r')}\, dr'} + \frac{1}{2} \left(\sqrt{\pi} \phi_+(0) - \frac{i A_+}{2 \sqrt{\pi}}\right)\right) e^{-\frac{i \pi}{4} - \int_0^r \sqrt{\lambda_+(r')} \, dr'}\Bigg].
\end{aligned}
\end{equation}
In either case (above or below threshold), the small-$r$ wavefunctions are in a form that will match smoothly onto the WKB wavefunctions.
\subsection{Approximate Large-$r$ Wavefunctions}
\label{sec:larger}
\indent \indent At large $r$, we can approximate the Yukawa potential as a purely exponential potential of the form $V_0 \,e^{\mu r}$. We impose conditions on $V_0$ and $\mu$ by requiring that the exponential potential mimic the Yukawa for $r> r_M$, where $r_M$ is the matching radius which we have chosen such that $V(r_M) = \text{max}\left(\frac{\ed^2}{2}, \ephi^2\right)$. The potentials should match at $r_M$ so $\frac{e^{-\ephi r_M}}{r_M} = V_0 e^{-\mu \,r_M}$. We also require that $\int_{r_M}^\infty e^{-\ephi r} \, dr = \int_{r_M}^\infty  r\, V_0\, e^{-\mu r}\, dr$, which comes from solving the Lippman-Schwinger form of the Schr{\" o}dinger equation and requiring that the rescaled wavefunctions from both potentials match to first order in the coupling constant $\alpha$ (following \cite{Cassel:2009wt}). The parameters $\mu$ and $ V_0$ are therefore given by
\begin{equation}
\begin{aligned}
&\mu = \ephi \left(\frac{1}{2} + \frac{1}{2} \sqrt{1 + \frac{4 }{\ephi r_M}}~ \right) &&& V_0= \frac{ e^{\ephi r_M \big(-\frac{1}{2} + \frac{1}{2} \sqrt{1 + \frac{4 }{\ephi r_M}}\, \big) }}{r_M} .
\end{aligned}
\end{equation}

The wavefunctions for an exponential potential can be solved for exactly in terms of $_0F_3$ hypergeometric functions \cite{PhysRevA.49.265} as follows:
\begin{equation}
\begin{aligned}
&\psi_1 = C_1\, \left(\frac{V_0 e^{-\mu r} }{4 \mu^2}\right)^{\frac{i \epsilon_v}{\mu}}\,  _0F_3 \left[\{\}, \left\{1 +\frac{i \epsilon_v}{\mu},\, \frac{1}{2} + \frac{i( \epsilon_v -  \epsilon_\Delta)}{2 \mu},\, \frac{1}{2} + \frac{i( \epsilon_v +  \epsilon_\Delta)}{2 \mu} \right \}, \left(\frac{V_0 e^{-\mu r} }{4 \mu^2}\right)^2 \right] \\
&+ C_2 \,\left(\frac{V_0 e^{-\mu r} }{4 \mu^2}\right)^{-\frac{i \epsilon_v}{\mu}}\,  _0F_3 \left[\{\}, \left\{1 - \frac{i \epsilon_v}{\mu},\, \frac{1}{2} - \frac{i( \epsilon_v +  \epsilon_\Delta)}{2 \mu},\, \frac{1}{2} - \frac{i( \epsilon_v -  \epsilon_\Delta)}{2 \mu} \right \}, \left(\frac{V_0 e^{-\mu r} }{4 \mu^2}\right)^2 \right]\\
& + \left(\frac{V_0 e^{-\mu r} }{4 \mu^2}\right)\, \Bigg( C_3 \,\left(\frac{V_0 e^{-\mu r} }{4 \mu^2}\right)^{\frac{i \eD }{ \mu}} \,  _0F_3 \left[\{\}, \left\{\frac{3}{2} - \frac{i( \epsilon_v - \eD)}{2\mu},\, \frac{3}{2} +\frac{i( \epsilon_v +  \epsilon_\Delta)}{2 \mu},\, 1 + \frac{i \epsilon_\Delta }{ \mu} \right \}, \left(\frac{V_0 e^{-\mu r} }{4 \mu^2}\right)^2 \right]\\
& C_4\, \left(\frac{V_0 e^{-\mu r} }{4 \mu^2}\right)^{-\frac{i \eD }{ \mu}} \,  _0F_3 \left[\{\}, \left\{\frac{3}{2} - \frac{i( \epsilon_v + \eD)}{2\mu},\, \frac{3}{2} +\frac{i( \epsilon_v -  \epsilon_\Delta)}{2 \mu},\, 1 - \frac{i \epsilon_\Delta }{ \mu} \right \}, \left(\frac{V_0 e^{-\mu r} }{4 \mu^2}\right)^2 \right] \Bigg)\\
\end{aligned}
\end{equation}
\begin{equation}
\begin{aligned}
\\
&\psi_2 = -\left(\frac{V_0 e^{-\mu r} }{4 \mu^2}\right) \Bigg( \frac{C_1 \left(\frac{V_0 e^{-\mu r} }{4 \mu^2}\right)^{\frac{i \ev}{\mu}}}{\big( \frac{1}{2} + \frac{i \ev}{2 \mu}\big)^2 +\frac{\eD^2}{4 \mu^2}} \,  _0F_3 \left[\{\}, \left\{\frac{3}{2} + \frac{i( \epsilon_v - \eD)}{2\mu},\, \frac{3}{2} +\frac{i( \epsilon_v +  \epsilon_\Delta)}{2 \mu},\, 1 + \frac{i \ev }{ \mu} \right \}, \left(\frac{V_0 e^{-\mu r} }{4 \mu^2}\right)^2 \right] \\
& +\frac{C_2 \left(\frac{V_0 e^{-\mu r} }{4 \mu^2}\right)^{-\frac{i \ev}{\mu}}}{\big( \frac{1}{2} - \frac{i \ev}{2 \mu}\big)^2 +\frac{\eD^2}{4 \mu^2}} \,  _0F_3 \left[\{\}, \left\{\frac{3}{2} - \frac{i( \epsilon_v + \eD)}{2\mu},\, \frac{3}{2} -\frac{i( \epsilon_v -  \epsilon_\Delta)}{2 \mu},\, 1 - \frac{i \ev }{ \mu} \right \}, \left(\frac{V_0 e^{-\mu r} }{4 \mu^2}\right)^2 \right] \Bigg)\\
& - C_3 \left(\frac{V_0 e^{-\mu r} }{4 \mu^2}\right)^{\frac{i \eD}{ \mu}} \left(\frac{\ev^2}{4 \mu^2} + \left(\frac{1}{2} + \frac{i \eD}{2 \mu}\right)^2\right) \,   \\& \times \,  _0F_3 \left[\{\}, \left\{1 +\frac{i \epsilon_\Delta}{\mu},\, \frac{1}{2} - \frac{i( \epsilon_v -  \epsilon_\Delta)}{2 \mu},\, \frac{1}{2} + \frac{i( \epsilon_v +  \epsilon_\Delta)}{2 \mu} \right \}, \left(\frac{V_0 e^{-\mu r} }{4 \mu^2}\right)^2 \right]\\
& - C_4 \left(\frac{V_0 e^{-\mu r} }{4 \mu^2}\right)^{-\frac{i \eD}{ \mu}} \left(\frac{\ev^2}{4 \mu^2} + \left(\frac{1}{2} - \frac{i \eD}{2 \mu}\right)^2\right)\\ &\times \,      _0F_3 \left[\{\}, \left\{1 -\frac{i \epsilon_\Delta}{\mu},\, \frac{1}{2} - \frac{i( \epsilon_v +  \epsilon_\Delta)}{2 \mu},\, \frac{1}{2} + \frac{i( \epsilon_v -  \epsilon_\Delta)}{2 \mu} \right \}, \left(\frac{V_0 e^{-\mu r} }{4 \mu^2}\right)^2 \right],
\end{aligned}
\end{equation} where we have defined $\eD \equiv \sqrt{\ev^2 - \ed^2}\,$. The wavefunctions are expressed in terms of four linearly-independent solutions, corresponding to ingoing or outgoing particles in the ground or excited states. In particular,
\begin{itemize}
\item the $C_1$ term represents an ingoing wave in the ground state,
\item the $C_2$ term represents an outgoing wave in the ground state,
\item the $C_3$ term represents an ingoing wave in the excited state,
\item the $C_4$ term represents an outgoing wave in the excited state.
\end{itemize}
\subsection{WKB Approximation for the Intermediate-$r$ Wavefunctions}
\label{sec:WKB}
To match the large-$r$ wavefunctions to the small-$r$ wavefunctions, one can use the WKB approximation to propagate the known wavefunctions of the exponential potential into the transition region. We write the large-$r$ WKB solutions as
\begin{equation}
\label{largewkb}
\phi_\pm = \frac{1}{\abs{\tilde{\lambda}_\pm}^{1/4}} \left(E_\pm e^{\int_0^r \sqrt{\tilde{\lambda}_\pm(r')} dr'} + F_\pm e^{-\int_0^r \sqrt{\tilde{\lambda}_\pm(r')} dr'} \right).
\end{equation}
where $\tilde{\lambda}_\pm$ are the eigenvalues of the matrix Schr{\"o}dinger equation with an exponential potential rather than a Yukawa.

In order to match the WKB solution with the exact solution for the large-$r$ exponential potential, we define the following convenient quantities:
\begin{equation}
\begin{aligned}
\label{gammas}
&\eta \equiv \frac{1}{4} \left[\frac{\epsilon_v^2}{\mu^2} + \left(1 + \frac{i \epsilon_\Delta}{\mu}\right)^2\right]\\
&\Gamma_v \equiv \Gamma \left(1 + \frac{i \epsilon_v}{\mu}\right) \Gamma \left( \frac{i \epsilon_v - i \epsilon_\Delta}{2 \mu} + \frac{1}{2}\right) \Gamma \left( \frac{i \epsilon_v + i \epsilon_\Delta}{2 \mu} + \frac{1}{2}\right) \\ 
& \Gamma_\Delta  \equiv   \Gamma \left(1 + \frac{i \epsilon_\Delta}{\mu}\right) \Gamma \left( \frac{i \epsilon_\Delta - i \epsilon_v}{2 \mu} + \frac{1}{2}\right) \Gamma \left( \frac{i \epsilon_v + i \epsilon_\Delta}{2 \mu} + \frac{1}{2}\right). 
\end{aligned}
\end{equation}
Then, deriving expressions for the WKB coefficients $E_\pm$ and $F_\pm$ to match onto the exponential wavefunctions is a matter of using the asymptotic behavior of the $_0F_3$ hypergeometric functions in the $r\rightarrow -\infty$ limit (see \cite{Slatyer:2009vg} for details), and using the WKB approximation again to propagate these asymptotic solutions into the matching region. We find that
\begin{equation}
\begin{aligned}
\label{larger}
&E_+=0\\
&F_+  = - \frac{\sqrt{\mu}}{(2 \pi)^{3/2}} e^{\,\int^{r_s}_0 \sqrt{\tilde{\lambda}_+(r')}\, d r'}e^{ \frac{2 i\,\sqrt{ V_0\, e^{- \mu r_s} }}{\mu}  } \left( C_1 \Gamma_v + C_2 \Gamma_v^* + C_3\, \eta\, \Gamma_\Delta + C_4 \,\eta^*\,\Gamma_\Delta^*\right)\\
&E_- = \frac{\sqrt{\mu}}{(2 \pi)^{3/2}} e^{\,\int_{r_s}^0 \sqrt{\tilde{\lambda}_-(r')} \,d r'+ \frac{i \pi}{4}}e^{ -\frac{2 i\,\sqrt{ V_0\, e^{- \mu r_s} }}{\mu}  } \left(C_1 \Gamma_v \, e^{- \frac{\pi \epsilon_v}{\mu}} + C_2 \Gamma_v^* \, e^{\frac{\pi \epsilon_v }{\mu}}  - C_3\,\eta\, \Gamma_\Delta e^{ -\frac{\pi \epsilon_\Delta}{\mu}} - C_4 \,\eta^*\,\Gamma_\Delta^* e^{\frac{\pi \epsilon_\Delta}{\mu}}  \right)\\
& F_- = \frac{\sqrt{\mu}}{(2 \pi)^{3/2}} e^{\int^{r_s}_0 \sqrt{\tilde{\lambda}_-(r')} \,d r'- \frac{i \pi}{4} }e^{ \frac{2 i\,\sqrt{ V_0\, e^{- \mu r_s} }}{\mu}  }\left(C_1 \Gamma_v e^{\frac{\pi \epsilon_v}{\mu}} + C_2 \Gamma_v^* e^{-\frac{\pi \epsilon_v}{\mu}} -C_3 \,\eta\, \Gamma_\Delta e^{\frac{\pi \epsilon_\Delta}{\mu}}  -C_4\,\eta^*\, \Gamma_\Delta ^* e^{-\frac{\pi \epsilon_\Delta}{\mu}}\right)\\
\end{aligned}
\end{equation}
where $r_s$ is some (possibly negative) radius chosen such that (a) $V_0 e^{-\mu r} \gg \mu^2$ (required for the asymptotic expansion to be valid) and (b) the potential dominates the kinetic and mass-splitting terms, i.e. $V_0\,e^{-\mu r_s} \gg \ev^2,\, \ed^2$. There is no reason not to choose $r_s$ arbitrarily large and negative, however, and as discussed in the main text, we generally take $r_s \rightarrow -\infty$. (Of course, this does not correspond to a physical region of real space, but this matching is simply a mathematical trick to translate the hypergeometric functions into a form that facilitates matching to the previously derived WKB solutions.)

Note that this procedure sets $E_+=0$. This originates from the neglect of a particular (exponentially suppressed) term in the asymptotic expansion for the hypergeometric functions. However, this term can gain an exponentially large prefactor when $\ev \gtrsim \mu$, as discussed in \cite{Slatyer:2009vg}; we shall explore the consequences of neglecting this term further in Appendix \ref{sec:scattering}.

In order to match the WKB solution with the small-$r$ solution below threshold, we equate \eqref{largewkb} with \eqref{smallwkb1}, which gives
\begin{equation}
\label{below}
\begin{aligned}
& E_+ = \frac{A_+}{2 \sqrt{\pi}} ~ e^{\int_0^{r_M} \left( \sqrt{\lambda_+} -\sqrt{\tilde{\lambda}_+} \right)dr'}\\
&F+ =  \left(\sqrt{\pi} \phi_+(0) - \frac{i A_+}{2 \sqrt{\pi}} \right) e^{-\int_0^{r_M} \left( \sqrt{\lambda_+} -\sqrt{\tilde{\lambda}_+} \right)dr'}\\
& E_- = -\frac{(-1)^{1/4}}{2 \sqrt{\pi}} \left(A_- + i \pi \phi_-(0) \right) e^{\int_0^{r_M} \left( \sqrt{\lambda_-} -\sqrt{\tilde{\lambda}_-} \right)dr'}\\
&F_- = \frac{i (-1)^{1/4}}{2 \sqrt{\pi}} \left(A_- - i \pi \phi_-(0) \right) e^{-\int_0^{r_M} \left( \sqrt{\lambda_-} -\sqrt{\tilde{\lambda}_-} \right)dr'}\\
\end{aligned}
\end{equation}
and similarly, above threshold, equating \eqref{largewkb} with \eqref{smallwkb2} gives\\
\begin{equation}
\label{above}
\begin{aligned}
E_+ &= \frac{i A_+}{2 \sqrt{\pi}}  \left[ e^{2 \int_0^{r^*} \sqrt{\lambda_+} \, dr- \int_0^{r_M} \left( \sqrt{\lambda_+} -\sqrt{\tilde{\lambda}_+} \right)dr}  -e ^{2 \int_0^{r^\dagger} \sqrt{\tilde{\lambda}_+} \, dr+ \int_0^{r_M} \left( \sqrt{\lambda_+} -\sqrt{\tilde{\lambda}_+} \right)dr} \right] \\
&+ \frac{1}{2} \left(\sqrt{\pi} \phi_+(0) - \frac{i A_+}{2 \sqrt{\pi}}\right) \left[e^{- \int_0^{r_M} \left( \sqrt{\lambda_+} -\sqrt{\tilde{\lambda}_+} \right)dr} + e^{2 \int_0^{r^\dagger} \sqrt{\tilde{\lambda}_+} \, dr \,-\, 2 \int_0^{r^*} \sqrt{\lambda_+} \, dr + \int_0^{r_M} \left( \sqrt{\lambda_+} -\sqrt{\tilde{\lambda}_+} \right)dr}\right]\\
F_+ &= \, \frac{1}{2i} \Bigg\{ \frac{i A_+}{2 \sqrt{\pi}} \left[ e^{2 \int_0^{r^*} \sqrt{\lambda_+} \, dr \, -\, 2\int_0^{r^\dagger} \sqrt{\tilde{\lambda}_+} \, dr \,-\, \int_0^{r_M} \left( \sqrt{\lambda_+} -\sqrt{\tilde{\lambda}_+} \right)dr} + e^{\int_0^{r_M} \left( \sqrt{\lambda_+} -\sqrt{\tilde{\lambda}_+} \right)dr} \right]\\
& +\frac{1}{2} \left(\sqrt{\pi} \phi_+(0) - \frac{i A_+}{2\sqrt{\pi}}\right) \left[ e^{- \int_0^{r_M} \left( \sqrt{\lambda_+} -\sqrt{\tilde{\lambda}_+} \right)dr \, -\, 2\int_0^{r^\dagger} \sqrt{\tilde{\lambda}_+} \, dr} - e^{-2 \int_0^{r^*} \sqrt{\lambda_+} \, dr + \int_0^{r_M} \left( \sqrt{\lambda_+} -\sqrt{\tilde{\lambda}_+} \right)dr} \right] \Bigg\},
\end{aligned}
\end{equation}
where $r^\dagger$ is the radius above threshold at which the eigenvalue of the repulsed eigenstate of the exponential potential passes through zero, defined by $V_0\,e^{-\mu r^\dagger} = \ev \sqrt{\ev^2-\ed^2\,}$.

We can then equate the coefficients from matching the WKB solution onto the small-$r$ solution with the coefficients from matching the WKB solution with the large-$r$ solutions. After imposing appropriate boundary conditions, we can then determine the coefficients for the various physical solutions to the Schr{\"o}dinger equation. For instance, we can calculate the coefficients for the repulsed and attracted eigenstates for the small-$r$ solution. We can also determine the coefficients for the ingoing and outgoing spherical waves in the ground or excited states for the large-$r$ solution.
\section{Extracting Scattering Amplitudes from the Wavefunctions}
\label{sec:scattering}
\subsection{Matching the Wavefunctions Using the Boundary Conditions}
As mentioned in Section~\ref{sec:scat}, we require that $\phi_+(0) = \phi_-(0)=0$, since $\phi(r)$ is the radial wavefunction rescaled by $r$ and the wavefunctions should be regular at the origin. For utility, we then define the following useful phases:
\begin{equation}
\label{eq:phi}
\begin{aligned}
&i \theta \equiv \int_0^{r_M} \sqrt{\lambda_+} dr + \int_{r_M}^{r_s} \sqrt{\tilde{\lambda}_+} dr + \frac{2 i\,\sqrt{ V_0\,  }}{\mu}\,e^{- \mu r_s/2} \\
&i \varphi \equiv \int_0^{r_M} \sqrt{\lambda_-} dr + \int_{r_M}^{r_s} \sqrt{\tilde{\lambda}_-} dr + \frac{2 i\,\sqrt{ V_0\, }}{\mu}\,e^{- \mu r_s/2},
\end{aligned}
\end{equation}
where $r_s$ is some radius chosen such that $V_0\,e^{-\mu r_s} \gg \ev^2,\, \ed^2$, as in \eqref{larger}. Then, for the below-threshold case, equating \eqref{larger} with \eqref{below} at the matching radius, $r_M$ gives 
\begin{equation}
\begin{aligned}
&E_+=\frac{A_+}{2 \sqrt{\pi}} e^{i \theta} = 0\\
&F_+=\frac{i A_+}{2 \sqrt{\pi}} e^{-i \theta} = \frac{\sqrt{\mu}}{(2\pi)^{3/2}} \left[ C_1 \Gamma_v+C_2 \Gamma_v^* + C_3 \,\eta\, \Gamma_\Delta+ C_4 \,\eta^*\,\Gamma_\Delta^*   \right]\\
&E_- =-\frac{A_- \,e^{i \varphi} }{2 \sqrt{\pi}}  = \frac{\sqrt{\mu}}{(2 \pi)^{3/2}} \left[ C_1 \Gamma_v e^{- \frac{\pi \epsilon_v}{\mu}}+ C_2 \Gamma_v^* e^{ \frac{\pi \epsilon_v}{\mu}}  - C_3 \,\eta\,\Gamma_\Delta e^{ - \frac{\pi \epsilon_\Delta}{\mu}}- C_4\,\eta^*\, \Gamma_\Delta^* e^{ \frac{\pi \epsilon_\Delta}{\mu}}    \right]\\
&F_-=- \frac{ A_- \,e^{-i \varphi}}{2 \sqrt{\pi}}   = \frac{\sqrt{\mu}}{(2 \pi)^{3/2}} \left[  C_1 \Gamma_v e^{ \frac{\pi \epsilon_v}{\mu}} +C_2 \Gamma_v^* e^{- \frac{\pi \epsilon_v}{\mu}}- C_3\,\eta\, \Gamma_\Delta e^{ \frac{\pi \epsilon_\Delta}{\mu}}  - C_4 \,\eta^*\, \Gamma_\Delta^* e^{- \frac{\pi \epsilon_\Delta}{\mu}} \right],\\
\label{efeqs}
\end{aligned}
\end{equation}
where we have defined $\eD \equiv \sqrt{\ev^2 - \ed^2}$ and $\Gamma_v$ and $\Gamma_\Delta$ are defined in \eqref{gammas}.
Similarly, above threshold, equating \eqref{larger} with \eqref{above} gives
\begin{equation}
\begin{aligned}
E_+&= \frac{i A_+}{2 \sqrt{\pi}}  \Bigg[\left( e^{2 \int_0^{r^*} \sqrt{\lambda_+} \, dr- \int_0^{r_M} \left( \sqrt{\lambda_+} -\sqrt{\tilde{\lambda}_+} \right)dr}  -e ^{2 \int_0^{r^\dagger} \sqrt{\tilde{\lambda}_+} \, dr+ \int_0^{r_M} \left( \sqrt{\lambda_+} -\sqrt{\tilde{\lambda}_+} \right)dr} \right) \\
&- \frac{1}{2}  \left(e^{- \int_0^{r_M} \left( \sqrt{\lambda_+} -\sqrt{\tilde{\lambda}_+} \right)dr} + e^{2 \int_0^{r^\dagger} \sqrt{\tilde{\lambda}_+} \, dr \,-\, 2 \int_0^{r^*} \sqrt{\lambda_+} \, dr + \int_0^{r_M} \left( \sqrt{\lambda_+} -\sqrt{\tilde{\lambda}_+} \right)dr}\right) \Bigg] = 0\\
F_+& = \frac{i A_+}{4 \sqrt{\pi}}  \Bigg\{ \left[ e^{2 \int_0^{r^*} \sqrt{\lambda_+} \, dr \, -\, 2\int_0^{r^\dagger} \sqrt{\tilde{\lambda}_+} \, dr \,-\, \int_0^{r_M} \left( \sqrt{\lambda_+} -\sqrt{\tilde{\lambda}_+} \right)dr} + e^{\int_0^{r_M} \left( \sqrt{\lambda_+} -\sqrt{\tilde{\lambda}_+} \right)dr} \right]\\
& -\frac{1}{2}\left[ e^{- \int_0^{r_M} \left( \sqrt{\lambda_+} -\sqrt{\tilde{\lambda}_+} \right)dr \, -\, 2\int_0^{r^\dagger} \sqrt{\tilde{\lambda}_+} \, dr} - e^{-2 \int_0^{r^*} \sqrt{\lambda_+} \, dr + \int_0^{r_M} \left( \sqrt{\lambda_+} -\sqrt{\tilde{\lambda}_+} \right)dr} \right] \Bigg\}\\
& = \frac{\sqrt{\mu}}{(2\pi)^{3/2}} \left[ C_1 \Gamma_v+C_2 \Gamma_v^* + C_3\,\eta\, \Gamma_\Delta+ C_4 \,\eta^*\,\Gamma_\Delta^*   \right]\\
\end{aligned}
\end{equation}

Both above and below threshold, the $E_+$ equation gives us $A_+ = 0$, which is akin to neglecting the contribution from the repulsed eigenstate at small radii. This indicates that repulsive scattering occurs most significantly at large $r$, where the exponential part of the Yukawa dominates the behavior of the wavefunction. This makes intuitive sense since we are interested in the low-velocity limit, which means that incoming particles must climb up to the classically disallowed region of a repulsive potential in order to even reach the small-$r$ region. Recall that the result $E_+=0$ was a consequence of assuming $\ev \lesssim \mu$: when this assumption is not satisfied, we cannot expect $A_+=0$ on physical grounds.

Since $A_+ = 0$, the $F_+$ equation gives \begin{equation}\label{sum} C_1 \Gamma_v+C_2 \Gamma_v^* + C_3 \,\eta\, \Gamma_\Delta+ C_4 \,\eta\, \Gamma_\Delta^* =0 ,\end{equation} both above and below  threshold. 
\subsection{General Scattering Amplitudes} 
Once we have imposed the relevant boundary conditions on the large-$r$ wavefunctions, we can extract the scattering amplitudes by reading off the coefficients of the outgoing solutions. Since the hypergeometric functions asymptote to 1 as $r \rightarrow \infty$, the large-$r$ ground and excited wavefunctions approach ingoing and outgoing spherical waves:
\begin{equation}
\begin{aligned}
& \psi_1 =C_1 \left(\frac{V_0}{4 \mu^2}\right)^{i \epsilon_v/\mu} e^{-i \epsilon_v r} + C_2 \left(\frac{V_0}{4 \mu^2}\right)^{-i \epsilon_v / \mu} e^{\,i \epsilon_v r}  \\\\
& \psi_2 = - C_3\,\eta \left(\frac{V_0}{4 \mu^2}\right) ^{ i \epsilon_\Delta /\mu} e^{-i \epsilon_\Delta r}- C_4 \,\eta^* \left(\frac{V_0}{4 \mu^2}\right) ^{- i \epsilon_\Delta /\mu} e^{i \epsilon_\Delta r}  .
\end{aligned}
\end{equation}

More generally, consider wavefunctions for nondegenerate states $X$ and $Y$ given by 
\begin{equation}
\begin{aligned}
& \psi_X = (A+B) e^{i k r} - A e^{-i k r} \\
& \psi_Y  = C e^{i k' r}
\label{xandy}
\end{aligned}
 \end{equation}
where the $A$ terms represent the unscattered wavefunction in state $X$, the $B$ term represents the elastically scattered wavefunction in state $X$, and the $C$ term represents the inelastically scattered wavefunction in state $Y$ (hence the wavenumber $k'$ as distinct from $k$.) In this case, conservation of probability current dictates that $k \abs{A}^2 = k \abs{A + B}^2 + k' \abs{C}^2$. We can reformulate these wavefunctions (recall that all wavefunctions used in this paper are rescaled by $r$) in the context of a 3-dimensional scattering problem as an ingoing cylindrical wave and a scattered outgoing spherical wave: \begin{equation} \label{griff} 
\psi = N \left(r \colvec{e^{i k z}}{0} + \colvec{f_X(\theta) e^{i k r}}{f_Y(\theta) e^{i k'r}}\right)
 = N \left(\frac{1}{2 i k } \colvec{e^{i k r} - e^{- i kr}}{0} +  \colvec{f_X(\theta) e^{i k r}}{f_Y(\theta) e^{i k'r}}\right)
 \end{equation}
where the second equality follows because for the s-wave, \begin{equation}e^{i kz} \rightarrow j_0 (kr) P_0\left(\cos \theta \right) = \frac{\sin kr }{kr}. \end{equation} In general to get a differential cross section, we relate the incident probability flux through an area to the scattered outgoing probability flux through a solid angle:
\begin{equation}
\begin{aligned}
\frac{d P_\text{in}}{d \sigma\, dt} &= \abs{\psi_\text{in}}^2  v_\text{in} = \abs{N}^2 \frac{\hbar\, k_\text{in}}{m},\\
\frac{d P_\text{out}}{r^2 d \Omega \,dt} &= \abs{\psi_\text{out}}^2  v_\text{out} = \frac{\abs{N}^2 \abs{f(\theta)}^2}{r^2} \frac{\hbar\, k_\text{in}}{m}\\ \Rightarrow \frac{d\sigma}{d\Omega} &= \frac{k_\text{out}}{k_\text{in}} \abs{f(\theta)}^2
\end{aligned}
\end{equation}
Equating the top row of \eqref{griff} with $\psi_X$ from \eqref{xandy} gives \begin{equation}\begin{aligned} &A = \frac{N}{2 i k}, \\&B = N f_X(\theta)\\ \Rightarrow& \frac{d\sigma_\text{elastic}}{d \Omega} = \abs{f_X(\theta)}^2 = \frac{1}{4 k^2} \frac{\abs{A}^2}{\abs{B}^2}\\\Rightarrow &\sigma_\text{elastic} =  \frac{\pi}{ k^2} \frac{\abs{B}^2}{\abs{A}^2}.\end{aligned} \label{eq:AB}\end{equation} 
Similarly, equating the bottom row of \eqref{griff} with $\psi_Y$ from \eqref{xandy} gives \begin{equation}\begin{aligned}&C = N f_Y(\theta)\\\Rightarrow& \frac{d\sigma_\text{inelastic}}{d \Omega} = \frac{k'}{k} \abs{f_Y(\theta)}^2 = \frac{k'}{4 k^3} \frac{\abs{C}^2}{\abs{A}^2}\\ \Rightarrow & \sigma_\text{inelastic} = \frac{\pi k'}{ k^3} \frac{\abs{C}^2}{\abs{A}^2}\end{aligned}\label{eq:C}\end{equation}

If we apply the analogy to our wavefunctions for the case where we begin purely in the ground state (which corresponds to setting $C_3$ to zero), then the elastic scattering cross section is
\begin{equation}
\label{elground}
\sigma_\text{elastic} = \frac{\pi}{\ev^2} \,\abs{C_2  \left(\frac{V_0}{4 \mu^2}\right)^{-i \epsilon_v/\mu} +  \left(\frac{V_0}{4 \mu^2}\right)^{i \epsilon_v/\mu}}^2
\end{equation}
and the inelastic scattering cross section is
\begin{equation}
\label{inelground}
\sigma_\text{inelastic} = \frac{\pi \epsilon_\Delta}{\epsilon_v^3} \abs{C_4\,\eta^*}^2.
\end{equation}

Similarly, for the case where we begin purely in the excited state (which corresponds to setting $C_1$ to zero), then the elastic scattering cross section is
\begin{equation}
\label{elexcited}
\sigma_\text{elastic}= \frac{\pi \abs{C_4\, \eta^*\left(\frac{V_0}{4 \mu^2}\right) ^{- i \epsilon_\Delta /\mu}   +  \eta\left(\frac{V_0}{4 \mu^2}\right) ^{ i \epsilon_\Delta /\mu} }^2}{\eD^2 \abs{\eta}^2}
\end{equation}
and the inelastic scattering cross section is 
\begin{equation}
\label{inelexcited}
\sigma_\text{inelastic} = \frac{\pi \epsilon_v \abs{C_2}^2}{\epsilon_\Delta^3 \abs{\eta}^{2}}~.
\end{equation}
\subsection{Deriving the Scattering Cross Sections for our Model}
\subsubsection{Incoming in the Ground State}
We will impose boundary conditions such that the ingoing wave is purely in the ground state, which implies that $C_3 = 0$. We are free to set $C_1 = 1$ up to some overall normalization. 
Dividing the $E_-$ equation by the $F_-$ equation of \eqref{efeqs} yields
\begin{equation}
e^{2 i \varphi} = \frac{ \Gamma_v e^{- \frac{\pi \epsilon_v}{\mu}}+ C_2 \Gamma_v^* e^{ \frac{\pi \epsilon_v}{\mu}} - C_4 \,\eta^*\,\Gamma_\Delta^* e^{ \frac{\pi \epsilon_\Delta}{\mu}} }
{ \Gamma_v e^{ \frac{\pi \epsilon_v}{\mu}} +C_2 \Gamma_v^* e^{- \frac{\pi \epsilon_v}{\mu}}  - C_4\,\eta^*\, \Gamma_\Delta^* e^{- \frac{\pi \epsilon_\Delta}{\mu}}}\\ \\
\end{equation}
and combining this with \eqref{sum} gives
\begin{equation}
\begin{aligned}
 C_4 
&=\frac{ -2\, \Gamma_v\, \text{sinh}\left(\frac{\pi \epsilon_v }{ \mu}\right)  \left[1 +e^{2 i \varphi} \right]}{\eta^*\,\Gamma_\Delta^*\left[ \left( e^{ \frac{\pi \epsilon_\Delta}{\mu}} +e^{ \frac{\pi \epsilon_v}{\mu}}\right)-  e^{2 i \varphi} \left( e^{ - \frac{\pi \epsilon_\Delta}{\mu}} +  e^{ - \frac{\pi \epsilon_v}{\mu}}\right) \right] }.\\
 C_2  &= \frac{\Gamma_v}{\Gamma_v^*}\left[\frac{ 2\, \text{sinh}\left(\frac{\pi \epsilon_v }{ \mu}\right)  \left[1 +e^{2 i \varphi} \right]}{ \left( e^{ \frac{\pi \epsilon_\Delta}{\mu}} +e^{ \frac{\pi \epsilon_v}{\mu}}\right)-  e^{2 i \varphi} \left( e^{ - \frac{\pi \epsilon_\Delta}{\mu}} +  e^{ - \frac{\pi \epsilon_v}{\mu}}\right)  } -1\right]
\end{aligned}
\end{equation}
So by \eqref{elground} the elastic scattering cross section is 
\begin{equation}
\sigma_\text{elastic} = \frac{\pi}{\ev^2}\, \abs{ 1 + \left(\frac{V_0}{4\mu^2}\right)^{-\frac{2 i\epsilon_v}{\mu}} \left(\frac{\Gamma_v}{\Gamma_v^*} \right) 
 \left[\frac{\cosh\left(\frac{\pi (\epsilon_\Delta + \epsilon_v)}{2 \mu} \right) \sinh\left(\frac{\pi (\epsilon_v - \epsilon_\Delta)}{2\mu} + i \varphi \right)}{\cosh\left( \frac{\pi (\epsilon_v - \epsilon_\Delta)}{2 \mu}\right) \sinh \left(\frac{\pi (\epsilon_\Delta + \epsilon_v)}{2 \mu} - i \varphi \right)} \right]}^2
 \end{equation}
and by \eqref{inelground}, the inelastic scattering cross section is 
\begin{equation}
\sigma_\text{inelastic}= \frac{2\pi\, \text{cos}^2\varphi ~ \text{sinh} \left( \frac{\pi \epsilon_v}{\mu}\right) \text{sinh} \left( \frac{\pi \epsilon_\Delta}{\mu}\right)}{ \ev^2 \cosh^2 \left( \frac{\pi (\epsilon_\Delta - \epsilon_v)}{2 \mu}\right) \left(\cosh \left(\frac{\pi (\epsilon_v + \epsilon_\Delta)}{\mu}\right) - \cos(2 \varphi)\right)}.
\end{equation}

\subsubsection{Incoming in the Excited State}
We will impose boundary conditions such that the ingoing wave is purely in the excited state, which implies that $C_1 = 0$. We are free to set $C_3 = 1$ up to some overall normalization. 
Dividing the $E_-$ equation by the $F_-$ equation of \eqref{efeqs} yields
\begin{equation}
e^{2 i \varphi} = \frac{ C_2 \Gamma_v^* e^{ \frac{\pi \epsilon_v}{\mu}}- \eta\, \Gamma_\Delta e^{ - \frac{\pi \epsilon_\Delta}{\mu}} - C_4\,\eta^*\, \Gamma_\Delta^* e^{ \frac{\pi \epsilon_\Delta}{\mu}} }
{C_2 \Gamma_v^* e^{- \frac{\pi \epsilon_v}{\mu}}  - \eta\,\Gamma_\Delta e^{  \frac{\pi \epsilon_\Delta}{\mu}}- C_4\,\eta^*\, \Gamma_\Delta^* e^{- \frac{\pi \epsilon_\Delta}{\mu}}}\\ \\
\end{equation}
and combining this with \eqref{sum} gives
\begin{equation}
\begin{aligned}
 C_4 &=
\frac{\eta\,\Gamma_\Delta  \left[ e^{2 i \varphi} \left( e^{ \frac{\pi \epsilon_\Delta}{\mu}} +  e^{ - \frac{\pi \epsilon_v}{\mu}}\right) - \left( e^{ \frac{\pi \epsilon_v}{\mu}} +  e^{ - \frac{\pi \epsilon_\Delta}{\mu}}\right) \right]}{\eta^*\,\Gamma_\Delta^*\left[ \left(e^{ \frac{\pi \epsilon_\Delta}{\mu}} + e^{ \frac{\pi \epsilon_v}{\mu}}\right) - e^{2 i \varphi}  \left( e^{- \frac{\pi \epsilon_\Delta}{\mu}} +  e^{ - \frac{\pi \epsilon_v}{\mu}}\right)\right] }\\
 C_2  &= \frac{ -2\, \eta\,\Gamma_\Delta \, \text{sinh}\left(\frac{\pi \epsilon_\Delta }{ \mu}\right)  \left[1 +e^{2 i \varphi} \right]}{\Gamma_v^*\left[ \left( e^{ \frac{\pi \epsilon_\Delta}{\mu}} +e^{ \frac{\pi \epsilon_v}{\mu}}\right)-  e^{2 i \varphi} \left( e^{ - \frac{\pi \epsilon_\Delta}{\mu}} +  e^{ - \frac{\pi \epsilon_v}{\mu}}\right) \right] }
\end{aligned}
\end{equation}
So by \eqref{elexcited},
the elastic scattering cross section is 
\begin{equation}
\sigma_\text{elastic} =\frac{\pi}{\eD^2} \left|1 + \left(\frac{V_0}{4\mu^2}\right)^{-\frac{2 i\epsilon_\Delta}{\mu}} \left(\frac{ \Gamma_\Delta}{ \Gamma_\Delta^*} \right)   \left[\frac{\cosh\left(\frac{\pi (\epsilon_\Delta + \epsilon_v)}{2 \mu} \right) \sinh\left(\frac{\pi (\epsilon_\Delta - \epsilon_v)}{2\mu} + i \varphi \right)}{\cosh\left( \frac{\pi (\epsilon_\Delta - \epsilon_v)}{2 \mu}\right) \sinh \left(\frac{\pi (\epsilon_\Delta + \epsilon_v)}{2 \mu} - i \varphi \right)} \right] \right|^2
\end{equation}
and by \eqref{inelexcited}, the inelastic scattering cross section is 
\begin{equation}
\sigma_\text{inelastic}= \frac{2\pi\, \text{cos}^2\varphi ~ \text{sinh} \left( \frac{\pi \epsilon_v}{\mu}\right) \text{sinh} \left( \frac{\pi \epsilon_\Delta}{\mu}\right)}{\eD^2  \cosh^2 \left( \frac{\pi (\epsilon_\Delta - \epsilon_v)}{2 \mu}\right) \left(\cosh \left(\frac{\pi (\epsilon_v + \epsilon_\Delta)}{\mu}\right) - \cos(2 \varphi)\right)}.
\end{equation}
\subsubsection{Relation to the Transfer Cross Section}
\label{app:transferxsec}
When considering the effects of DM scattering on structure formation in general, the total cross section may acquire an unphysical forward divergence. Generally, the literature has instead employed the transfer cross section $\sigma_T$, which determines the longitudinal momentum transfer:
\begin{equation}
\sigma_T = \int{d\Omega \,(1 - \cos\theta)\, \frac{d \sigma}{d \Omega}}.
\end{equation}
Since our differential cross sections are angle-independent, we can pull those out of the integral. Since the $\cos\theta$ term is orthogonal to the $\sin\theta$ term in the $d\Omega$ Jacobian, the remaining integral just gives
\begin{equation}
\sigma_T = 4 \pi\, \frac{d \sigma}{d \Omega},
\end{equation}
which is the same as the cross section that we computed. (Note that \cite{Tulin:2013teo} argues for the use of the viscosity cross section instead, but computes $\sigma_T$ in order to make contact with the literature -- we follow their approach, albeit we reiterate that for the $s$-wave component of the amplitude, the distinction is a trivial one.)

\section{The Perturbative Regime}
\label{app:born}

The Born approximation can be straightforwardly applied to this multi-state system, by writing the three-dimensional Schr{\"o}dinger equation in the form:

\begin{equation} \Psi(\vec{r}) = \Psi_0(\vec{r}) - \frac{m_\chi}{4 \pi} \int \frac{d^3 \vec{r_0}}{|\vec{r} - \vec{r_0}|}  \left( \begin{matrix} e^{i k |\vec{r} - \vec{r_0}|} & 0 \\ 0 & e^{i k^\prime |\vec{r} - \vec{r_0}|} \end{matrix}\right) \cdot \bar{V}(\vec{r_0}) \cdot \Psi(\vec{r_0}).\end{equation}

Here $\Psi(\vec{r}) = \left( \begin{matrix} \Psi_1(\vec{r_0}) \\ \Psi_2(\vec{r_0}) \end{matrix}\right)$ is the full, three-dimensional physical wavefunction for the two-body state; for the spherically symmetric potential and $s$-wave scattering we consider, it is related to the wavefunction $\psi(r)$ by $\psi(r) = r \Psi(\vec{r})$. $\Psi_0(\vec{r})$ is the equivalent wavefunction for the unperturbed system with no potential, satisfying $\nabla^2 \Psi_0(\vec{r}) = - \left( \begin{matrix} k^2 & 0 \\ 0 & k^{\prime 2} \end{matrix}\right)  \cdot \Psi_0(\vec{r})$. (Note, since $\Psi(\vec{r})$ describes a two-body state, we have replaced the mass in the Schr{\"o}dinger equation with the reduced mass $m_\chi/2$.) $\bar{V}(\vec{r})$ is the matrix potential given in Eq. \ref{schrod}, but now with the diagonal $2 \delta c^2$ term omitted (we instead incorporate the effect of this term in the difference between $k^\prime$ and $k$). Here we use units where $c=\hbar=1$.

The first-order Born contribution to the scattering amplitude is solely inelastic, since the matrix $\bar{V}$ is purely off-diagonal. For an initial state consisting of a plane wave in a single state (purely for notational purposes, we here choose the upper row of the matrix to correspond to the initial state), we obtain:

\begin{align}  \Psi(\vec{r}) & = \left( \begin{matrix} A e^{i k z} \\ 0 \end{matrix} \right) - \frac{m_\chi}{4 \pi} \int \frac{d^3 \vec{r_0}}{|\vec{r} - \vec{r_0}|}  \left( \begin{matrix} e^{i k |\vec{r} - \vec{r_0}|} & 0 \\ 0 & e^{i k^\prime |\vec{r} - \vec{r_0}|} \end{matrix}\right) \cdot \bar{V}(\vec{r_0}) \cdot \left( \begin{matrix} A e^{i k z_0} \\ 0 \end{matrix} \right) \nonumber \\
 & = \left( \begin{matrix} A e^{i k z} \\ 0 \end{matrix} \right) - \frac{m_\chi}{4 \pi} \int \frac{d^3 \vec{r_0}}{|\vec{r} - \vec{r_0}|}  e^{i k^\prime |\vec{r} - \vec{r_0}|} \left(-\frac{\alpha}{r} e^{-m_\phi r} \right) \left(\begin{matrix} 0 \\ A e^{i k z_0} \end{matrix}\right).\end{align}
 
 Writing $\vec{k} = k \hat{r}$, $\vec{k^\prime} = k^\prime \hat{r}$, $\vec{p} = k \hat{z}$, $\vec{p^\prime} = k^\prime \hat{z}$, and making the long-distance approximation $e^{i k |\vec{r} - \vec{r_0}|} / |\vec{r} - \vec{r_0}| \approx (e^{i k r}/r) e^{-i \vec{k} \cdot \vec{r_0}}$, we obtain:
 
 \begin{align}  \Psi(\vec{r}) & \approx A \left[ \left( \begin{matrix} e^{i k z} \\ 0 \end{matrix} \right) + \frac{\alpha m_\chi}{4 \pi} \frac{e^{i k^\prime r}}{r} \left( \begin{matrix} 0 \\ 1 \end{matrix} \right)\int d^3 \vec{r_0}  \left(\frac{1}{r} e^{-m_\phi r} \right) e^{i (\vec{p} - \vec{k}^\prime) \cdot \vec{r_0}} \right] \nonumber \\
 & = A \left[ \left( \begin{matrix} e^{i k z} \\ 0 \end{matrix} \right) + \frac{\alpha m_\chi}{4 \pi} \frac{e^{i k^\prime r}}{r} \left( \begin{matrix} 0 \\ 1 \end{matrix} \right) \frac{4 \pi}{|\vec{p} - \vec{k^\prime}|^2 + m_\phi^2} \right]. \end{align}

We can read off the scattering amplitudes as (in the notation of Appendix \ref{sec:scattering}):

\begin{equation} f_X = 0, \quad f_Y = \frac{\alpha m_\chi}{|\vec{p} - \vec{k^\prime}|^2 + m_\phi^2}.\end{equation}
 
Now $|\vec{p} - \vec{k}^\prime|^2 = k^2 + (k^\prime)^2 - 2 k k^\prime \cos \theta$, where $\theta$ describes the angle between $\hat{r}$ and $\hat{z}$, and as in Appendix \ref{sec:scattering} we have:

\begin{equation} \frac{d\sigma}{d\Omega} = \frac{k^\prime}{k} f_Y^2 = \frac{k^\prime}{k} \frac{\alpha^2 m_\chi^2}{\left(k^2 + (k^\prime)^2 - 2 k k^\prime \cos\theta + m_\phi^2\right)^2}.\end{equation}

Performing the angular integral  yields:

\begin{equation} \sigma = \frac{k^\prime}{k} \frac{4 \pi \alpha^2 m_\chi^2}{\left(m_\phi^2 + (k^\prime)^2 - k^2\right)^2 + 4 k^2 m_\phi^2}.\end{equation}

This result applies to both particles initially in the ground or excited states, with the proper choices of $k$ and $k^\prime$. For upscattering, $k = m_\chi v$ and $k^\prime = \sqrt{m_\chi^2 v^2 - 2 \delta m_\chi}$, where $v$ is the speed of a single particle in the center-of-mass frame in the initial state. If we use the convention in the body of the text where $v$ always refers to the speed of the particle in the ground state, then for downscattering these choices of $k$ and $k^\prime$ are simply reversed. (If instead we take $v$ to be the speed of the particle in the initial state, for downscattering we have $k = m_\chi v$ and $k^\prime = \sqrt{m_\chi^2 v^2 + 2 \delta m_\chi}$.)

Thus we obtain the upscattering cross section,

\begin{equation} \sigma =  \frac{4 \pi \alpha^2 m_\chi^2 \sqrt{1 - \frac{2\delta}{m_\chi v^2}}}{m_\phi^4 \left[\left(1 - 2 \delta m_\chi/m_\phi^2\right)^2 + 4 m_\chi^2 v^2/m_\phi^2 \right]} \label{eq:bornup} \end{equation}

and the downscattering cross section (taking $v$ to be the speed of the ground-state particle):

\begin{equation} \sigma =  \frac{4 \pi \alpha^2 m_\chi^2}{ \sqrt{1 - \frac{2\delta}{m_\chi v^2}} m_\phi^4 \left[\left(1 - 2 \delta m_\chi/m_\phi^2\right)^2 + 4 m_\chi^2 v^2/m_\phi^2 \right]}. \label{eq:borndown} \end{equation}

The corresponding dimensionless cross sections are respectively:

\begin{equation} \sigma = \frac{\epsilon_\Delta}{\epsilon_v} \frac{4 \pi}{\epsilon_\phi^4 (  1 - \epsilon_\delta^2/\epsilon_\phi^2)^2 + 4 \epsilon_v^2/\epsilon_\phi^2}, \quad  \sigma = \frac{\epsilon_v}{\epsilon_\Delta} \frac{4 \pi}{\epsilon_\phi^4 (  1 - \epsilon_\delta^2/\epsilon_\phi^2)^2 + 4 \epsilon_v^2/\epsilon_\phi^2}.\end{equation}

The lowest-order contribution to elastic scattering comes via the second term in the Born series. For the initial condition above, the contribution to the wavefunction is:

\begin{align} \Psi_\text{2nd-order}(\vec{r}) & = \left(\frac{\alpha m_\chi}{4\pi} \right)^2 \int \frac{d^3 \vec{r_0}}{|\vec{r} - \vec{r_0}|}   \left( \begin{matrix} e^{i k |\vec{r} - \vec{r_0}|} & 0 \\ 0 & e^{i k^\prime |\vec{r} - \vec{r_0}|} \end{matrix}\right) \cdot \bar{V}(\vec{r_0}) \nonumber \\
& \cdot \int \frac{d^3 \vec{r}_1}{|\vec{r_0} - \vec{r_1}|} \left( \begin{matrix} e^{i k |\vec{r_0} - \vec{r_1}|} & 0 \\ 0 & e^{i k^\prime |\vec{r_0} - \vec{r_1}|} \end{matrix}\right) \cdot \bar{V}(\vec{r_1}) \cdot \Psi_0(\vec{r_1}) \nonumber \\
& = A \left(\frac{\alpha m_\chi}{4\pi} \right)^2 \left(\begin{matrix} 1 \\ 0 \end{matrix} \right) \int \frac{d^3 \vec{r_0}}{|\vec{r} - \vec{r_0}|}  e^{i k |\vec{r} - \vec{r_0}|}  \frac{e^{-m_\phi r_0}}{r_0} \int \frac{d^3 \vec{r}_1}{|\vec{r_0} - \vec{r_1}|} e^{i k^\prime |\vec{r_0} - \vec{r_1}|}  \frac{e^{-m_\phi r_1}}{r_1} e^{i k z_1}.  \end{align}

Again making the long-distance approximation we can read off the scattering amplitudes (at this order) as:

\begin{equation} f_X = \left(\frac{\alpha m_\chi}{4\pi} \right)^2  \int d^3 \vec{r_0}  e^{-i \vec{k} \cdot \vec{r_0}} \frac{e^{-m_\phi r_0}}{r_0} \int \frac{d^3 \vec{r}_1}{|\vec{r_0} - \vec{r_1}|} e^{i k^\prime |\vec{r_0} - \vec{r_1}|}  \frac{e^{-m_\phi r_1}}{r_1} e^{i k z_1}, \quad f_Y = 0.\end{equation}

In the limit where the initial particle is very slow-moving, so we can take $k=0$ (but not $k^\prime=0$), this amplitude has a simple analytic form: for the integral over $\vec{r_1}$, we can choose coordinates where $\vec{r_0}$ lies along the $z$-axis to simplify the angular integral, obtaining:

\begin{equation} \int \frac{d^3 \vec{r}_1}{|\vec{r_0} - \vec{r_1}|} e^{i k^\prime |\vec{r_0} - \vec{r_1}|}  \frac{e^{-m_\phi r_1}}{r_1} = \frac{4\pi}{m_\phi^2 + (k^\prime)^2} \frac{e^{i k^\prime r_0} - e^{-m_\phi r_0}}{r_0}. \end{equation}

The amplitude in the $k=0$ limit then becomes,

\begin{equation} f_X = \left(\frac{\alpha m_\chi}{4\pi} \right)^2 \frac{(4\pi)^2}{m_\phi^2 + (k^\prime)^2} \left( \frac{1}{m_\phi - i k^\prime} - \frac{1}{2 m_\phi} \right).\end{equation}

For scattering where the initial particles are in the excited state, $k^\prime = \sqrt{2 \delta m_\chi}$ for $k=0$, and the amplitude is given by:

\begin{equation} \frac{d\sigma}{d\Omega} = |f_X|^2 = \frac{\alpha^4}{4 m_\phi^2} \left( \frac{m_\chi^2}{m_\phi^2 + 2 \delta m_\chi} \right)^2 .\end{equation}

For scattering where the initial particles are in the ground state and $k=0$, $k^\prime = i \sqrt{2 \delta m_\chi}$, and we obtain:

\begin{equation} \frac{d\sigma}{d\Omega} = |f_X|^2 = \frac{\alpha^4}{4 m_\phi^2} \left( \frac{m_\chi^2}{m_\phi^2 - 2 \delta m_\chi} \right)^2 \left(\frac{m_\phi - \sqrt{2 \delta m_\chi}}{m_\phi + \sqrt{2 \delta m_\chi}} \right)^2 = \frac{\alpha^4}{4 m_\phi^2}  \frac{m_\chi^4}{\left(m_\phi + \sqrt{2 \delta m_\chi}\right)^4} .\end{equation}

Thus the cross sections for ground-state elastic scattering and excited-state elastic scattering, in the limit of slow-moving initial particles, are respectively,

\begin{equation} \sigma = \frac{\pi \alpha^4 m_\chi^4}{m_\phi^2 (m_\phi + \sqrt{2 \delta m_\chi})^4}, \quad \sigma = \frac{\pi \alpha^4 m_\chi^4}{m_\phi^2 (m_\phi^2 + 2 \delta m_\chi)^2}. \label{eq:bornelastic} \end{equation}

The corresponding dimensionless cross sections are:

\begin{equation} \sigma = \frac{\pi}{\epsilon_\phi^6} \frac{1}{\left(1 + \epsilon_\delta/\epsilon_\phi\right)^4}, \quad \sigma = \frac{\pi}{\epsilon_\phi^6} \frac{1}{\left(1 + \epsilon_\delta^2/\epsilon_\phi^2\right)^2}. \end{equation}

Note that these second-order cross sections become comparable to (or larger than) the first-order inelastic scattering cross sections for $\epsilon_\phi \lesssim 1$, signaling the breakdown of the Born approximation and the need to transition to the resonant regime described in the main text.

\section{Comparing our $\delta \rightarrow 0$ Limit with Previous Results}
\label{app:degeneratelimit}
An analytic approximate form for the phase shift due to elastic scattering, for both attractive and repulsive potentials, has previously been presented in the literature \cite{Tulin:2013teo}. We show here how to recover the analogous result in our approximation.

In the $\delta \rightarrow 0$ limit, the potential matrix can be diagonalized, yielding the exact eigenstate basis $\psi_+ = \frac{1}{\sqrt{2}} \left(-1, 1\right)$, $\psi_- = \frac{1}{\sqrt{2}} \left(1, 1\right)$ (as in Eq. \ref{eq:approxev}). Since the potential is now diagonal, scatterings from $\psi_+$ to the $\psi_-$ (and vice versa) do not occur: the two eigenstates are decoupled. The $\psi_+$ and $\psi_-$ eigenstates experience, respectively, a repulsive and attractive potential.

For the $-$ state, let the scattering solution $\phi_-(r)$ (for the coefficient of the $\psi_-$ eigenvector) have the asymptotic form $\phi_-(r) = e^{i (k r + 2 \delta_-)} + e^{-i k r}$. The phase shift $\delta_-$ characterizes the scattering amplitude, which is given by $\left|1 - e^{2 i\delta_-}\right|^2$. Likewise, for the $+$ state, let the phase shift be $\delta_+$.

Since the differential equation is linear, any linear combination of these solutions ($A \phi_-(r) \psi_- + B \phi_+(r) \psi_+$) is also a solution. In particular, if we set $A=B=1/\sqrt{2}$ and $A=-B=1/\sqrt{2}$, we obtain the two solutions:
\begin{equation} \psi(r) =  \left( \begin{matrix}e^{ikr} \left(\frac{e^{2 i \delta_-} - e^{2 i \delta_+}}{2}\right) \\ e^{ikr} \left(\frac{e^{2 i \delta_-} + e^{2 i \delta_+}}{2}  \right) + e^{- i k r} \end{matrix} \right), \quad \psi(r) =  \left( \begin{matrix}e^{ikr} \left(\frac{e^{2 i \delta_-} + e^{2 i \delta_+}}{2}\right) + e^{-i k r} \\ e^{ikr} \left(\frac{e^{2 i \delta_-} - e^{2 i \delta_+}}{2}  \right) \end{matrix} \right). \end{equation}
These correspond to the cases we studied above, where the particles are initially purely in the ground or excited states. So we see that by calculating the phase shifts for these initial conditions (given by the $A$, $B$ and $C$ coefficients in Eq. \ref{eq:AB} and Eq. \ref{eq:C}), we can recover the values for $\delta_-$ and $\delta_+$, and vice versa. Our cross sections are given by:
\begin{equation} \sigma\mathrm{(ground \rightarrow ground)} = \frac{\pi}{k^2} \left|1 - \frac{e^{2 i \delta_-} + e^{2 i \delta_+}}{2} \right|^2, \quad \sigma\mathrm{(ground \rightarrow excited)} = \frac{\pi}{k^2} \left|\frac{e^{2 i \delta_-} - e^{2 i \delta_+}}{2}  \right|^2, \end{equation}
and in this case, since $\delta=0$, swapping the identifications of ``ground'' and ``excited'' states has no effect. Note that the sum of these cross sections gives $\sigma_\mathrm{tot} = \frac{\pi}{k^2} \left(\left|1 - e^{2 i\delta_-}\right|^2 + \left|1 - e^{2 i\delta_+}\right|^2 \right) = \sigma_- + \sigma_+$, as it must --- the total scattering rate cannot depend on the choice of basis.

In the limit where the phase shifts are small (which we will see is the case at low velocities and away from resonances), we can expand:
\begin{equation} \sigma\mathrm{(ground \rightarrow ground)} \approx \frac{\pi}{k^2} \left| \delta_- + \delta_+ \right|^2, \quad \sigma\mathrm{(ground \rightarrow excited)} = \frac{\pi}{k^2} \left| \delta_- -  \delta_+  \right|^2. \end{equation}
The authors of \cite{Tulin:2013teo} define a quantity $a$ which corresponds to our $\epsilon_v$, and a quantity $c = 1/(\kappa \epsilon_\phi)$, where $\kappa$ is set to 1.6.

The phase shifts derived for the repulsive and attractive case by \cite{Tulin:2013teo} in the low-velocity limit are given by:
\begin{equation} \delta_- \approx - [2 \gamma + \psi(1 + \sqrt{c}) + \psi(1 - \sqrt{c})] a c, \quad \delta_+ \approx - [2 \gamma + \psi(1 + i \sqrt{c}) + \psi(1 - i \sqrt{c})] a c, \end{equation}
where $\gamma$ is the Euler-Mascheroni constant and $\psi(z)$ is the digamma function. Note these phase shifts become small in the low-velocity limit due to the scaling with $a$, as claimed above.

The asymptotic expansions of the digamma function, as $|z| \rightarrow \infty$, are $\psi(z) \approx \ln(z) + i \pi (i \cot(\pi z) - 1) \left \lfloor \,\abs{\arg(z)}/\pi \right \rfloor$ for $z$ not a negative integer. Thus in the large-$c$ limit (corresponding to $\epsilon_\phi \ll 1$, which is necessary for our approximations to hold), and neglecting terms of $\mathcal{O}(1/c)$ and higher, these phase shifts approach:
\begin{equation} \delta_- =  -\left[2 \gamma + \ln(c) + \pi \cot(\pi \sqrt{c}) \right] a c, \quad \delta_+ = -\left[2 \gamma + \ln(c) \right] a c.\end{equation}
So the cross sections for ground-ground and ground-excited scattering should be set by:
\begin{align}  \sigma(\text{ground}\rightarrow \text{ground}) & =  \pi \left[4 \gamma + 2 \ln(c) + \pi \cot(\pi \sqrt{c}) \right]^2 c^2,  \nonumber \\ 
 \sigma(\text{ground}\rightarrow \text{excited}) & = \pi^3 \cot^2(\pi \sqrt{c}) c^2.\end{align}
(Note the prefactor $1/k^2$ has canceled out the factors of $a$ in the phase shifts.)

We now take the same limits (first $\delta \rightarrow 0$, and then $v \rightarrow 0$) in our semi-analytic approximation. Setting $\ev = \eD$ we obtain: 
\begin{equation}
\Gamma_v \rightarrow \sqrt{\pi} ~\Gamma\left(1 + \frac{i \ev}{\mu}\right) \Gamma\left(\frac{1}{2} + \frac{i \ev} {2}\right) = \pi \,2^{-\frac{2 i \ev}{\mu} } \Gamma\left( 1 + \frac{i \ev}{\mu}\right), 
\end{equation}
where the second equality comes from the gamma function identities $\Gamma(1 + z) = z\, \Gamma(z)$, $\Gamma(1-z)\,\Gamma(z) = \pi / \sin(\pi z)$, and $\Gamma(z) \Gamma(z+1/2) = 2^{1 - 2z} \sqrt{\pi}\, \Gamma(2z)$. So the elastic scattering cross section to first order in $\ed$ (for both the ground and excited state, since they are now degenerate) becomes:
\begin{equation}
\sigma_\text{elastic}= \frac{\pi}{\ev^2}\abs{ 1 + \frac{\Gamma\left(1 + \frac{i \epsilon_v}{\mu} \right)}{\Gamma\left(1 - \frac{i \epsilon_v}{\mu} \right)} \left(\frac{i \sin \varphi \cosh\left(\frac{\pi \epsilon_v}{\mu}\right)}{\sinh\left(\frac{\pi \epsilon_v}{ \mu} - i \varphi \right)}\right) \left(\frac{V_0}{\mu^2} \right)^{-2 i\epsilon_v/\mu} }^2.
\end{equation}
Off-resonance, in the limit as $\epsilon_v \rightarrow 0$, $\sinh\left(\frac{\pi \epsilon_v}{ \mu} - i \varphi \right) \rightarrow - i \sin \varphi$, and the scattering amplitude approaches zero. More precisely, a Taylor expansion yields:
\begin{equation}\sigma_\text{elastic} \rightarrow \frac{\pi}{\mu^2} \left[ 2\gamma + 2 \ln\left(\frac{V_0}{\mu^2} \right) + \pi \cot \varphi \right]^2,\end{equation}
where $\gamma$ is the Euler-Mascheroni constant.
On-resonance, where $\sin \varphi = 0$, the elastic scattering amplitude is simply 1.

Meanwhile, for the inelastic case, setting  $\ev = \eD$ yields
\begin{equation}
\sigma_\text{inelastic}= \frac{2 \pi\, \text{cos}^2\varphi ~ \text{sinh}^2 \left( \frac{\pi \epsilon_v}{\mu}\right) }{\ev^2\left( \cosh \left(\frac{2 \pi \epsilon_v }{\mu}\right) - \cos(2 \varphi)\right)}.
 \end{equation}
 Off-resonance, as $\epsilon_v \rightarrow 0$, this probability approaches \begin{equation}\sigma_\text{inelastic} \rightarrow \pi \left( \frac{\pi  \cot\varphi}{\mu} \right)^2.\end{equation} On-resonance, where $\cos(2\varphi) = 1$, the inelastic scattering amplitude simply approaches 1.
 
 We see that these would agree precisely with the approximate forms of the cross sections derived from the results of \cite{Tulin:2013teo} if we made the replacements:
 
 \begin{equation} \mu \rightarrow 1/c, \quad \varphi \rightarrow \pi \sqrt{c}, \quad \ln (V_0/\mu) \rightarrow \gamma.\end{equation} 
 
These replacements are parametrically correct --- $\mu \sim \epsilon_\phi \sim 1/c$ up to $\mathcal{O}(1)$ factors, likewise $V_0 \sim \mu$ up to $\mathcal{O}(1)$ corrections. The results are most sensitive to the identification $\varphi \rightarrow \pi \sqrt{c}$, since this sets the resonance positions: taking $c = 1/(\kappa \epsilon_\phi)$, and our approximate expression $\varphi \sim \sqrt{2 \pi/\epsilon_\phi}$, we see that they agree exactly if $\kappa = \pi/2 \approx 1.57$. The value of $\kappa = 1.6$ chosen by \cite{Tulin:2013teo} therefore leads to percent-level agreement in the resonance positions.

Perfect agreement between the two analyses should not be expected, since they use different potentials (albeit with similar properties), but our approach agrees both qualitatively and quantitatively with the results of \cite{Tulin:2013teo} in the region of parameter space where they can both be used.
\section{The High-Velocity Regime}
\label{beyond}
\subsection{Higher Partial Waves}

\begin{figure}[h!]
\vspace{0.2cm}\center{\includegraphics[width=0.49\textwidth]{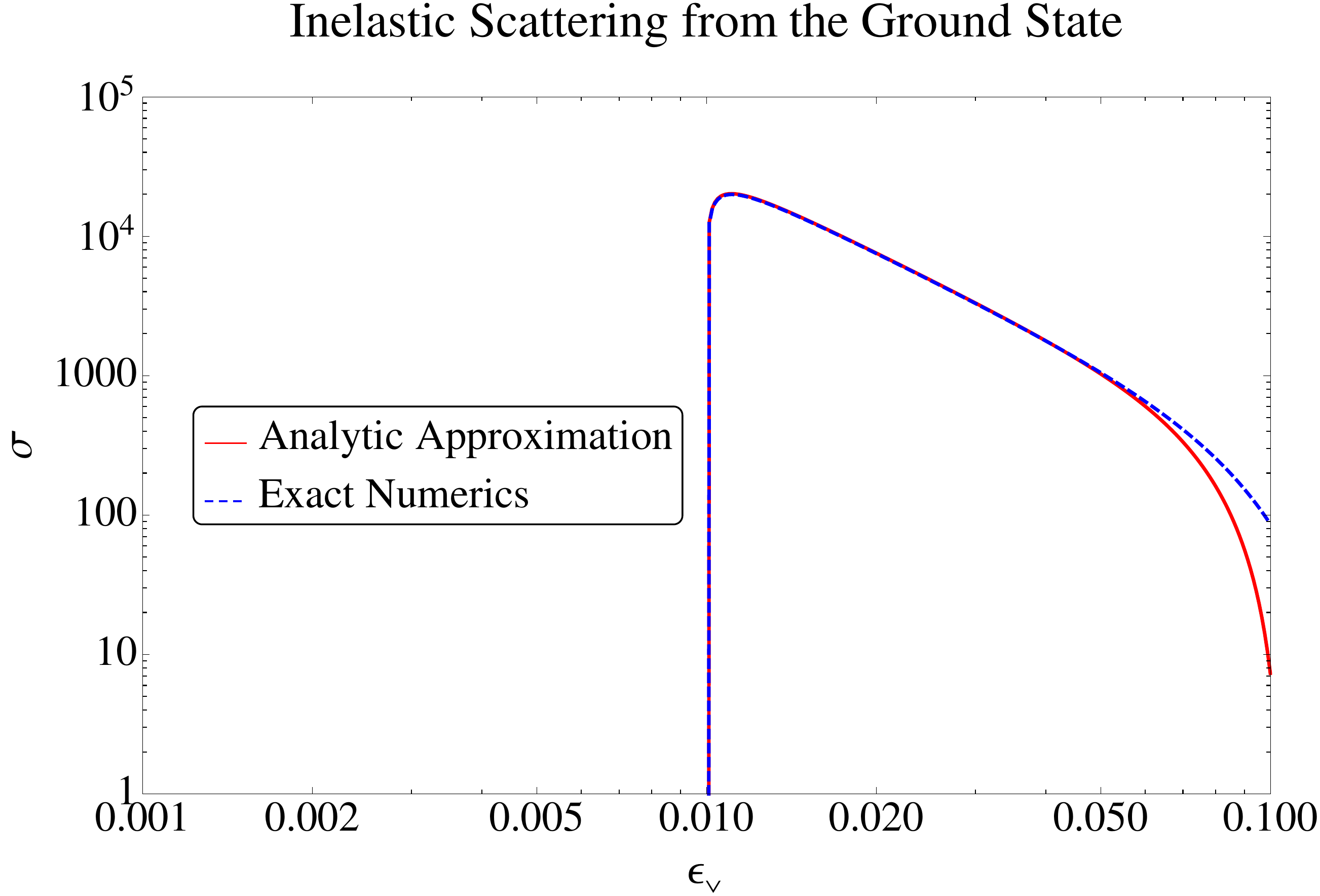}
\includegraphics[width=0.49\textwidth]{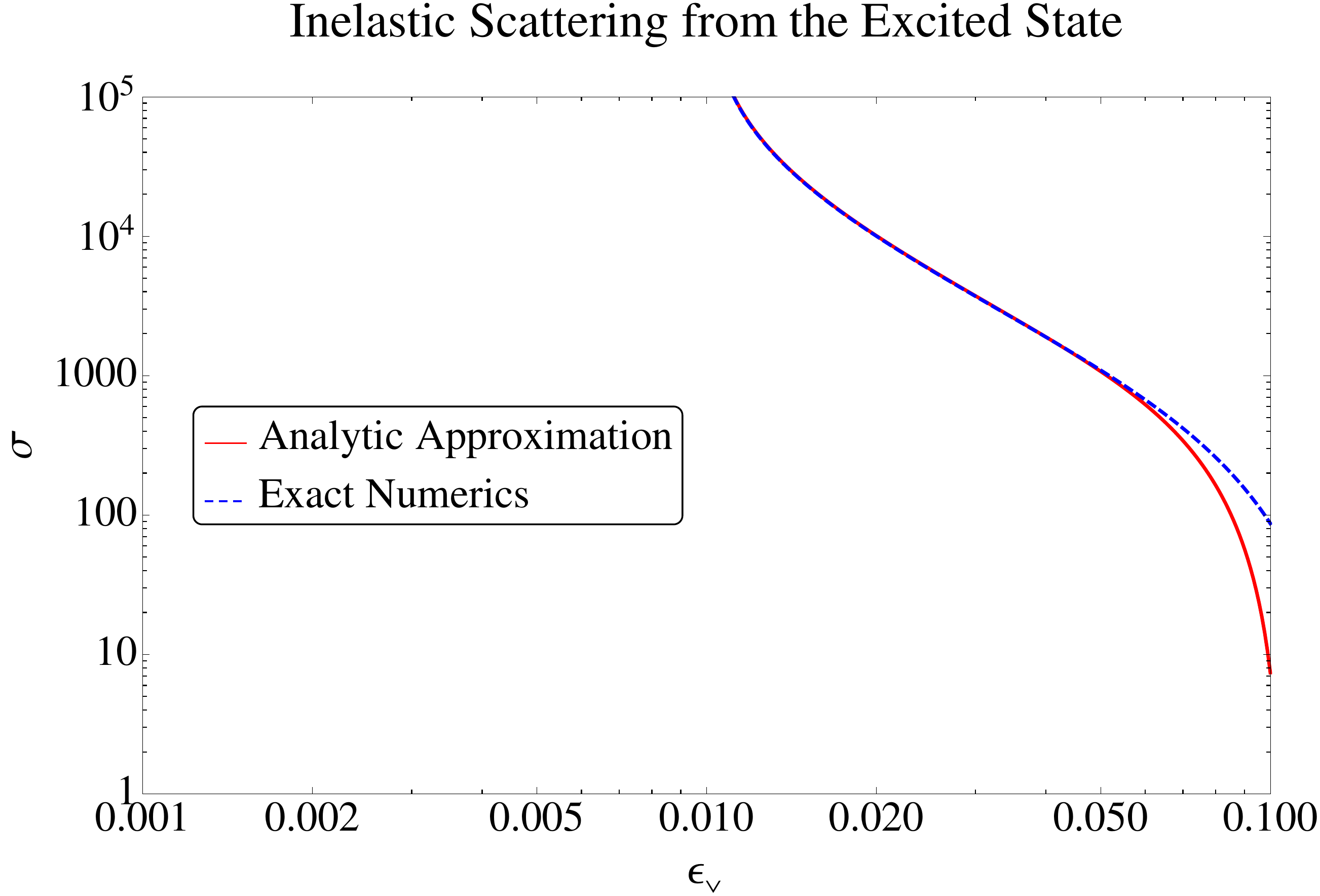}}\\
\caption{A comparison of the numerical vs analytic results for $s$-wave inelastic scattering, for upscattering (left) and downscattering (right), with $\ed$ = 0.01 and $\ephi$ = 0.04. The two diverge for $\ev \gtrsim \ephi$, as expected.}
\label{fig:inel}
\end{figure} 

Figure \ref{fig:inel} shows a particular slice through the parameter space of Figure \ref{fig:comp1}, for inelastic scattering, but extending to $\ev > \ephi$. As expected, our analytic approximation breaks down in this regime (note that $\mu$ and $\epsilon_\phi$ are generally equal up to a $\mathcal{O}(1)$ factor). Conveniently, $\epsilon_v \lesssim \epsilon_\phi$ is also precisely the condition for $s$-wave scattering to dominate over the higher partial waves. Consequently, while a more careful treatment of the matching between the WKB and large-$r$ regimes (see Figure \ref{fig:matching}) might allow extension of our approximation for the $s$-wave to the region with $\epsilon_v \gtrsim \epsilon_\phi$, at that point it would be necessary to include the higher partial waves as well.

This can be easily seen by comparing the relevant length scales: for the $\ell$th partial wave, the vacuum solution is proportional to the Bessel function $j_\ell(\epsilon_v r)$, which peaks when $\ell \sim \epsilon_v r$, i.e. $r \sim \ell/\epsilon_v$. In order for scattering of the $\ell$th partial wave to be significant, this peak must lie within the range of the potential, i.e. $r \lesssim 1/\epsilon_\phi$, and so we must have $\ell \lesssim \epsilon_v/\epsilon_\phi$. If $\epsilon_v/\epsilon_\phi \lesssim 1$, then only the $s$-wave term can penetrate the potential far enough to experience significant scattering.

In the case where $\delta$ is non-zero, this argument still holds --- for particles in the excited state, the asymptotic wave function is now $j_\ell(\sqrt{\epsilon_v^2 - \epsilon_\delta^2})$, but since $\sqrt{\epsilon_v^2 - \epsilon_\delta^2} < \epsilon_v$, requiring $\epsilon_v \ll \epsilon_\phi$ is certainly \emph{sufficient} to ensure that the potential cuts off at smaller $r$ than the peak of the higher-$\ell$ wavefunctions. Since whenever particles in the excited state are present and the scattering rate is significant, their downscatterings will populate the ground state, we will generally consider $\epsilon_v \lesssim \epsilon_\phi$ to be both a necessary and sufficient condition for our approximate solution to be useful.

%
%
\subsection{The Adiabatic Regime}
\label{adiaba}
However, there is a regime where $\epsilon_v \gtrsim \epsilon_\phi$ but our approximate solution remains valid, although the $s$-wave does not generally dominate scattering in this part of parameter space, and so we caution that our $s$-wave result should not be used as a proxy for the total scattering cross section. However, it can be used as a lower bound.

As discussed in Appendix \ref{sec:scattering}, our method essentially neglects repulsive scattering at small distances, which is valid when the range of the potential is relatively short and so the scattering wavefunction for the repulsed eigenstate is peaked outside its range (note this is the same reason we can ignore the higher partial waves in this regime).

There is another regime where this approximation is valid, for a different reason. Suppose the system starts with both dark matter particles in the ground state (i.e. the state of lowest energy). At short distances, the lowest-energy eigenstate is the one that experiences an attractive potential (corresponding to the $+-$ two-body state at high energies). If the transition from long distances to short distances is adiabatic --- i.e. this transition occurs slowly relative to the scale associated with the splitting between the eigenstates --- then particles in the lowest-energy eigenstate at long distances will find themselves entirely in the attracted eigenstate at short distances, in analogy to the adiabatic theorem, and ignoring the repulsed eigenstate will be valid because it will simply never be populated. 

The splitting between the eigenstates corresponds to an energy scale of $\epsilon_\delta^2$ in our dimensionless coordinates, and hence to a time scale of $1/\epsilon_\delta^2$; the corresponding distance scale, for an inward-moving wavepacket, would be $\sim \epsilon_v/\epsilon_\delta^2$. The rotation of the eigenstates (from mass eigenstates to gauge eigenstates of the high-energy potential), as described in Section \ref{sec:modelintro}, occurs when $V(r) = e^{-\ephi r}/r$ becomes comparable to $\epsilon_\delta^2/2$. If the cause of this transition is the exponential cutoff, i.e. $r \sim 1/\ephi$, then the transition occurs over a radius $\Delta r \sim 1/\ephi$; if $\ephi \lesssim \epsilon_\delta^2$ then it occurs when $V(r) \sim 1/r$ and over a range $\Delta r \sim 1/\epsilon_\delta^2$. \begin{figure}[htb]
\center{\includegraphics[width=0.5\textwidth]{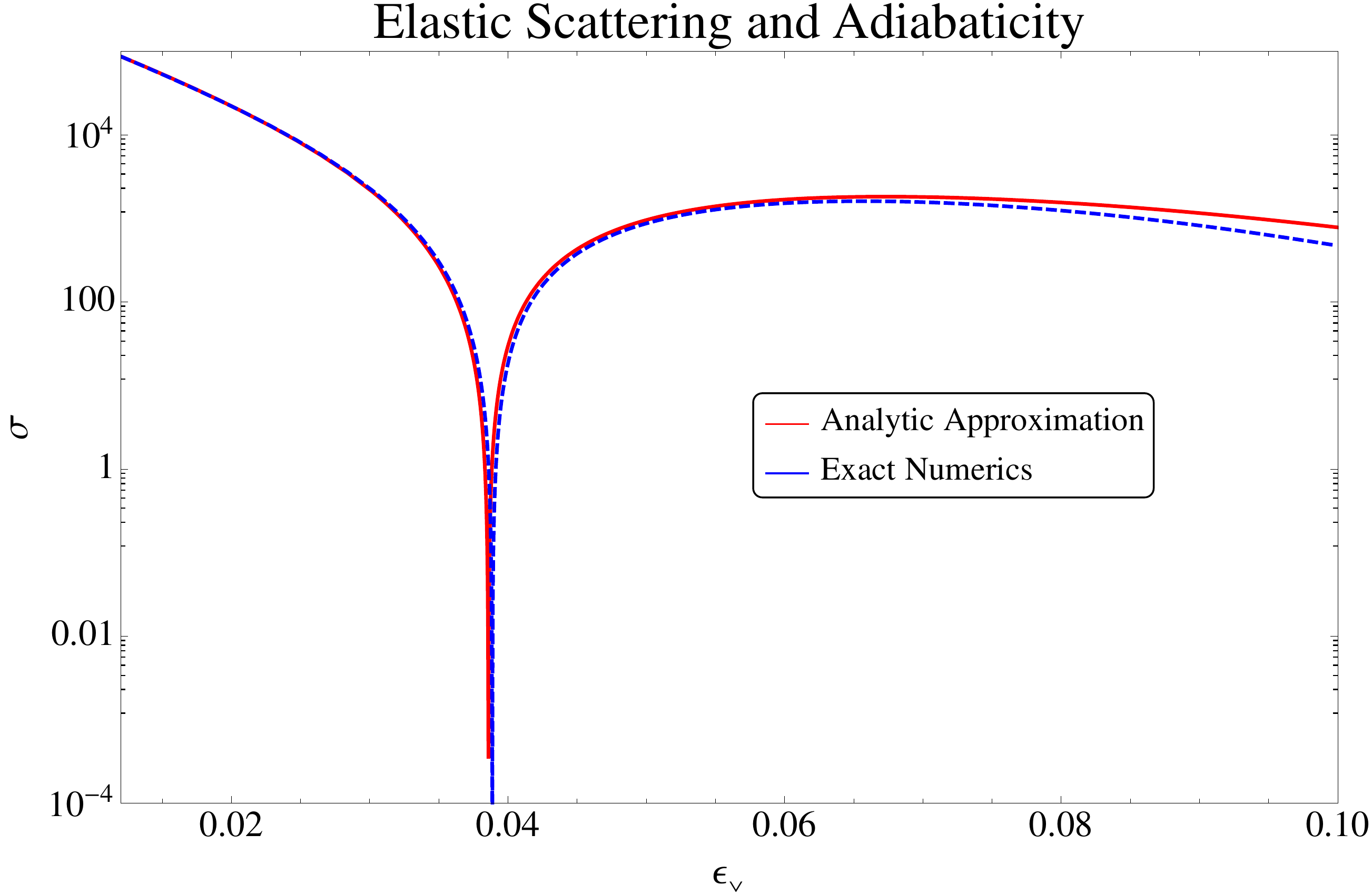}}
\caption{A scan through $\ev$ with $\ephi= 0.03$ and $\ed = 0.04$. This demonstrates the shift from the transition to small-$r$ being adiabatic vs. nonadiabatic. We can see the breakdown near $\ed^2 \sim \ev \ephi$, which happens near $\ev \sim 0.06$.}
\label{fig:adiabatic}
\end{figure}

So the criterion for adiabaticity is $\epsilon_v/\epsilon_\delta^2 \lesssim 1/\ephi$ if $\ephi \gtrsim \epsilon_\delta^2$, or $\epsilon_v/\epsilon_\delta^2 \lesssim 1/\epsilon_\delta^2$ otherwise. In the first case, the transition is adiabatic for $\ev \ephi \lesssim \ed^2$; in the second case adiabaticity always holds for $\epsilon_v \lesssim 1$ (however, note that $\ephi \lesssim \epsilon_\delta^2$ is a regime where the approximations we use are known to be less accurate \cite{Slatyer:2009vg}), which by the condition $\ephi \lesssim \ed^2$ implies $\ephi \ev \lesssim \ed^2$. We can summarize this by saying adiabaticity holds if and only if $\ephi \ev \lesssim \ed^2$, provided our other assumptions hold (that is, $\ev, \ed, \ephi \lesssim 1$).

This mechanism is also responsible for the enhancement in the annihilation rate noted in \cite{Slatyer:2009vg} for the case with a mass splitting, compared to the case where the mass splitting is negligible relative to the kinetic energy and can be ignored; under adiabatic conditions, the presence of the mass splitting causes particles initially in the ground state to transition into a purely attracted state, rather than a equal linear combination of attracted and repulsed states.

This argument cannot be applied to scattering from the excited state or into the excited state, as if the excited state is populated then this implies the repulsed eigenstate will also be populated and cannot be ignored. But provided we are only interested in elastic scattering from the ground state (i.e. for the below-threshold case $\epsilon_v \lesssim \epsilon_\delta$), we expect our results to be accurate (for the $s$-wave) even when $\epsilon_v \gtrsim \epsilon_\phi$, in the event that $\ephi \ev \lesssim \ed^2$.

\bibliographystyle{unsrt}
\bibliography{InelasticDM_v9}
\end{document}